\documentclass[preprints,rreview,accept,oneauthor]{mdpi}  
\firstpage{1} 
\pubvolume{1}
\issuenum{1}
\articlenumber{0}
\pubyear{2025}
\copyrightyear{2025}
\datereceived{ } 
\daterevised{ } 
\dateaccepted{ } 
\datepublished{ } 
\hreflink{https://doi.org/} 

\usepackage{amscd,amsmath,amssymb,amsfonts,xspace,mathrsfs,upgreek}
\usepackage{dsfont}
\usepackage{mathtools}
\usepackage{bbm}
\usepackage[bbgreekl]{mathbbol}

\def\bea{\begin{eqnarray}}
	\def\eea{\end{eqnarray}}
\def\be{\begin{equation}}
	\def\ee{\end{equation}}
\def\ba{\begin{align}}
	\def\ea{\end{align}}
\def\bse{\begin{subequations}}
	\def\ese{\end{subequations}}

\newtheorem{conj}{Conjecture}

\numberwithin{equation}{section}
\numberwithin{theorem}{section}
\numberwithin{proposition}{section}
\numberwithin{corollary}{section}
\numberwithin{lemma}{section}
\numberwithin{conj}{section}
\numberwithin{assumption}{section}
\numberwithin{remark}{section}
\numberwithin{example}{section}
\numberwithin{definition}{section}

\def\det{\,{\rm det}\, }
\def\diag{{\rm diag}}
\def\sign{{\rm sgn}}

\def\Ch{{\rm Ch}}

\def\Sym{\,{\rm Sym}\, }

\def\Span{\,{\rm Span}\, }

\def\Re{\,{\rm Re}\,}

\DeclareMathOperator{\Td}{Td}
\DeclareMathOperator{\ch}{ch}

\DeclareMathOperator{\Erf}{Erf}
\DeclareMathOperator{\Erfc}{Erfc}
\newcommand{\Tr}{\mbox{Tr}}
\newcommand{\Pf}{\mbox{Pf}}
\newcommand{\sgn}{\mbox{sgn}}
\newcommand{\rank}{\mbox{rank}}

\newcommand{\CP}{\IC P^1}

\def\({\left(}
\def\){\right)}
\def\[{\left[}
\def\]{\right]}
\def\<{\left\langle}
\def\>{\right\rangle}
\def\hf{{1\over 2}}
\def\haf{\textstyle{1\over 2}}
\newcommand{\p}{\partial}

\newcommand\PT{\operatorname{PT}}
\newcommand\DT{\operatorname{DT}}
\newcommand\GV{\operatorname{GV}}
\newcommand\PE{\operatorname{PE}}
\newcommand{\GVg}[2][Q]{{\GV}^{(#2)}_{#1}}

\newcommand{\vth}{\vartheta}

\def\vph{\varphi}

\newcommand{\eps}{\epsilon}
\newcommand{\veps}{\varepsilon}

\DeclareMathAlphabet{\mathbb}{U}{bbold}{m}{n}
\DeclareSymbolFont{bbold}{U}{bbold}{m}{n}
\SetSymbolFont{bbold}{bold}{U}{bbold}{m}{n}
\DeclareMathSymbol{\bbGamma}{\mathord}{bbold}{'000}
\DeclareMathSymbol{\bbDelta}{\mathord}{bbold}{'001}
\DeclareMathSymbol{\bbVarTheta}{\mathord}{bbold}{'002}
\DeclareMathSymbol{\bbLambda}{\mathord}{bbold}{'003}
\DeclareMathSymbol{\bbTheta}{\mathord}{bbold}{'004}
\DeclareMathSymbol{\bbPi}{\mathord}{bbold}{'005}
\DeclareMathSymbol{\bbSigma}{\mathord}{bbold}{'006}
\DeclareMathSymbol{\bbUpsilon}{\mathord}{bbold}{'007}
\DeclareMathSymbol{\bbPhi}{\mathord}{bbold}{'010}
\DeclareMathSymbol{\bbPsi}{\mathord}{bbold}{'011}
\DeclareMathSymbol{\bbOmega}{\mathord}{bbold}{'012}
\DeclareMathSymbol{\bbalpha}{\mathord}{bbold}{"0B}
\DeclareMathSymbol{\bbbeta}{\mathord}{bbold}{"0C}
\DeclareMathSymbol{\bbgamma}{\mathord}{bbold}{"0D}
\DeclareMathSymbol{\bbdelta}{\mathord}{bbold}{"0E}
\DeclareMathSymbol{\bbespilon}{\mathord}{bbold}{"0F}
\DeclareMathSymbol{\bbzeta}{\mathord}{bbold}{"10}
\DeclareMathSymbol{\bbeta}{\mathord}{bbold}{"11}
\DeclareMathSymbol{\bbtheta}{\mathord}{bbold}{"12}
\DeclareMathSymbol{\bbiota}{\mathord}{bbold}{"13}
\DeclareMathSymbol{\bbkappa}{\mathord}{bbold}{"14}
\DeclareMathSymbol{\bblambda}{\mathord}{bbold}{"15}
\DeclareMathSymbol{\bbmu}{\mathord}{bbold}{"16}
\DeclareMathSymbol{\bbnu}{\mathord}{bbold}{"17}
\DeclareMathSymbol{\bbxi}{\mathord}{bbold}{"18}
\DeclareMathSymbol{\bbpi}{\mathord}{bbold}{"19}
\DeclareMathSymbol{\bbrho}{\mathord}{bbold}{"1A}
\DeclareMathSymbol{\bbsigma}{\mathord}{bbold}{"1B}
\DeclareMathSymbol{\bbtau}{\mathord}{bbold}{"1C}
\DeclareMathSymbol{\bbupsilon}{\mathord}{bbold}{"1D}
\DeclareMathSymbol{\bbphi}{\mathord}{bbold}{"1E}
\DeclareMathSymbol{\bbchi}{\mathord}{bbold}{"1F}
\DeclareMathSymbol{\bbpsi}{\mathord}{bbold}{"20}
\DeclareMathSymbol{\bbomega}{\mathord}{bbold}{"7F}
\DeclareMathSymbol{\bbell}{\mathord}{bbold}{"40}
\newcommand{\BBsymbol}[1]{%
	\ifcat#1a\mathbbm{#1}\else
	\ifx#1\Gamma\bbGamma\fi
	\ifx#1\Delta\bbDelta\fi
	\ifx#1\VarTheta\bbVarTheta\fi
	\ifx#1\Lambda\bbLambda\fi
	\ifx#1\Theta\bbTheta\fi
	\ifx#1\Pi\bbPi\fi
	\ifx#1\Sigma\bbSigma\fi
	\ifx#1\Upsilon\bbUpsilon\fi
	\ifx#1\Phi\bbPhi\fi
	\ifx#1\Psi\bbPsi\fi
	\ifx#1\Omega\bbOmega\fi
	\ifx#1\alpha\bbalpha\fi
	\ifx#1\beta\bbbeta\fi
	\ifx#1\gamma\bbgamma\fi
	\ifx#1\delta\bbdelta\fi
	\ifx#1\espilon\bbespilon\fi
	\ifx#1\zeta\bbzeta\fi
	\ifx#1\eta\bbeta\fi
	\ifx#1\theta\bbtheta\fi
	\ifx#1\iota\bbiota\fi
	\ifx#1\kappa\bbkappa\fi
	\ifx#1\lambda\bblambda\fi
	\ifx#1\mu\bbmu\fi
	\ifx#1\nu\bbnu\fi
	\ifx#1\xi\bbxi\fi
	\ifx#1\pi\bbpi\fi
	\ifx#1\rho\bbrho\fi
	\ifx#1\sigma\bbsigma\fi
	\ifx#1\tau\bbtau\fi
	\ifx#1\upsilon\bbupsilon\fi
	\ifx#1\phi\bbphi\fi
	\ifx#1\chi\bbchi\fi
	\ifx#1\psi\bbpsi\fi
	\ifx#1\omega\bbomega\fi
	\ifx#1\ell\bbell\fi
	\fi
}

\newcommand{\de}{\mathrm{d}}

\newcommand{\I}{\mathrm{i}}

\newcommand{\rmR}{\mathrm{R}}

\newcommand{\cD}{\mathcal{D}}
\newcommand{\cF}{\mathcal{F}}

\newcommand{\cC}{\mathcal{C}}
\newcommand{\cS}{\mathcal{S}}
\newcommand{\cG}{\mathcal{G}}
\newcommand{\cB}{\mathcal{B}}

\newcommand{\cM}{\mathcal{M}}

\newcommand{\cN}{\mathcal{N}}
\newcommand{\cE}{\mathcal{E}}
\newcommand{\cX}{\mathcal{X}}

\newcommand{\cR}{\mathcal{R}}
\newcommand{\cT}{\mathcal{T}}
\newcommand{\cJ}{\mathcal{J}}
\newcommand{\cZ}{\mathcal{Z}}
\newcommand{\cI}{\mathcal{I}}
\newcommand{\cO}{\mathcal{O}}

\newcommand{\cH}{\mathcal{H}}

\newcommand{\cA}{\mathcal{A}}

\newcommand{\pbbm}{\mathbbm{p}}
\newcommand{\xbbm}{\mathbbm{x}}
\newcommand{\ybbm}{\mathbbm{y}}
\newcommand{\vbbm}{\mathbbm{v}}
\newcommand{\wbbm}{\mathbbm{w}}
\newcommand{\ubbm}{\mathbbm{u}}
\newcommand{\zbbm}{\mathbbm{z}}

\newcommand{\kbbm}{\mathbbm{k}}
\newcommand{\gbbm}{\mathbbm{g}}

\newcommand{\abbm}{\mathbbm{a}}
\newcommand{\bbbm}{\mathbbm{b}}
\newcommand{\ebbm}{\mathbbm{e}}

\newcommand{\Asf}{{\sf A}}

\newcommand{\asf}{{\sf a}}

\newcommand{\zsf}{{\sf z}}

\newcommand{\IA}{\mathds{A}}
\newcommand{\ID}{\mathds{D}}

\newcommand{\IT}{\mathds{T}}
\newcommand{\IR}{\mathds{R}}
\newcommand{\IC}{\mathds{C}}
\newcommand{\IZ}{\mathds{Z}}

\newcommand{\IN}{\mathds{N}}
\newcommand{\IB}{\mathds{B}}
\newcommand{\IF}{\mathds{F}}
\newcommand{\IH}{\mathds{H}}

\newcommand{\IP}{\mathds{P}}

\def\scR{\mathscr{R}}

\def\Bv{\mathscr{B}}
\def\Ev{\mathscr{E}}
\def\Fv{\mathscr{F}}

\def\Zv{\mathscr{Z}}

\def\Dv{\mathscr{D}}

\def\frb{\mathfrak{b}}

\def\frm{\mathfrak{m}}
\def\frh{\mathfrak{h}}
\def\frr{\mathfrak{r}}

\def\frp{\mathfrak{p}}

\def\frt{\mathfrak{t}}
\def\frz{\mathfrak{z}}

\newcommand{\bfv}{{\boldsymbol v}}

\newcommand{\bfk}{{\boldsymbol k}}

\newcommand{\bfp}{{\boldsymbol p}}
\newcommand{\bfq}{{\boldsymbol q}}
\newcommand{\bfr}{{\boldsymbol r}}
\newcommand{\bfs}{{\boldsymbol s}}

\newcommand{\bfz}{{\boldsymbol z}}
\newcommand{\bfx}{{\boldsymbol x}}
\newcommand{\bfy}{{\boldsymbol y}}

\newcommand{\bftet}{{\boldsymbol \theta}}
\newcommand{\bfmu}{{\boldsymbol \mu}}
\newcommand{\bfnu}{{\boldsymbol \nu}}

\newcommand{\bfD}{{\boldsymbol D}}
\newcommand{\bfLam}{{\boldsymbol \Lambda}}

\newcommand{\bfgam}{{\boldsymbol \gamma}}

\newcommand{\tc}{\tilde c}

\newcommand{\tp}{\tilde p}
\newcommand{\tq}{\tilde q}
\newcommand{\tlh}{\tilde h}

\newcommand{\tmu}{\tilde\mu}
\newcommand{\tnu}{\tilde \nu}

\newcommand{\bftmu}{\tilde\bfmu}
\newcommand{\bftnu}{\tilde\bfnu}
\def\tfrm{\tilde\frm}

\def\ba{\bar a}

\def\by{\bar y}
\def\bz{\bar z}
\def\bw{\bar w}

\def\bv{\bar v}

\def\btau{\bar \tau}

\def\bOm{\overline\Omega}

\def\hc{\hat c}

\def\hq{\hat q}

\def\hM{\hat M}

\def\hatt{\hat t}

\def\hmu{\hat\mu}
\def\hPhi{\hat\Phi}

\def\hgam{\hat\gamma}
\def\bfhgam{\hat\bfgam}
\def\bfhmu{\hat\bfmu}
\def\hcC{\hat \cC}
\def\hcD{\hat \cD}

\def\hubbm{\hat\ubbm}
\def\hvbbm{\hat\vbbm}
\def\hwbbm{\hat\wbbm}

\def\htmu{\,\hat{\tilde{\!\mu}}}
\def\bfhtmu{\,\hat{\tilde{\!\bfmu}}}

\def\chg{\check g}

\def\chS{\check S}

\def\cmu{\check\mu}

\def\whvph{\widehat\vph}

\def\Fcl{F^{\rm cl}}

\def\CY{\mathfrak{Y}}

\def\base{S}

\def\ver{\mathfrak{v}}
\def\vert{v}
\def\vth{\vartheta}

\def\cXcl{\cX^{\rm cl}}

\def\hi#1{h^{(#1)}}

\def\han{h^{\rm (an)}}
\def\hh{h^{(0)}}

\def\ths#1{\theta^{(#1)}}

\def\vthls#1{\vartheta^{(#1)||}}

\def\vthA#1{\varTheta^{(#1)}}

\def\di#1{d_{\Nr_#1}}

\def\cbfr{c_\bfr}

\def\gi#1{g^{(#1)}}

\def\girf#1{g^{(#1)\rm{ref}}}

\def\chgirf#1{\chg^{(#1)\rm{ref}}}
\def\vwgi#1{\mathfrak{g}_{#1}}

\def\cGr{\cG^{\rm (ref)}}
\def\cXr{\cX^{\rm (ref)}}
\def\hcXr{\hat\cX^{\rm (ref)}}

\def\cJr{\cJ^{\rm (ref)}}

\def\Ef{\Ev^{(0)}}
\def\Efrf{\Ev^{(0){\rm ref}}}

\def\cGr{\cG^{\rm (ref)}}
\def\cXr{\cX^{\rm (ref)}}

\def\Er{\Ev^{\rm (ref)}}
\def\cJr{\cJ^{\rm (ref)}}

\def\Kr{K^{\rm (ref)}}

\def\Ep{\Ev^{(+)}}
\def\Eprf{\Ev^{(+){\rm ref}}}

\def\trmRi#1{\tilde \rmR^{(#1)}}

\def\rmRi#1{\rmR^{(#1)}}

\def\rmRirf#1{\rmR^{(#1)\rm ref}}
\def\trmRirf#1{\tilde\rmR^{(#1)\rm ref}}

\def\scRrf{\scR^{\rm ref}}

\def\hr{h^{\rm ref}}
\def\thr{\tlh^{\rm ref}}

\def\whhr{\widehat h^{\rm ref}}

\def\whchgirf#1{\lefteqn{\widehat \chg}\hphantom{\chg}^{(#1)\rm{ref}}}

\def\whh{\widehat h}

\def\whhi#1{\whh^{(#1)}}

\def\whpsi{\widehat \psi}

\def\whgi#1{\widehat g^{(#1)}}

\def\whPhi{\widehat\Phi}

\def\EPhi{\Phi^{\,\Ev}}

\def\bfLami#1{\bfLam^{(#1)}}
\def\tbfLami#1{\tilde\bfLam^{(#1)}}

\def\bbLami#1{\bbLambda^{(#1)}}

\def\whcZ{\widehat\cZ}

\def\whE{\widehat{E}}

\def\Gi#1{G^{(#1)}}
\def\cGi#1{\cG^{(#1)}}
\def\Mi#1{M^{(#1)}}

\def\Fvi#1{\Fv^{(#1)}}
\def\whFv{\widehat\Fv}

\def\OmMSW{\Omega^{\rm MSW}}
\def\bOmMSW{\bOm^{\rm MSW}}
\def\Omi#1{\Omega^{(#1)}}
\def\bOmi#1{\bOm^{(#1)}}

\def\Latc{\Lambda^\parallel}
\def\Latp{\Lambda^\perp}

\def\mupr{\mu^\parallel}

\def\vu{\mathfrak{u}}

\newcommand{\gtr}{g_{{\rm tr},n}}

\newcommand{\q}{\mathrm{q}}

\def\Nr{r}

\def\Ms{s}

\def\rr{\mathrm{r}}
\def\sp{\xi}

\newcommand{\glueg}{\gbbm}

\newcommand{\gluegp}{\glueg^\perp}
\newcommand{\gluegl}{\glueg^{||}}
\def\gluegi#1{\glueg^{(#1)}}

\def\nA{n_\infty}
\def\nX{n_{\rm fr}}
\def\nI{n_0}
\def\kk{K}
\def\dd{d}

\def\vt{v}
\def\vtA{\vt_{A}}
\def\vtI{\vt_{I}}
\def\vtX{\vt_{X}}

\def\nv{v_0}

\def\aD{\overline{\rm D6}}
\def\tors{\mathsf{d}_{\rm tors}}
\def\gmax{g_{\rm max}}

\def\vsp{\vspace{0.2cm}}

\Title{Mock modularity at work, or black holes in a forest}

\TitleCitation{Mock modularity at work}

\Author{Sergei Alexandrov $^{1}$\orcidA{}}

\AuthorNames{Sergei Alexandrov}

\isAPAStyle{%
       \AuthorCitation{Lastname, F., Lastname, F., \& Lastname, F.}
         }{%
        \isChicagoStyle{%
        \AuthorCitation{Lastname, Firstname, Firstname Lastname, and Firstname Lastname.}
        }{
        \AuthorCitation{Alexandrov, S.}
        }
}

\address{%
$^{1}$ \quad Laboratoire Charles Coulomb (L2C), Universit\'e de Montpellier,
CNRS, F-34095, Montpellier, France; sergey.alexandrov@umontpellier.fr}

\corres{Correspondence: sergey.alexandrov@umontpellier.fr}

\abstract{Mock modular forms, first invented by Ramanujan, provide a beautiful generalization of the usual modular forms. 
In recent years, it was found that they capture generating functions of the number of microstates
of BPS black holes appearing in compactifications of string theory with 8 and 16 supercharges.
This review describes these results and their applications which range from the actual computation
of these generating functions for both compact and non-compact compactification manifolds 
(encoding, respectively, Donaldson-Thomas and Vafa-Witten topological invariants)
to the construction of new non-commutative structures on moduli spaces of Calabi-Yau threefolds.
}

\begin{document}


\tableofcontents

\section{Introduction}

One of the great achievements of string theory is the understanding of the nature of the microstates responsible for
the black hole entropy. For various types of black holes appearing in string theory compactifications, 
this allowed to reproduce the celebrated Bekenstein-Hawking area law 
with the precise coefficient \cite{Strominger:1996sh,Maldacena:1996gb,Maldacena:1997de}.
It is even more remarkable that, at least for some black holes preserving sufficient amount of supersymmetry,
string theory is able to compute the number of black hole microstates {\it exactly}
(see, e.g., \cite{Dijkgraaf:1996it,Dabholkar:2004yr,Dabholkar:2005by,Gaiotto:2006wm,Sen:2007qy,Benini:2015eyy})!
The resulting integer numbers contain highly valuable information for quantum gravity because
they should be obtainable by summing all quantum corrections, perturbative and non-perturbative,
to the macroscopic Bekenstein-Hawking formula. Many quantum corrections have indeed been computed and matched 
against the exact microscopic counting (see \cite{Cassani:2025sim} for a recent review)
and, amazingly, for the simplest black holes even the precise integer numbers have been recently reproduced \cite{Iliesiu:2022kny},
improving earlier results in \cite{Dabholkar:2010uh,Dabholkar:2014ema}.

In most cases where one has access to exact black hole degeneracies,
this holds for a family of black holes labeled by a number of charges and
what one really computes are their {\it generating functions}.
For example, if one considers a family labeled by a single charge $n$ bounded from below,
it is natural to introduce a function
\be  
h(\tau)=\sum_{n\ge n_{\rm min}} \Omega(n) \q^n, 
\qquad \q=e^{2\pi\I \tau},
\label{exp-hn}
\ee  
where $\Omega(n)$ is the number of microstates for the black hole of charge $n$.
A remarkable fact is that such generating functions typically turn out to be given by {\it modular forms}
(see, e.g., \cite{Murthy:2023mbc}), 
i.e. they transform nicely (see \S\ref{subsec-modular} for the precise definition) 
under the following fractional linear transformation 
\be  
\tau\ \mapsto\ \frac{a\tau+b}{c\tau+d}\, , 
\qquad
\(\begin{array}{cc} a & b \\ c & d \end{array}\)\in SL(2,\IZ).
\label{transtau}
\ee 
Modular forms have been studied since the nineteenth century and are known 
to have very stringent properties. For example, the fact that for large charges $\Omega(n)$ behaves as an exponential 
of the black hole area can be seen as a simple consequence of the growth property 
satisfied by the Fourier coefficients of (weakly holomorphic) modular forms.

In string compactifications with many supercharges it is actually quite natural to expect the appearance of modular forms.
Indeed, such compactifications are typically constructed using a torus $T^2$ as a compact submanifold.
The torus has a complex structure parametrized by a complex parameter $\tau$ living in the upper half-plane $\IH$, 
and $SL(2,\IZ)$ is its modular group identifying tori with the complex structures related by \eqref{transtau}.
This means that all physical results depending on $\tau$ must be invariant under $SL(2,\IZ)$.
Of course, this does not imply yet the modularity of the generating function $h(\tau)$ because
its argument is a formal expansion parameter and {\it a priory} has nothing to do with the complex structure of the torus.
Nevertheless, in practice it does and the observed modular behavior is not an accident.

If one reduces the number of supersymmetries, one encounters new interesting phenomena.
First, in compactifications with 16 supercharges, such as type II string theory on  $K3\times T^2$,
the generating function of degeneracies of black holes preserving only 4 supercharges, known as $\frac14$-BPS states, 
turns out to belong to the class of Siegel modular forms \cite{Dijkgraaf:1996it}.
Such functions transform nicely under a large symmetry group $Sp(2,\IZ)$.
Despite the existence of some ideas in the literature \cite{Gaiotto:2005hc}, the origin of this extended symmetry
remains rather mysterious.

In fact, this generating function hides another beautiful structure which was revealed in \cite{Dabholkar:2012nd}.
To explain it, let us recall that the Fourier coefficients of our generating functions typically count
{\it all} black holes of a given charge, including those which can be thought of as bound states.
The latter are known as multi-centered black holes, in contrast to single-centered ones,
and are full-fledged solutions of supergravity \cite{Denef:2000nb}. Of course, as any bound state, 
they are stable in some region of the parameter space, but can decay after crossing certain stability walls,
which is known as the {\it wall-crossing} phenomenon. From this point of view, single-centered black holes are special
since they never decay. In the context of $\cN=4$ supergravity in four dimensions, 
single-centered $\frac14$-BPS black holes are called {\it immortal dyons}
(because they must have both electric and magnetic charges non-vanishing).
So, what was found in \cite{Dabholkar:2012nd} is that the generating functions of degeneracies 
of the immortal dyons, which can be extracted from the Siegel modular form, are not modular, but {\it mock modular}!
Since mock modularity will be the central topic of this review, let us briefly unveil what hides behind this notion.
More details will be given in \S\ref{subsec-mock}.

Mock modularity has its origin in the work of Srinivasa Ramanujan during the last year (1919–1920) of his life,
which was found in his last letter to G.H. Hardy and in his famous lost notebook.
There he put forward and analyzed several functions which he called {\it mock theta functions}.
As the name suggests, he found that they are similar to ordinary theta functions, but not quite.
Although it was clear that there was something special about these functions, one had to wait more than 80 years until 
their general theory was constructed by S. Zwegers \cite{Zwegers-thesis} (see also \cite{MR2605321}). 
According to this theory, mock theta functions are particular examples of {\it mock modular forms}.
The latter are similar to modular forms, but different from them by failing to satisfy the modular transformation property.
However, the anomaly, which measures how much they fail, has a special form being determined by another modular form
called {\it shadow}. Equivalently, the form of the anomaly ensures that it can be canceled by adding 
a non-holomorphic term, also determined by the shadow, to the mock modular form, 
producing the function known as {\it modular completion}.
Thus, a mock modular form is holomorphic, but has a modular anomaly, while its completion is modular,
but has a holomorphic anomaly.

Mock modular forms give rise to a natural and rich generalization of usual modular forms, 
which is still quite restrictive. In other words, if one knows that a function is mock modular with a given shadow,
it is sufficient to find just a few data (for example, its first few Fourier coefficients), 
to fully determine the function. For the immortal dyons this is not really a problem since 
their degeneracy can be calculated starting from the known Siegel form.
However, in other cases mock modularity provides an invaluable tool to find the objects of interest.
 
Keeping this in mind, let us further reduce supersymmetry and consider compactifications with 8 supercharges
which are obtained by putting type II string theory on a Calabi-Yau (CY) threefold $\CY$.
In the type IIA formulation, supersymmetric black holes at the microscopic level are described as 
bound states of D6, D4, D2 and D0 branes wrapping non-contractible cycles of $\CY$
and characterized by an electro-magnetic charge $\gamma=(p^0,p^a,q_a,q_0)$ with $a=1,\dots,b_2(\CY)$,
where the components of the charge vector play the role of the respective brane charges.
The BPS index counting the number of states of these black holes is known to coincide with 
the generalized Donaldson-Thomas (DT) topological invariant of $\CY$.
In mathematical language, the bound state of D-branes corresponds to a complex of coherent sheaves on $\CY$ and
the D6-brane charge $p^0$ is its rank.

Since there is not any torus in the structure of a generic CY, 
one could think that the modular symmetry is not relevant for the above black holes.
However, it turns out that it is, but only for a particular class corresponding to D4-D2-D0 bound states,
i.e. with vanishing D6-brane charge. 
(As should be clear from above, for this class the black hole entropy is captured by {\it rank 0} DT invariants.)
The point is that in the dual M-theory picture, D4-D2-D0 BPS states, with D4-brane wrapped on a 4-cycle $\cD$,
are realized by M5-brane wrapped on $\cD\times S^1$. In the limit of large $S^1$, the world-volume theory on 
the M5-brane reduces to a superconformal field theory (SCFT) in two dimensions, first considered in \cite{Maldacena:1997de}.
This theory allows us to define a modified elliptic genus \cite{Maldacena:1999bp}, a torus partition function with
certain insertions ensuring that only contributions of BPS states survive cancellations between bosons and fermions
\cite{Witten:1986bf}. On one hand, it contains information about the BPS spectrum, i.e. the number of BPS states,
and on the other hand, being defined on a torus, it is expected to be a modular form. 
This is why the generating functions of D4-D2-D0 BPS indices, or rank 0 DT invariants, are also expected 
to exhibit modular properties \cite{Gaiotto:2006wm,deBoer:2006vg,Denef:2007vg}.

The precise modular properties of these generating functions have been derived only recently 
\cite{Alexandrov:2012au,Alexandrov:2016tnf,Alexandrov:2018lgp} using a different picture 
and turned out to be very intricate and beautiful.
In fact, the main goal of this review is to explain these properties, their origin and 
the results produced on their basis. 
Before entering mathematical details, let us summarize here the main points for the ease of reading.

The main qualitative result is that the modular properties of the generating functions of D4-D2-D0 BPS indices
with a fixed D4-brane charge $p^a$, 
which will be denoted by\footnote{The meaning of index $\mu$ will be explained in \S\ref{sec-genfun}.} 
$h_{p,\mu}$, crucially depend on this charge or, more precisely, 
on its {\it degree of reducibility}. It is equal to the maximal number of 4-cycles into 
which the wrapped cycle $\cD$ can be decomposed, $\cD=\sum_{i=1}^r \cD_i$, 
where some $\cD_i$ may represent the same cycle.
If the cycle $\cD$ is irreducible, i.e. $r=1$, then the corresponding generating function is modular.
If $r=2$, then it is mock modular. And if  $r>2$, it is described by a generalization of mock modularity
known as mock modular forms of higher depth.

Physically, the degree of reducibility can be thought of as the maximal number of constituents
forming a bound state that can contribute to a given BPS index.
This makes it clear why in compactifications with $\cN=4$ supersymmetry only usual mock modular forms appear, while
in $\cN=2$ story one finds this intricate pattern: it is well known that in the former case
there are only bound states with two constituents, whereas in the latter any number of constituents is possible.

It is important that one knows not only the qualitative behavior of the generating functions under modular transformations,
but also the precise form of their modular anomaly conveniently encoded in an explicit expression
for their modular completion $\whh_{p,\mu}$ \cite{Alexandrov:2018lgp,Alexandrov:2024jnu}.
It can be written as 
\be
\whh_{p,\mu}(\tau,\btau)=h_{p,\mu}(\tau)+ \sum_{n=2}^{r}\sum_{\sum_{i=1}^n p_i=p}
\sum_{\{\mu_i\}}
\rmRi{\{p_i\}}_{\mu,\{\mu_i\}}(\tau, \btau)
\prod_{i=1}^n h_{p_i,\mu_i}(\tau),
\label{exp-whh-new}
\ee
where the second sum goes over all ordered decompositions of the D4-brane charge into charges with non-negative components
and can be thought of as a sum over possible bound states.
The main non-trivial ingredient of this formula is the function $\rmRi{\{p_i\}}_{\mu,\{\mu_i\}}$
which will  be defined in \S\ref{sec-anomaly}. 
Here we just mention the fact that it is given by a sum over various types of trees,
which was used in the title to intrigue the reader (see Fig. \ref{fig-table} for an illustration of relevant trees).

Since the r.h.s. of \eqref{exp-whh-new} depends on the generating functions for smaller charges corresponding to constituents,
the set of equations for different D4-brane charges gives rise to
an iterative system of anomaly equations on the functions $h_{p,\mu}$.
This system has a very rich and universal structure because, although it was originally derived 
in a concrete setup (compactification on a compact CY with D4-brane wrapping an ample divisor),
it turns out that it is applicable or has a simple extension to much more general situations.
For instance, it is still valid for certain degenerations and, in particular, for non-compact CY manifolds \cite{Alexandrov:2019rth}.
Since non-compact CYs can be used to geometrically engineer gauge theories, 
in favorable circumstances their BPS spectrum can also be constrained by our anomaly equations. 
So in \cite{Alexandrov:2020bwg,Alexandrov:2020dyy}, they have been used to solve Vafa-Witten (VW) topological theory 
with gauge group $U(r)$ on various rational surfaces, for arbitrary rank $r$.
Besides, the system of the anomaly equations \eqref{exp-whh-new} has a natural generalization 
that includes {\it refinement} \cite{Alexandrov:2019rth}, 
a one-parameter deformation corresponding physically to switching on the $\Omega$-background 
\cite{Moore:1998et,Nekrasov:2002qd}. 
In turn, this generalization allows us to put the compactifications with higher supersymmetry 
with different numbers of preserved supercharges into the same single framework, 
so that most of the known modularity results on the generating functions of BPS indices, 
including the mock modularity of the immortal dyons \cite{Dabholkar:2012nd}, turn out to be consequences
of this extended system \cite{Alexandrov:2020qpb}.

Finally, and perhaps most importantly, the constraints of modularity can be used 
as a tool to find explicit expressions for the generating functions $h_{p,\mu}$
and thereby to determine the exact degeneracies of BPS black holes.
In fact, this program was initiated long ago in 
\cite{Gaiotto:2006wm} where it was applied to the generating function $h_{1,\mu}$
of D4-D2-D0 BPS indices with unit D4-brane charge on the quintic threefold,
and then extended to a few other one-parameter CYs in \cite{Gaiotto:2007cd,Collinucci:2008ht,VanHerck:2009ww}.
In this case there is not any modular anomaly yet since the generating function must be a modular form,
and the main difficulty consists in computing its polar terms, 
given by the Fourier coefficients in \eqref{exp-hn} with negative $n$, 
which together with modularity are enough to determine the full function.
However, the two approaches used in that works are hardly generalizable and 
even produced mutually inconsistent results. A more systematic approach has been proposed recently in 
\cite{Alexandrov:2023zjb} and allowed to resolve the previous inconsistencies and to compute
$h_{1,\mu}$ for a dozen one-parameter CY threefolds \cite{Alexandrov:2023zjb,McGovern:2024kno}.

The modular anomaly starts playing a crucial role when one goes to higher D4-brane charges.
Then the polar terms are not enough to uniquely fix the generating functions and one should follow a two-step procedure:
i) first, solve the modular anomaly, which gives a unique solution up to the addition of a pure modular form
(modular ambiguity); ii) fix the ambiguity by computing the polar terms.
For one-parameter CY threefolds, the first step has been realized for $r=2$ in \cite{Alexandrov:2022pgd}
and for arbitrary $r$ in \cite{Alexandrov:2024wla}.
The second step has been done so far only for two CY threefolds and for a D4-brane charge equal to 2 
\cite{Alexandrov:2023ltz}. As a result, for the first time we got access to charge 2 states on 
CY threefolds without any additional structure that are organized in a mock modular form.
Furthermore, the computation of Fourier coefficients beyond polar terms following the approach of \cite{Alexandrov:2023zjb}
provided an impressive test of (mock) modularity, which still remains conjectural from the mathematical 
point of view.\footnote{See \cite{talkSheshmani} for an attempt to rigorously prove it
in the simplest pure modular case on the quintic threefold.}

Finally, one should mention another important implication of the above results 
(see \S\ref{subsec-topinv} for more details). 
The method of \cite{Alexandrov:2023zjb} to compute polar terms relies on the knowledge of 
Gopakumar-Vafa (GV) invariants, which can be mapped to rank 0 DT invariants 
using the MNOP formula \cite{gw-dt,gw-dt2} and wall-crossing relations.
The GV invariants in turn are computed using the direct integration approach to 
topological string theory on compact threefolds,
which is based on solving a holomorphic anomaly equation for its partition function 
\cite{Bershadsky:1993ta,Huang:2006hq,Grimm:2007tm}.
The problem however is that the solution has a holomorphic ambiguity that needs to be fixed,
precisely as our modular anomaly equations fix the generating functions only up to a modular ambiguity.
Fortunately, there are several well-known conditions that can be used for this purpose. But their number grows slower 
with genus than the number of parameters to be fixed, so that at some maximal genus the method does not work anymore.
However, once a generating function of rank 0 DT invariants is found, one can invert the relations mentioned above
and find new GV invariants that can serve as new conditions for fixing the holomorphic ambiguity.
Thus, the two systems of anomaly equations work together and help each other to overcome their own limitations.

The organization of the review is the following. 
In the next two sections we provide the mathematical background needed to understand the results presented below.
First, in \S\ref{sec-modular} we introduce modular and mock modular forms. 
Then in \S\ref{sec-indef} we describe an important class of functions known as indefinite theta series,
which provide the simplest example of mock modular forms and play an important role in our construction.
In \S\ref{sec-genfun} we define the main object of interest --- generating functions of D4-D2-D0 BPS indices.
In \S\ref{sec-anomaly} we present the main result about their modular behavior,
while in \S\ref{sec-ext} we discuss its various extensions including some degenerate cases, non-compact CYs and the refinement.
In \S\ref{sec-solution} we explain a solution to the system of anomaly equations,
and in \S\ref{sec-application} we present various applications including the computation of topological invariants
on compact CY threefolds, the solution of Vafa-Witten theory and an extension to higher supersymmetry.
We conclude with a discussion of open issues in \S\ref{sec-concl}.

\section{Modular and mock modular forms}
\label{sec-modular}

\subsection{Modular forms}
\label{subsec-modular}

Let us first review the definition of a standard modular form and then incorporate 
various (also standard) generalizations that are required in physical applications.
For a more in-depth presentation, one can consult \cite{Ap-book,book-modular}.

In this review we deal only with modular forms of $SL(2,\IZ)$ represented by $2\times 2$
matrices as in \eqref{transtau} with integer coefficients and determinant equal to 1. 
We denote the modular parameter by $\tau=\tau_1+\I \tau_2$ and take it to be a complex number belonging 
to the upper half-plane $\IH$ defined by the condition $\tau_2>0$.
Then one has
{\Definition\label{def-mod} 
A function $h(\tau)$ is a modular form of weight $w$ if it is holomorphic on $\IH$, bounded as $\tau_2\to\infty$ 
and transforms as  
\be  
h\( \frac{a\tau+b}{c\tau+d}\)=(c\tau+d)^w h(\tau)\, .
\label{transmod}
\ee 
} 
It immediately follows that $h(\tau)$ is periodic under $\tau\mapsto \tau+1$
and hence has the following Fourier expansion
\be  
h(\tau)=\sum_{n=0}^\infty h_n \q^n ,
\qquad 
\q= e^{2\pi\I \tau}.
\label{Fourier}
\ee 

{\Example
For integer $k>1$, the Eisenstein series
\be  
E_{2k}(\tau)=\frac{1}{2\zeta(2k)}\sum_{(m,n)\in\IZ^2\setminus \{(0,0)\}}\frac{1}{(m\tau+n)^{2k}}
\label{Eisenstein2k}
\ee 
is a modular form of weight $2k$. The overall coefficient given by the Riemann zeta function
is introduced so that to have the constant term of the Fourier series equal to 1.
\vsp}

In fact, the definition \ref{def-mod} is so restrictive that all such modular forms can be generated by 
just two Eisenstein series, $E_4$ and $E_6$. Namely, each modular form of weight $w$ has a unique expansion as
\be  
h(\tau)=\sum_{4k+6l=w\atop k,l\geq 0} c_{k,l} E_4^{k}(\tau) E_6^{l}(\tau).
\ee 
Therefore, it is natural to relax the definition in several ways.

First, we relax the behavior at infinity and instead require $h$ to have at most polynomial growth in $\q^{-1}$
as $\tau_2\to\infty$. 
Provided $h$ still satisfies \eqref{transmod}, it is called {\it weakly holomorphic} modular form
and, if $h(\tau)=O(\q^{-N})$, its Fourier expansion \eqref{Fourier} acquires
additional terms with $-N\leq n <0$.

{\Example\label{ex-Delta}
	The inverse discriminant function $\Delta^{-1}(\tau)$, where $\Delta= (E_4^3-E_6^2)/1728$, is 
a weakly holomorphic modular form of weight $-12$. Its Fourier expansion starts with 
$\Delta^{-1}(\tau)=\q^{-1}+24+\cdots$.
\vsp}

Second, one can relax the transformation property \eqref{transmod} by allowing a phase factor
depending on the group element,
which is called {\it multiplier system} of the modular form.
The modified modular transformation is then given by 
\be  
h\( \frac{a\tau+b}{c\tau+d}\)=(c\tau+d)^w M(\rho) h(\tau)\, ,
\qquad
\rho=\(\begin{array}{cc} a & b \\ c & d \end{array}\)\in SL(2,\IZ).
\label{transmod-M}
\ee
It is clear that $M(\rho)$ should satisfy a cocycle condition, which makes it similar to a character of $SL(2,\IZ)$.
Moreover, since $SL(2,\IZ)$ is generated by two elements,
\be  
T=\(\begin{array}{cc} 1 & 1 \\ 0 & 1 \end{array}\) 
\qquad \mbox{and} \qquad
S=\(\begin{array}{cc} 0 & -1 \\ 1 & 0 \end{array}\),
\ee 
the multiplier system is completely characterized by specifying $M(T)$ and $M(S)$.
The advantage of allowing for a multiplier system is that now the Fourier expansion of a modular form 
does not need to be in integer powers of $\q$ and the weight $w$ can be half-integer.

{\Example
	The Dedekind eta function 
\be  
\eta(\tau)=\Delta^{\frac{1}{24}}(\tau)=\q^{\frac{1}{24}}\prod_{n=1}^\infty (1-\q^n)
=\q^{\frac{1}{24}}\sum_{n\in\IZ} (-1)^n \q^{\hf(3n^2-n)}
\ee 
is a modular form of weight 1/2 and multiplier system
\be  
\Mi{\eta}(T)=e^{\frac{\pi\I}{12}},
\qquad
\Mi{\eta}(S)=e^{-\frac{\pi\I}{4}}.
\label{Meta}
\ee 
}

Another important generalization is to {\it vector valued modular forms}. In this case, instead of 
a single function $h(\tau)$, one considers a vector of functions $h_\mu(\tau)$ where the index $\mu$ 
takes a finite number of values. Then the only modification to be done in the above equations
is that the multiplier system becomes matrix valued, $M_{\mu\nu}(\rho)$, so that the modular transformation reads
\be  
h_\mu\( \frac{a\tau+b}{c\tau+d}\)=(c\tau+d)^w \sum_\nu M_{\mu\nu}(\rho) h_\nu(\tau)\, .
\label{transmod-vec}
\ee
Besides, the Fourier expansion now has the following form
\be  
h_\mu(\tau)=\sum_{n\in \IN-\Delta_\mu} h_{\mu,n} \q^n,
\label{Fourier-gen}
\ee 
where $\Delta_\mu$ can be rational numbers.

{\Example\label{ex-theta}
The two-component vector 
\be 
\vth_\mu(\tau)=(\theta_3(2\tau),\theta_2(2\tau))=\sum_{n\in\IZ+\mu/2} \q^{n^2},
\qquad\mu=0,1,
\label{defthetamu}
\ee 
where $\theta_2$ and $\theta_3$ are the standard Jacobi theta functions,
is a vector valued modular form of weight 1/2 and multiplier system
\be
\Mi{\vth}_{\mu \nu}(T) =
e^{\frac{\pi \I}{2} \,\mu^2}\delta_{\mu \nu},
\qquad
\Mi{\vth}_{\mu \nu}(S)=
\frac{e^{-\frac{\pi\I}{4}}}{\sqrt{2}} \, (-1)^{\mu \nu} .
\label{mult-theta}
\ee
}

Finally, one can drop the holomorphicity condition and define modular forms of mixed weight $(w,\bw)$.
Thus, in the most general case we have the following 
{\Definition\label{def-modgen} 
	$h(\tau,\btau)$ is a vector valued modular form of weight $(w,\bw)$ and multiplier system $M_{\mu\nu}$ 
		if it satisfies 
	\be  
	h_\mu\( \frac{a\tau+b}{c\tau+d}\, , \, \frac{a\btau+b}{c\btau+d}\)
	=(c\tau+d)^w (c\btau+d)^{\bw} \sum_\nu M_{\mu\nu}(\rho) h_\nu(\tau,\btau)\, .
	\label{transmod-nh}
	\ee 
} 

{\Example
A trivial example of a non-holomorphic modular form is given by $\tau_2$ which has weight $(-1,-1)$.
\vsp}	
	
It is important that as long as one remains in the realm of (weakly) holomorphic modular forms,
for a given weight and multiplier system, the transformation property \eqref{transmod-vec} restricts them to
form a finite-dimensional space. In particular, for weakly holomorphic modular forms of negative weight
the dimension of this space is bounded from above by the number of {\it polar terms}, i.e. 
the terms in the Fourier expansion in \eqref{Fourier-gen} with negative power $n$ 
\cite{Bantay:2007zz,Manschot:2007ha,Manschot:2008zb}.
In the vector valued case, one should sum up the number of polar terms for all (independent) components.

\subsection{Mock modular forms}
\label{subsec-mock}

The next level of generalization is provided by (vector valued weakly) holomorphic {\it mock modular forms}
which spoil the transformation property \eqref{transmod-vec}, but in a very specific way controlled by another modular form.
More precisely, if $h_\mu$ is mock modular of weight $w$, it should satisfy
\be 
h_\mu\( \frac{a\tau+b}{c\tau+d}\)=(c\tau+d)^w \sum_\nu M_{\mu\nu}(\rho) \left(h_\nu(\tau)
- \int_{-d/c}^{-\I\infty} \frac{\overline{g_\nu(\bar z)}}{(\tau-z)^w}\,\de z\right),
\label{transf-mock}
\ee 
where $g_\mu$ is a modular form of weight $2-w$, called the {\it shadow} of $h_\mu$.
It is easy to see that the anomalous term can be canceled by adding to $h_\mu$ a {\it non-holomorphic} contribution,
known as the Eichler or period integral of the shadow,
\be 
g_\mu^*(\tau,\btau)= \int_{\bar\tau}^{-\I\infty}\frac{\overline{g_\mu(\bar z)}}{(\tau-z)^w}\,\de z.
\label{def-Eichler}
\ee
The resulting non-holomorphic function 
\be 
\whh_\mu(\tau,\btau)= h_\mu(\tau)- g_\mu^*(\tau,\btau)
\label{def-compl}
\ee
is called the {\it modular completion} of $h_\mu$ and transforms as a usual modular form of weight $(w,0)$.
It contains all interesting information: on one hand, the original mock modular form can be obtained from it
by taking the limit $\btau\to\infty$ keeping $\tau$ fixed and, on the other hand, the shadow can be extracted
by taking the non-holomorphic derivative
\be 
g_\mu(\tau)=\overline{(2\I\tau_2)^w \p_{\btau}\whh_\mu}.
\ee 
In fact, the completion provides an alternative and somewhat more convenient way of defining mock modular forms.

{\Definition
A (vector valued weakly) holomorphic function $h_\mu$ is a mock modular form of weight $w$ with shadow $h_\mu$, 
if its completion defined by \eqref{def-compl} transforms as a usual modular form of the same weight.
}

{\Example\label{ex-E2}
The simplest example of a mock modular form is provided by the quasi-modular Eisenstein series $E_2(\tau)$.
It can be defined by the same formula \eqref{Eisenstein2k} as other Eisenstein series, specified to $k=1$, 
but in contrast to them, the double sum is not absolutely convergent, which is the origin of a modular anomaly.
Alternatively, it can be expressed as a logarithmic derivative of the discriminant function $\Delta(\tau)$
\be 
E_2(\tau)=\frac{1}{2\pi\I}\, \p_\tau\log\Delta(\tau)=1-24\sum_{n=1}^\infty \frac{n \q^n}{1-\q^n}\, .
\label{defE2}
\ee 
From the modular transformation of $\Delta(\tau)$, it is immediate to derive the transformation of $E_2(\tau)$:
\be
E_2\(\frac{a\tau+b}{c \tau +d}\)  = \(c\tau+d\)^2 \(E_2(\tau) + \frac{6}{\pi\I}\, \frac{c}{c\tau+d}\),
\label{modtr-E2}
\ee
which fits the transformation of a generic mock modular form with the shadow taken to be constant,
$g=6\I/\pi$. It is also easy to check that the following non-holomorphic function
\be 
\widehat E_2(\tau,\btau)=E_2(\tau)-\frac{3}{\pi\tau_2}
\ee 
transforms as a standard modular form and fits the definition of the completion \eqref{def-compl} with the same shadow.
}

{\Example\label{ex-Hurwitz}
The $n$-th Hurwitz class number $H(n)$ is defined as the number of $PSL(2,\IZ)$-equivalence classes of integral
binary quadratic forms of discriminant $n$, divided by the number of their automorphisms.
Setting also $H(0)=-1/12$, they can be organized into a generating series.
However, it does not transform properly under the full $SL(2,\IZ)$ group. 
Therefore, it is more convenient to split it into a two-component vector
\be
\label{H01}
\begin{split}
	H_0(\tau) =&\, \sum_{n\ge 0} H(4n)\q^n
	=-\frac{1}{12} + \frac{1}{2}\,\q+\q^2+\frac{4}{3}\,\q^3+\frac{3}{2}\, \q^4+2 \q^5
	+\dots\, ,
	\\
	H_1(\tau) =&\, \sum_{n> 0} H(4n-1)\q^n
	=\q^{\frac34} \left( \frac{1}{3}+\q+\q^2+2\q^3+\q^4+3 \q^5
	+ \dots \right),
\end{split}
\ee
where $H(4n+1)$ and $H(4n+2)$ do not appear because they all vanish.
It has been discovered in \cite{Zagier:1975} that $H_\mu$ is a vector valued mock modular form 
of weight 3/2 with the shadow proportional to the theta series $\vth_\mu$ \eqref{defthetamu}.
This example is highly important because this function, after multiplication by 3, 
turns out to coincide with the generating series of $SU(2)$ Vafa-Witten invariants
on $\IP^2$ \cite{Vafa:1994tf}, which is one of the first examples of the appearance of mock modularity in physics. 
\vsp}

Although mock modular forms are much more general than modular forms, they are still severely restricted 
by their transformation property. For example, a weakly holomorphic mock modular form of negative weight
is completely determined by its polar terms, similarly to its pure modular cousin. 
The difference is that, in general, the standard modularity requires the polar coefficients to satisfy certain constraints,
whereas they can be chosen freely to generate a mock modular form \cite{Manschot:2007ha}.
In this sense, mock modular forms are even ``more natural" than usual ones.

Before we move on, we need to introduce one more generalization.
{\Definition 
$h_\mu$ is a mixed mock modular form of weight $w$ if its modular completion has the form
\be 
\whh_\mu= h_\mu-\sum_{j,\alpha} f_{j,\mu\alpha}\, g_{j,\alpha}^*,
\label{def-compl-moxed}
\ee
where $f_{j,\mu\alpha}$ and $g_{j,\alpha}$ are holomorphic modular forms of weight $w+r_j$ and $2+r_j$, respectively.
\vsp}

In other words, for mixed mock modular forms the shadow is allowed to be a sum of products 
of holomorphic and anti-holomorphic functions.
It is clear that one can generate infinitely many examples of such functions by simply taking products
of mock modular forms with usual modular forms. 
In fact, most of the mock modular forms appearing in this review will be of the mixed type.

Note also that for this class of functions, the knowledge of only polar terms is not sufficient anymore to 
fix them uniquely. As we will see, in addition one should know the precise modular anomaly encoded in the shadow.

\subsection{Higher depth mock modular forms}

Mixed mock modular forms are at the basis of another huge generalization which is known as 
{\it higher depth mock modular forms} \cite{bringmann2017higher}. 
This class of functions is defined iteratively in depth.
Namely, we take the usual modular and mock modular forms as objects of depth 0 and 1, respectively.
Then we take  

{\Definition
$h_\mu$ is a depth $r$ mock modular form of weight $w$
if the anti-holomorphic derivative of its modular completion has the form
\be 
\p_{\btau}\whh^{(r)}_\mu= \sum_{j,\alpha} \tau_2^{r_j}\, \whh^{(r-1)}_{j,\mu\alpha}\, \overline{g_{j,\alpha}},
\label{def-hdepth}
\ee
where $g_{j,\alpha}$ are modular forms of weight $2+r_j$, while $\whh^{(r-1)}_{j,\mu\alpha}$ are 
completions of mock modular forms of depth $r-1$ and weight $w+r_j$.
\vsp}

We will provide an important example of higher depth mock modular forms in \S\ref{sec-indef}.
Besides, as already mentioned in the Introduction, the generating functions of D4-D2-D0 BPS indices 
in CY string compactifications, coinciding with rank 0 DT invariants, also turn out to belong to this class.
Thus, the higher depth modularity is not an abstract generalization, 
but captures the modular properties of important physical and mathematical objects.

\subsection{Jacobi forms and their variations}
\label{subsec-Jacobi}

We finish our presentation of modular functions by introducing {\it Jacobi forms}, 
which carry dependence on an additional complex variable $z$ \cite{MR781735}. 
Besides the modular weight $w$, their transformation properties are characterized 
also by a number $m$ known as {\it index}. Their precise definition is as follows:

{\Definition
A holomorphic function $\vph(\tau,z)$ on $\IH\times \IC$ is a Jacobi form of weight $w$ and index $m$ if it satisfies 
\begin{subequations}
	\bea
	\vph(\tau,z+a\tau+b)&=&
	e^{-2\pi\I m \( a^2\tau + 2 a z\)} \,\vph(\tau,z),
	\qquad a,b\in \IZ,
	\label{Jacobi-ell}
	\\
	\vph\(\frac{a\tau+b}{c\tau+d}, \frac{z}{c\tau+d}\)
	&=& (c\tau+d)^w\, e^{\frac{2\pi\I m c z^2}{c\tau+d}} \vph(\tau,z).
	\label{Jacobi-mod}
	\eea
	\label{Jacobi}
\end{subequations}
}

Of course, setting $z=0$, any Jacobi form gives rise to a modular form. However, even for non-vanishing $z$,
Jacobi forms are, in a sense, constructed out of modular forms. Indeed, using the ``elliptic property" \eqref{Jacobi-ell},
it is easy to show that $\vph(\tau,z)$ has the following theta expansion
\be  
\vph(\tau,z)=\sum_{\mu=0}^{2m-1} h_\mu(\tau) \ths{m}_\mu(\tau,z),
\label{thetadecomp}
\ee 
where 
\be
\ths{m}_\mu(\tau,z)
=\sum_{k\in 2m\IZ+\mu}\q^{\frac{k^2}{4m} }\,y^{k},
\qquad 
y=e^{2\pi\I z},
\label{deftheta}
\ee
is a unary theta series, while $h_\mu(\tau)$ is a vector valued modular form of weight $w-1/2$.
Thus, Jacobi forms carry essentially the same information as vector valued modular forms, but allow to encode it 
in a more compact and nice way.

The Jacobi forms introduced above are an extension of the modular forms from Definition \ref{def-mod}.
Similarly to the discussion in \S\ref{subsec-modular}, one can upgrade them 
to be vector valued, have a non-trivial multiplier system and carry a non-holomorphic dependence.
All these generalizations are obvious, so we do not provide the corresponding transformation properties. 

Furthermore, one can also allow for multiple elliptic parameters $z_i$ so that $\vph(\tau,\bfz)$
becomes a {\it multi-variable} Jacobi form with $\bfz=(z_1,\dots z_n)$.
In this situation, $z^2$ appearing in the exponential in \eqref{Jacobi-mod} should be replaced by 
$\bfz^2=\sum_{i,j=1}^n Q_{ij} z_i z_j$ where $Q_{ij}$ defines a scalar product in $\IC^n$.
Besides, the index now becomes matrix valued and equal to $mQ_{ij}$.
Such multi-variable Jacobi forms still have a theta expansion, but the unary theta series \eqref{deftheta}
is replaced by a theta series defined on $\IZ^n$ lattice with quadratic form $2mQ_{ij}$.

{\Example \label{ex-theta1}
The Jacobi theta function
\be
\label{free-theta-1}
\theta_1(\tau,z) = \sum_{k\in Z+\hf} \q^{k^2/2} (-y)^k
\ee
is a Jacobi form of weight 1/2, index 1/2 and the following multiplier system
\be
	\Mi{\theta_1}(T) = e^{\frac{\pi\I}{4}} ,
	\qquad
	\Mi{\theta_1}(S) = e^{-\frac{3\pi \I}{4}}.
\label{multi-theta-N}
\ee
}

Given the existence of the theta expansion, it is straightforward to introduce {\it mock Jacobi}
and even {\it higher depth mock Jacobi forms}. They can be defined as functions having a theta expansion \eqref{thetadecomp}
where $h_\mu$ is a mock or higher depth mock modular form. This also allows us to talk about modular completions 
of mock Jacobi forms $\whvph$ given by 
\be  
\whvph(\tau,\btau,z)=\sum_{\mu=0}^{2m-1} \whh_\mu(\tau,\btau) \ths{m}_\mu(\tau,z).
\label{thetadecomp-compl}
\ee 

Finally, if one drops the elliptic property \eqref{Jacobi-ell}, but keeps the modular transformation \eqref{Jacobi-mod},
one arrives at the definition of {\it Jacobi-like forms}.
They are not periodic in $z$, so that they do not need to depend on it through the exponential $y$ as in \eqref{deftheta}.
As a result, they do not have a theta expansion and hence cannot be reduced to a single modular form.
Instead, they give rise to an infinite set of modular forms through a Laurent expansion in $z$ \cite{Zagier:1994,Cohen1997}.
One way to construct them is provided by the following proposition \cite{Alexandrov:2024wla}

{\Proposition \label{prop-jacobi-like}
	Let $\vph_\mu(\tau,z)$ be a Jacobi-like form of modular weight $w$ and index $m$, and having a smooth limit at $z\to 0$.
We define the following differential operator
\be 
\cD_m^{(n)}=\sum_{k=0}^{\lfloor n/2\rfloor}c_{n,k}E_2^k(\tau) \,\p_z^{n-2k},
\label{defcDmn}
\qquad
c_{n,k}=\frac{n!\(\frac{2m}{3}\pi^2\)^k }{(2k)!!(n-2k)!}\, .
\ee
Then 
\be
\phi^{(n)}_\mu(\tau)\equiv \cD_m^{(n)}\vph_\mu(\tau,z)|_{z=0}	
\label{coeffJac}
\ee
are vector valued modular forms of weight $w+n$.
\vsp}

The differential operators $\cD_m^{(n)}$ and the modular forms generated by them will appear in \S\ref{sec-solution} 
in the construction of a solution of the modular anomaly equation that governs the generating functions of black hole degeneracies
in CY compactifications.

\section{Indefinite theta series}
\label{sec-indef}

In this section, we recall a few facts about an important class of functions having interesting modular properties, 
which are known as (generalized) {\it theta series}.
Usually, such a theta series is associated with a lattice $\bbLambda$ endowed 
with an integer valued quadratic form $\kbbm^2=\kbbm\star\kbbm$ 
and can be schematically written as
\be 
\vth_\bbmu(\tau;\bbLambda)=\sum_{\kbbm\in\bbLambda+\bbmu}\Phi(\kbbm,\tau_2)\, \q^{\hf \kbbm^2}\, ,
\label{gen-theta}
\ee 
where $\bbmu\in \bbLambda^*/\bbLambda$ is the so called {\it residue class} valued in the discriminant group of the lattice
and $\Phi$ is a function of at most polynomial growth which we will call a {\it kernel}.

In the simplest case, the quadratic form is positive definite so that the sum in \eqref{gen-theta} is absolutely convergent
for $\tau\in\IH$.
If in addition $\Phi=1$, the theta series is known to be a modular form 
of weight $d/2$ where $d=\dim\bbLambda$.\footnote{Strictly speaking, 
	this is true only for even lattices for which the quadratic form takes values in $2\IZ$. For odd lattices, $\Phi$
	should be chosen to be a sign factor to produce a modular form.} 
Ex. \ref{ex-theta} provides the simplest illustration of this situation.
By inserting the factor $e^{2\pi\I \kbbm\star\zbbm}$ depending on a vector of elliptic parameters $\zbbm$,
one can also convert the theta series into a (multi-variable) Jacobi form (see Ex. \ref{ex-theta1}).
This shows that, with a properly chosen kernel, the theta series satisfies standard modular transformation properties
of type \eqref{transmod-vec} or \eqref{Jacobi}.

The situation drastically changes once one allows the quadratic form to have an indefinite signature, 
say $(d-n,n)$ with $n\geq 1$. An immediate problem is that, for generic $\Phi$, the sum in \eqref{gen-theta} becomes divergent.
Thus, the kernel {\it must} be non-trivial just to have a well-defined theta series.

There are basically two ways to achieve the convergence.
The simplest way is to choose the kernel to be exponentially decaying in the ``dangerous" directions of the lattice
where the quadratic form is negative definite.
However, since the kernel is allowed to depend only on $\tau_2$, not on $\tau$, this can be done only 
at the cost of introducing non-holomorphicity. 
For example, one can take $\Phi=e^{2\pi\tau_2 \kbbm_-^2}$ where $\kbbm_\pm$ denote the projections of $\kbbm$, respectively,
on the positive and negative definite sublattices. The resulting $\vth_\bbmu$ is the Siegel theta series
\be 
\vth^{\rm (S)}_\bbmu(\tau)=\sum_{\kbbm\in\bbLambda+\bbmu} \q^{\hf \kbbm_+^2}\bar\q^{-\hf \kbbm_-^2},
\label{Siegel-ts}
\ee 
which is easily seen to be a non-holomorphic modular form of weight $(\frac{d-n}{2},\frac{n}{2})$.

However, often one has to deal with {\it holomorphic} indefinite theta series which can be obtained 
by the second way of achieving convergence. In this case, one takes the kernel so that it simply vanishes 
in the dangerous directions. Such behavior is easy to get if $\Phi$ is a combination of sign functions.
The simplest construction of this type is provided by the following theorem from \cite{Alexandrov:2020bwg} 
(generalizing results of \cite{Nazaroglu:2016lmr,Alexandrov:2017qhn,funke2017theta})\footnote{We warn the reader that 
in most of these references one uses a convention where there is a minus sign in the power of $\q$ in \eqref{gen-theta}.
This corresponds to flipping the overall sign of the quadratic form and hence all inequalities in Theorem \ref{th-converg} 
should be inverted. \label{foot-sign}}

{\Theorem\label{th-conv}
	Let the signature of the quadratic form be $(d-n,n)$ and
	\be
	\Phi(\kbbm)=\prod_{i=1}^n\Bigl(\sgn(\vbbm_{1,i}\star \kbbm)-\sgn(\vbbm_{2,i}\star\kbbm)\Bigr),
	\label{kerconverge}
	\ee
	where $\{\vbbm_{1,i}\}$, $\{\vbbm_{2,i}\}$ are two sets of $d$-dimensional vectors.
	Then the theta series \eqref{gen-theta} is convergent provided:
	\begin{enumerate}
		\item
		for all $i\in \Zv_{n}=\{1,\dots,n\}$,
		$\vbbm_{1,i}^2,\vbbm_{2,i}^2< 0$;
		\item
		for any subset $\cI\subseteq \Zv_{n}$ and any set of $s_i\in \{1,2\}$, $i\in\cI$,
		\be
		\mathop{\det}\limits_{i,j\in \cI}(-\vbbm_{s_i,i}\star \vbbm_{s_j,j})\geq 0;
		\label{condDel}
		\ee
		\item
		for all $\ell\in\Zv_n$ and any set of $s_i\in \{1,2\}$, $i\in\Zv_n\setminus\{\ell\}$,
		\be
		\vbbm_{1,\ell\perp\{s_i\}}\star \vbbm_{2,\ell\perp\{s_i\}}<0,
		\label{condscpr}
		\ee
		where $_{\perp\{s_i\}}$ denotes the projection on the subspace orthogonal to the span of
		$\{\vbbm_{s_i,i}\}_{i\in \Zv_n\setminus\{\ell\}}$.
	\end{enumerate}
	\label{th-converg}
\vsp}

Thus, to get a convergent holomorphic theta series, 
it is sufficient to take the kernel to be a product of differences of the sign functions
determined by a set of vectors of negative norm subject to a few conditions on their scalar products.
The number of factors should be equal to the number of negative eigenvalues of the quadratic form.
Importantly, if one considers theta series including an elliptic parameter (or a vector thereof),
one gets convergence even if some of the vectors $\vbbm_{s,i}$ are null, i.e. satisfy $\vbbm_{s,i}^2=0$,
provided they belong (possibly after a rescaling) to the lattice $\bbLambda$. 

Of course, the kernel \eqref{kerconverge} is not the only possibility to get a convergent indefinite theta series. 
For example, for $n=2$, in \cite{Alexandrov:2017qhn} 
an interesting cyclic combination of sign functions has been conjectured also to ensure the convergence.
The conjecture was given a geometric interpretation, proven and extended in \cite{funke2021casengon,Zhang_2024}.

An important consequence of having a non-trivial kernel is that it spoils modularity.
The theta series with a kernel constructed from sign functions transforms under $SL(2,\IZ)$ with a modular anomaly.
A remarkable fact is that any such theta series turns out to belong to the class of higher depth mock modular forms,
with the depth equal to $n$. 
In particular, for $n=1$ corresponding to the case of indefinite theta series of Lorentzian signature,
one obtains (mixed) mock modular forms, many of which are related to classic examples going back to Ramanujan.

As was explained in \S\ref{subsec-mock}, each mock modular form has a non-holomorphic modular completion.
So a natural and important question is: what are the completions of indefinite theta series?
For the case of the Lorentzian signature, the answer has been found in \cite{Zwegers-thesis}
and is extremely simple: the completion is given by a theta series with the kernel obtained by replacing each sign function
in \eqref{kerconverge} by the error function according to
\be  
\sgn(\vbbm\star \kbbm)\, \mapsto \Erf\(\sqrt{2\pi\tau_2}\, \frac{\vbbm\star \kbbm}{||\vbbm||}\),
\label{replaceLorentz}
\ee 
where $||\vbbm||=\sqrt{-\vbbm^2}$.
It turns out that for $n>1$ the recipe is very similar and can be formulated in terms of 
the generalized error functions described in appendix \ref{ap-generr}.
They can be seen as functions $\Phi_n^E(\{\vbbm_i\};\xbbm)$ of a vector $\xbbm$ and a set of $n$ vectors $\vbbm_i$,
which are given by a convolution of $\prod_{i=1}^n \sgn (\vbbm_i\star\,\xbbm)$ 
with a Gaussian kernel.
Then the recipe says \cite{Alexandrov:2016enp}:

\begin{quote}\it
To construct the completion of a theta series whose kernel is a combination of sign functions,
it is sufficient to replace each product of the sign functions according to the rule
\be 
\prod_{i=1}^n \sgn (\vbbm_i\star\,\kbbm)\ \mapsto \Phi_n^E\(\{\vbbm_i\};\sqrt{2\tau_2}\kbbm\).
\label{replace-alln}
\ee 
\end{quote}

Although it is easy to show {\it a posteriori} that the recipe \eqref{replace-alln} does produce 
a non-holomorphic modular form, to guess the functions $\Phi_n^E(\{\vbbm_i\};\xbbm)$ could have been an outstanding problem.
Fortunately, the guesswork was not required as string theory produced them for free!
The point is that such holomorphic indefinite theta series often arise in the analysis 
of Calabi-Yau compactifications because, as will be explained in \S\ref{sec-genfun}, 
the lattice $\Lambda=H_4(\CY,\IZ)$ has signature $(1,b_2(\CY)-1)$. However, typically in physics, modular symmetry
is more fundamental than holomorphicity (see, e.g., \cite{Dabholkar:2021lzt}). 
Therefore, the final physical results should be expressible
in terms of (possibly non-holomorphic) modular forms rather than holomorphic mock modular forms.
In particular, this implies that string theory should ``know" about the non-holomorphic modular completions
of indefinite theta series and they can be found by a (not necessarily simple) calculation.
This is precisely what was done in \cite{Alexandrov:2017qhn} and led to the introduction of 
the generalized error functions in \cite{Alexandrov:2016enp,Nazaroglu:2016lmr} and to the above construction
(see also \cite{funke2017mock}).

We finish this section by presenting a very general result on modularity of indefinite theta series 
which is a straightforward generalization of a Theorem proven by Vign\'eras in \cite{Vigneras:1977}.
As above, we take $\bbLambda$ to be a $d$-dimensional lattice equipped with an integer valued bilinear form.
But from this moment, to agree with most of the relevant literature (see footnote \ref{foot-sign}), 
we change conventions and take the bilinear form to be {\it opposite} to the one considered before. 
Hence, the associated quadratic form is assumed to have signature $(n,d-n)$ so that 
the case of convergent theta series with trivial kernel corresponds to a negative definite quadratic form. 
To avoid confusion with the previous conventions, we use the symbol $\ast$ instead of $\star$ for the bilinear form.
Besides, we take $\bbmu \in \bbLambda^*/\bbLambda$ to be a residue class and 
$\pbbm$ a characteristic vector satisfying 
$\kbbm\ast(\kbbm+\pbbm) =0 \mod 2$ for $\forall \kbbm\in \bbLambda$, 
which allows to deal with odd lattices. Finally, $\zbbm=\bbalpha-\tau\bbbeta\in\IC^d$
with $\bbalpha,\bbbeta\in\IR^d$ will be a vector of elliptic parameters.
Using these notations, we define
\be
\vth_{\bbmu}(\tau, \zbbm;\bbLambda, \Phi, \pbbm) = \sum_{\kbbm\in \bbLambda + \bbmu + \hf \pbbm} 
(-1)^{\pbbm\ast \kbbm} \Phi\(\sqrt{2\tau_2}\(\kbbm+\bbbeta \) \) \q^{-\hf \kbbm^2} 
e^{2\pi\I\zbbm\ast \kbbm}.
\label{gentheta}
\ee
The Vign\'eras theorem \cite{Vigneras:1977} asserts that if the kernel $\Phi(\xbbm)$ satisfies 
suitable decay properties as well as
the following differential equation 
\be
\[\p_\xbbm^2+2\pi (\xbbm\ast\p_\xbbm-\lambda)\]\Phi(\xbbm)=0,
\label{Vigdif}
\ee
where $\lambda$ is an integer parameter,
then the theta series is a vector valued (multi-variable) Jacobi form\footnote{More precisely, 
	the elliptic transformation \eqref{Jacobi-ell} can generate an additional sign factor 
	$(-1)^{\pbbm\ast(\abbm+\bbbm)}$.} 
with the following weight, index and multiplier system 
\be
\begin{split}
	w(\vth) &= \(\hf \,(d+\lambda), -\hf\, \lambda\),
	\qquad\qquad
	m(\vth) = -\hf\, \ast,
	\\
	\Mi{\vth}_{\bbmu \bbnu}(T) &= e^{-\pi \I\(\bbmu + \hf \pbbm\)^2 } \delta_{\bbmu\bbnu},
	\qquad
	\Mi{\vth}_{\bbmu \bbnu}(S) = \frac{e^{(2n-d)\frac{\pi\I}{4}}}{\sqrt{|\bbLambda^*/\bbLambda|}} \,
	e^{\frac{\pi \I}{2}\pbbm^2} e^{2\pi \I \bbmu \ast \bbnu},
\end{split}
\label{mult-genth}
\ee
where by $\ast$ in the formula for the index we mean the matrix representing the bilinear form.
If one takes $\zbbm=\bbtheta z$ where $\bbtheta\in \bbLambda$, so that the multi-variable Jacobi form is reduced 
to a usual Jacobi form, the index is a scalar and is given by
$
m(\vth)=-\hf\, \bbtheta^2.
$

\section{BPS indices and their generating functions}
\label{sec-genfun}

\subsection{BPS indices in type IIA/CY}

Let us now turn to physics and consider type IIA string theory compactified on a CY threefold $\CY$.
In four non-compact dimensions, one gets an effective theory given by $\cN=2$ supergravity coupled to 
$b_2=h^{1,1}(\CY)$ vector multiplets and $h^{2,1}(\CY)+1$ hypermultiplets.
The theory has $b_2+1$ abelian gauge fields which include the graviphoton belonging to the gravitational multiplet.

BPS states are labeled by an electro-magnetic charge which can be represented as a vector with $2b_2+2$ components
and is denoted by $\gamma=(p^0,p^a,q_a,q_0)$ where $a=1,\dots,b_2$ labels vector multiplets.
At strong string coupling these BPS states appear as black hole solutions of the effective theory, 
while at small string coupling they are realized as bound states of D6, D4, D2 and D0 branes 
wrapping 6, 4, 2 and 0-dimensional cycles of $\CY$, respectively. The components of the charge vector
correspond to the respective D-brane charges.
Note that the magnetic charges $p^0$ and $p^a$ are always integer, whereas the electric charges
are in general rational due to a non-trivial quantization condition \cite{Minasian:1997mm}.

The BPS states are counted (with sign) by a {\it BPS index} $\Omega(\gamma)$ which is stable 
under deformations of the string coupling and other hypermultiplet moduli. 
This is the reason why it takes the same values
in the two extreme regimes corresponding to four-dimensional black holes and 
to D-branes wrapped on the internal manifold. One can regard the latter picture as a microscopic realization
of the quantum states responsible for the black hole entropy and counted by the BPS index.\footnote{One could worry
	that since the BPS index counts bosons and fermions with different signs, it can differ from the actual degeneracy
	of BPS black holes. Fortunately, it was shown that in most situations this is not the case as 
	all relevant states are bosonic \cite{Sen:2009vz,Dabholkar:2010rm,Iliesiu:2022kny}.} 

However, in general, $\Omega(\gamma)$ does not remain constant under deformations of 
the (complexified) K\"ahler moduli of $\CY$ which parametrize the vector multiplet moduli space.
This moduli space is divided by {\it walls of marginal stability} 
into chambers with different values of $\Omega(\gamma)$, and the jump of the index between different chambers
is called {\it wall-crossing}. As explained in the Introduction, it happens due to the decay or formation of 
bound states contributing to the index. Its existence is responsible for many non-trivial phenomena 
and an extremely rich mathematical structure in theories with eight supercharges.

\subsection{DT invariants}
\label{subsec-DT}

From a mathematical viewpoint, BPS indices are a particular instance of 
{\it generalized Donaldson-Thomas invariants} \cite{Joyce:2008pc}
which compute the weighted Euler characteristic of the moduli space of coherent sheaves $E$ on $\CY$ 
having a Chern vector determined by the charge $\gamma$ and satisfying a certain stability condition.
In particular, the D6-brane charge $p^0$ determines the rank of the corresponding sheaf.
Although this relation is not crucial for understanding what follows (except \S\ref{subsec-topinv}), 
nevertheless we explain a few useful facts about the generalized DT invariants. 

First, we introduce a basis $(1,\omega_a,\omega^a,\omega_{\CY})$ of the even cohomology $H^{\rm even}(\CY)$
satisfying
\be 
\omega_a\wedge\omega_b=\kappa_{abc}\omega^c,
\qquad
\omega_a\wedge \omega^b =\delta_a^b\omega_{\CY},
\ee 
where $\kappa_{abc}$ are the triple intersection numbers of $\CY$, 
and combine the charge components into a differential form
\be 
\gamma(E)=p^0+p^a\omega_a-q_a\omega^a+q_0\omega_{\CY}.
\ee 
Then the D-brane charge and the Chern vector of the corresponding coherent sheaf satisfy
the following nice relation
\cite{Douglas:2006jp,Alexandrov:2010ca}
\be 
\gamma(E)=\ch(E)\sqrt{\Td(T\CY)},
\label{rel-gamCh}
\ee 
where $\Td(T\CY)$ is the Todd class of the tangent bundle.
Expanding the r.h.s. of \eqref{rel-gamCh} in the same cohomology basis, one finds explicit relations
between components
\be
p^0=\ch_0,
\qquad
p^a=\ch_1^a,
\qquad
q_a=-\ch_{2,a}-\frac{c_{2,a}}{24}\,\ch_0,
\qquad
q_0=\ch_3+\frac{c_{2,a}}{24}\,\ch_1^a,
\label{ch-Ch}
\ee 
where $c_{2,a}$ are the components of the second Chern class of $\CY$.

Second, to see why the physical BPS indices are only a particular instance of DT invariants,
note that, as mentioned above, the definition of the latter involves a stability condition.
This condition can be described\footnote{The central charge must be supplemented by the so called heart of a bounded $t$-structure
on the derived category of coherent sheaves. We will ignore this in our discussion.} 
by a {\it central charge} $Z_\gamma$ \cite{MR2373143}, 
which is a complex valued linear function on the charge lattice and hence can be represented as  
\be 
Z_\gamma=q_\Lambda X^\Lambda-p^\Lambda \cF_\Lambda ,
\ee 
where $\Lambda=(0,a)$ runs over $b_2+1$ values.
The vector $(X^\Lambda,\cF_\Lambda)$ parametrizes the space $\cS$ of stability conditions modulo an action of some symmetry group.
The point is that the stability conditions for which the DT invariants coincide with the BPS indices 
form only a Lagrangian subspace (with respect to the natural symplectic form $\de X^\Lambda\wedge \de\cF_\Lambda$)
of the full space $\cS$. They are called $\Pi$-stability \cite{Douglas:2000ah} and correspond 
to the slice $\cF_\Lambda=\p_{X^\Lambda} F(X)$ where $F(X)$ is the holomorphic prepotential
of the special K\'ahler geometry of the K\"ahler moduli space of $\CY$.
As expected, they are parametrized by the complexified K\'ahler moduli $z^a=b^a+\I t^a$.
 
Next, on the physical slice
DT invariants are known to be invariant under monodromy transformations around singularities in the moduli space:
\be 
\Omega(\cM\cdot \gamma; \cM\cdot z)=\Omega(\gamma;z),
\ee 
where $\cM$ represents a monodromy and we explicitly indicated the (piece-wise constant) dependence on the moduli.
It is important to take into account that the monodromy acts not only on charges, but also on the moduli, 
because its action can bring from one chamber to another.
A particularly important class of monodromies is around the large volume point $t^a=\infty$. They have 
a simple mathematical interpretation as tensoring the sheaf with a line bundle leading to the transformation
$\ch(E)\mapsto e^{\eps^a\omega_a}\ch(E)$ with $\eps^a\in\IZ$. 
In physics, it is known as ``spectral flow" transformation and 
it acts on the moduli by shifting their real part corresponding to the B-field, $b^a\mapsto b^a+\eps^a$.

In fact, the large volume region of the moduli space is where most computations 
are done and which we are really interested in. In this limit all quantum corrections to 
the prepotential become subleading and it has a simple cubic form
\be 
\Fcl(X)=-\frac{1}{6X^0}\,\kappa_{abc}X^a X^b X^c.
\label{Fcl}
\ee 
Furthermore, the generalized DT invariants of rank 1 (for sufficiently large negative $b^a$)
coincide with the ordinary DT invariants defined in \cite{Thomas:1998uj}, 
while for rank $-1$ (and sufficiently large positive $b^a$) they reproduce the 
Pandharipande-Thomas (PT) invariants \cite{pandharipande2009curve}.\footnote{Up to a torsion factor (see below \eqref{conseqTh1}),
	if we relax the condition on $\CY$ to be simply connected.}
Note that in these cases some stability walls extend to the large volume region.
This is why the limit $t^a\to \infty$ of $\Omega(\gamma;z)$ depends on the B-field.

Finally, an important fact is that the generalized DT invariants, and hence the physical BPS indices, 
satisfy universal wall-crossing formulas \cite{ks,Joyce:2008pc}
which are valid for any stability condition.
They allow to express $\Omega(\gamma;z)$ on one side of a wall from their values on the other side.
We will not present the general formula and restrict ourselves to the simplest case 
corresponding to the primitive wall-crossing \cite{Denef:2007vg} which describes 
the decay or formation of a bound state of two constituents with charges $\gamma_1$ and $\gamma_2$.
In this case the jump across the wall is given by
\be
\Delta\Omega(\gamma;z)
=(-1)^{\langle\gamma_1,\gamma_2\rangle-1}|\langle\gamma_1,\gamma_2\rangle|
\Omega(\gamma_1,z)\Omega(\gamma_2,z), 
\label{prim-wc}
\ee 
where the l.h.s is the difference between DT invariants for $\gamma=\gamma_1+\gamma_2$
on the two sides of the wall (in the chamber where the bound state exists minus where it does not), 
the DT invariants on the r.h.s. are evaluated at the wall, 
and $\langle\gamma_1,\gamma_2\rangle$ is the anti-symmetric Dirac product of charges,
equal to the Euler-Poincar\'e pairing of the corresponding Chern vectors,
\be
\langle\gamma_1,\gamma_2\rangle=p_2^\Lambda q_{1,\Lambda}-p_1^\Lambda q_{2,\Lambda}.
\label{defDirac}
\ee 
At least in principle, the knowledge of wall-crossing relations reduces the problem of finding the BPS spectrum
on the whole moduli space to finding it in a particular chamber. Of course, in practice, even going from one chamber 
to another might be non-trivial, not to mention that the chamber structure can itself be very intricate.

\subsection{Generating functions of D4-D2-D0 BPS indices}
\label{subsec-genfun}

Let us now specialize to the case of D4-D2-D0 bound states. 
Since $p^0=0$, the BPS indices counting them are the same as rank 0 DT invariants.
Our goal in this section will be to assemble these indices into generating functions that have a ``nice" 
behavior under modular transformations.

The first problem that we should solve to this end is due to wall-crossing:
given that the BPS indices take different values in different regions of the moduli space,
where should they be evaluated to exhibit modular properties? To answer this question, we first note that
each charge gives rise to a distinguished point in the moduli space --- the {\it attractor point}.
It is provided by the attractor mechanism of $\cN=2$ supergravity which states that, 
independently of the values of moduli $z^a$ at infinity, on the horizon of a single-centered black hole,
they take fixed values completely determined by the charge $\gamma$ \cite{Ferrara:1995ih}.
One can also show that in the vicinity of the attractor point there are no multi-centered 
black holes except the ones described by scaling solutions involving at least 3 constituents \cite{Denef:2007vg,Bena:2012hf}, 
so that the BPS index counts only the latter and single-centered ones \cite{Manschot:2009ia}.

However, different charges have different attractor points.
Besides, on general grounds one expects that modularity is closely related to the existence of a CFT description
\cite{Maldacena:1997de}. 
In our case it can be seen as a holographic description of the ${\rm AdS}_3\times S^2$ 
near horizon geometry in the M-theory picture.
In \cite{deBoer:2008fk} it was shown that the BPS indices counting states in such theory decoupled from the bulk
correspond to the DT invariants evaluated at the {\it large volume attractor point}.
The latter is defined as the attractor point for the classical prepotential \eqref{Fcl} and a D4-D2-D0 charge 
rescaled by an infinite factor,
\be
\label{lvolatt}
z^a_\infty(\gamma)
= \lim_{\lambda\to +\infty}\(-\kappa^{ab} q_b+\I\lambda p^a\),
\ee
where $\kappa^{ab}$ is the inverse of the quadratic form $\kappa_{ab}=\kappa_{abc}p^b$ which is
defined by the D4-brane charge and will play an important role from here to the end. 
Following \cite{Alexandrov:2012au}, we denote the resulting BPS indices by
\be 
\OmMSW(\gamma)=\Omega(\gamma,z_\infty(\gamma))
\label{defMSW}
\ee
and call them Maldacena-Strominger-Witten (MSW) invariants.

The definition \eqref{defMSW} ensures that $\OmMSW(\gamma)$ are invariant under monodromies
around the large volume or spectral flow transformations. 
For vanishing $p^0$, they leave the D4-brane charge $p^a$ invariant and act on the D2-D0 charges by
\be 
q_a\mapsto q_a-\kappa_{ab} \eps^b,
\qquad
q_0\mapsto q_0 -\eps^a q_a+\hf\, \kappa_{ab}\eps^a\eps^b.
\label{spectr-flow}
\ee 
The spectral flow parameter $\eps$ can be thought of as an element of the lattice $\Lambda_p=H_4(\CY,\IZ)$
where we put the index $p$ to indicate that the lattice is endowed with the quadratic form $\kappa_{ab}$
determined by $p^a$.
This suggests to decompose the D2-brane charge as
\be
q_a=\kappa_{ab}\eps^b +\mu_a+\hf\,\kappa_{ab}p^b,
\label{decomp-spfl}
\ee 
where $\mu\in \Lambda_p^*/\Lambda_p$ with $\Lambda_p^*=H_2(\CY,\IZ)$ 
is a residue class similar to those appearing in \S\ref{sec-indef}, and the last
term appears due to a non-trivial quantization condition of $q_a$. Then the spectral flow invariance of $\OmMSW(\gamma)$
implies that they depend only on the D4-brane charge $p^a$, residue class $\mu_a$ and invariant combination
of D2 and D0 charges 
\be
\hq_0=q_0-\hf\, \kappa^{ab}q_a q_b.
\label{def-hq0}
\ee 
Thus, we can set $\OmMSW(\gamma)=\Omega_{p,\mu}(\hq_0)$.

Another important observation made in \cite{Manschot:2010xp} is that modularity is expected to be manifest 
not for the integer valued BPS indices $\Omega(\gamma)$, but rather for their rational counterparts defined
for a generic charge $\gamma$ as 
\be 
\bOm(\gamma) = \sum_{d|\gamma} \frac{1}{d^2}\, \Omega(\gamma/d).
\label{def-bOm}
\ee 
Another advantage of the rational BPS indices is that they possess simpler properties under wall-crossing 
\cite{Manschot:2010qz,Alexandrov:2018iao}.

The final ingredient necessary for our construction is the Bogomolov-Gieseker bound \cite{bayer2016space}
which states that $\bOm_{r,\mu}( \hq_0)$ vanishes unless the invariant charge $\hq_0$ satisfies
\be
\hq_0 \leq \hq_0^{\rm max}=\frac{1}{24}\,(\kappa_{ab}p^ap^b+c_{2,a}p^a).
\label{qmax}
\ee
Combining it with the previous observations, we arrive at the following definition of generating series 
of D4-D2-D0 BPS indices
\be
h_{p,\mu}(\tau) =\sum_{\hq_0 \leq \hq_0^{\rm max}}
\bOm_{p,\mu}(\hq_0)\,\q^{-\hq_0 }\, .
\label{defhDT}
\ee
Since the index $\mu$ takes a finite number of values,
it can be thought of as a vector index and $h_p(\tau)$ as vector valued functions labeled by the D4-brane charge.
In general, the order of the discriminant group $\Lambda_p^*/\Lambda_p$ where $\mu$ takes values equals $|\det\kappa_{ab}|$, 
but half of it is redundant due to the symmetry under $\mu \mapsto -\mu$ following from 
dualization of the coherent sheaf induced by the D4-brane.

The generating series $h_{p,\mu}(\tau)$ \eqref{defhDT} will be the central object of our study in what follows.
In particular, we will be interested in their properties under modular transformations acting on $\tau$
and how these properties can be used to find the generating functions explicitly.

\section{Modular anomaly}
\label{sec-anomaly}

\subsection{Origin of modularity}

So far the expectation that the generating functions $h_{p,\mu}$ should possess some nice modular properties
was based on a connection with CFT. However, it is not very precise and at this point it is not known 
how it can be used to get modular properties of $h_{p,\mu}$ for generic D4-brane charge. 
Instead, there is another approach based on a target space picture that allows us not only to explain why 
these functions should be modular, but also to derive their precise behavior under modular transformations.

The idea is to compactify the original setup on an additional circle. After that we have type IIA string theory on 
$\CY\times S^1$ which leads to a three-dimensional effective theory. Since in three dimensions all vector fields 
can be dualized to scalars, the low energy effective theory can always be represented as some 
supersymmetric non-linear sigma model characterized by a moduli space $\cM$.
The effective action is determined by the metric on $\cM$. At classical level, it can be obtained by 
the standard Kaluza-Klein reduction of 4d $\cN=2$ supergravity coupled to vector multiplets.
However, at the quantum level, it receives quantum corrections, both perturbative and non-perturbative.
The latter appear as instanton effects due to BPS particles of the four-dimensional theory winding the circle.
Therefore, they are characterized by the electro-magnetic charge $\gamma$  
and weighted by the corresponding BPS indices $\Omega(\gamma)$, 
which roughly count the number of instantons of this charge.
Thus, we conclude that the metric on the moduli space $\cM$ depends on the BPS indices we are interested in.

Furthermore, we actually know this dependence explicitly!
First, one should note that from the ten-dimensional viewpoint
the instantons correcting the metric on $\cM$ arise as D-branes wrapping 
even-dimensional cycles on $\CY$ times the circle $S^1$.
In particular, we are interested in the instanton corrections generated by D4-branes on a divisor $\cD_p\subset\CY$ and $S^1$.
Second, applying T-duality along the compactification circle, one arrives at type IIB string theory
compactified on the same manifold $\CY\times S^1$. In this T-dual formulation, the moduli space $\cM$ is nothing but
the hypermultiplet moduli space of type IIB on $\CY$ since the additional circle compactification does not affect it. 
All D-instanton corrections to the metric on this moduli space have been computed in a series of works 
\cite{Alexandrov:2008gh,Alexandrov:2009zh,Alexandrov:2009qq,Alexandrov:2014sya} using a twistorial formalism
which allows us to deal with the complicated quaternion-K\"ahler geometry of $\cM$
(see \cite{Alexandrov:2011va} for a review).
In particular, for D3-instantons, the T-dual of the ones generated by D4-branes in type IIA, 
it is possible to obtain their corrections to the metric order by order in the expansion
in the instanton number \cite{Alexandrov:2017mgi}, 
and the BPS indices weighting them are the same as we had in the type IIA formulation.

Another important consequence of the duality with type IIB string theory is that it is known to be invariant 
under the S-duality group which coincides with $SL(2,\IZ)$.
This invariance manifests in the effective theory obtained by compactification 
as the existence of an $SL(2,\IZ)$ isometric action on the hypermultiplet moduli space.
Hence, the metric on $\cM$ must be invariant under $SL(2,\IZ)$!

\begin{figure}
	\isPreprints{\centering}{} 
	\includegraphics[width=10.2cm]{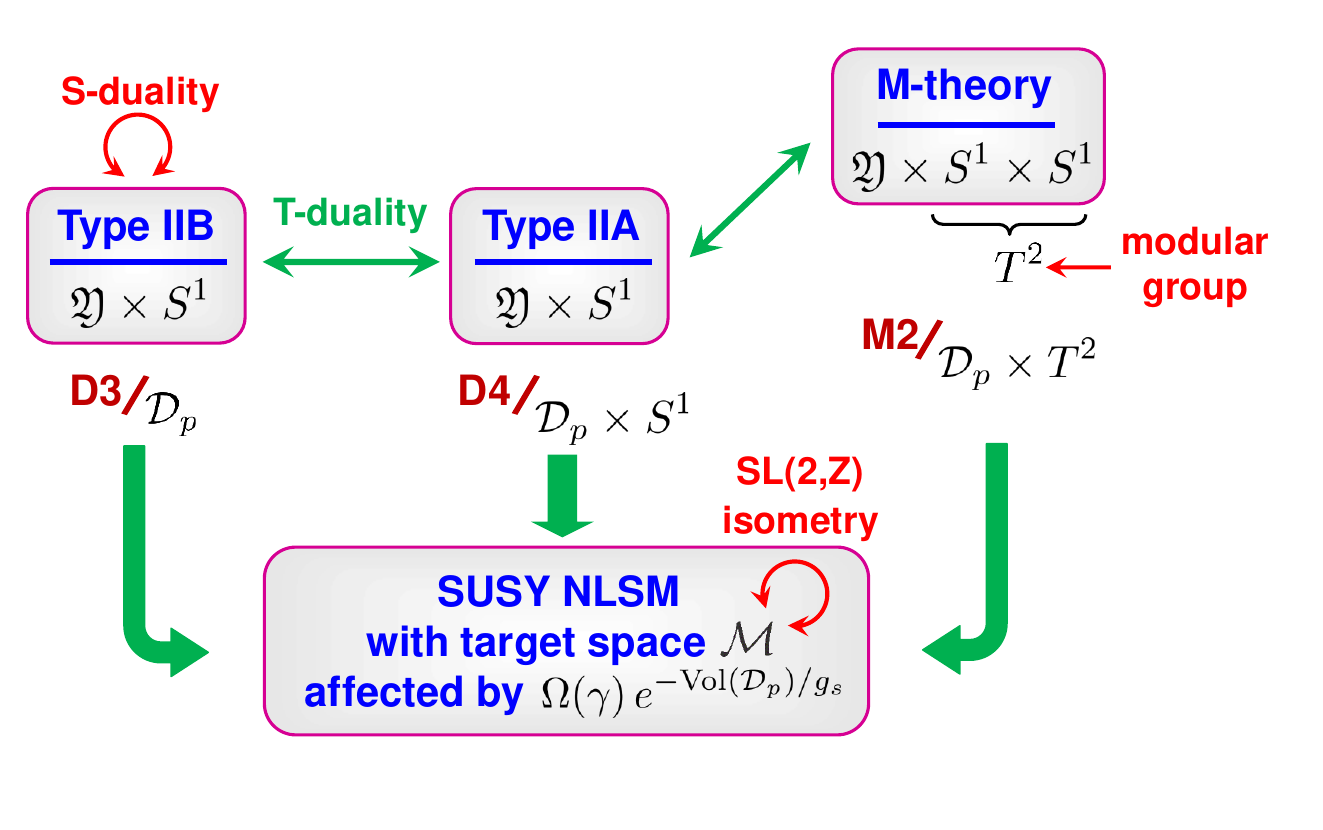}
	\vspace{-0.5cm}
	\caption{The scheme of dualities illustrating the origin of modularity of D4-D2-D0 BPS indices.\label{fig-scheme}}
\end{figure}

One arrives at the same conclusion if, instead of applying T-duality, one realizes type IIA 
as M-theory on a circle. Then our setup is equivalent to M-theory compactified on $\CY\times T^2$
and therefore it should be invariant under the modular group of the torus.
This torus can be seen as the geometric origin of the $SL(2,\IZ)$ isometry acting on $\cM$.
We illustrated the different duality frames and the corresponding brane wrappings responsible 
for the relevant instanton effects in three dimensions in Fig. \ref{fig-scheme}.

To summarize, there is a moduli space that, on one hand, has a known dependence on BPS indices
and, on the other hand, carries an isometric action of the modular group. So a natural question is 
whether this action is consistent with arbitrary values of the BPS indices or imposes some restrictions on them?
To answer this question, note that the action of S-duality on quantum effects in the type IIB frame 
splits into four orbits which, for large K\"ahler parameters, can be organized in a hierarchical order:
\begin{enumerate}
	\item 
	D(-1)-instantons mixed with perturbative $g_s$ and $\alpha'$-corrections;
	\item
	D1-instantons mixed with worldsheet instantons (known also as $(p,q)$-strings);
	\item 
	D3-instantons;
	\item 
	D5-instantons mixed with NS5-instantons  (known also as $(p,q)$-five-branes).
\end{enumerate}
One observes that in three of the four orbits $SL(2,\IZ)$ mixes different types of quantum corrections.
This fact has actually been used to find one type of corrections from a known other type 
just by applying the method of images
\cite{RoblesLlana:2006is,Alexandrov:2010ca,Alexandrov:2014rca,Alexandrov:2023hiv}.
This procedure is independent of the values of $\Omega(\gamma)$ and therefore does not constrain them.
At the same time, D3-instantons are invariant under S-duality which makes them very special.
They have been determined {\it without} imposing S-duality and nevertheless they must respect it.
Therefore, one can expect that the compatibility with S-duality imposes certain conditions on the BPS indices
associated with these instantons, which are precisely the BPS indices counting D4-D2-D0 bound states in 
the type IIA formulation. A remarkable fact is that these conditions can be derived explicitly using the knowledge 
of the instanton corrected metric on $\cM$.

\subsection{Sketch of the derivation}
\label{subsec-deriv}

A derivation of the modular constraints on the D4-D2-D0 BPS indices has been done, 
first, at one-instanton order in \cite{Alexandrov:2012au},
then at two-instanton order in \cite{Alexandrov:2016tnf} and, finally, at all orders 
in the instanton expansion in \cite{Alexandrov:2018lgp}.
As we will see, here the instanton order can be associated with the degree of reducibility
of the divisor wrapped by the D4-brane, i.e. the number $r$ of {\it irreducible} divisors
appearing in the decomposition $\cD_p=\sum_{i=1}^r \cD_{p_i}$, or simply $p^a=\sum_{i=1}^r p_i^a$.
The derivation was done assuming that $\cD_p$ is an {\it ample} divisor, i.e. the vector $p^a$ belongs 
to the K\"ahler cone.\footnote{In the basis of the even homology constructed from the generators of the K\"ahler cone, 
	the K\"ahler moduli are positive $t^a>0$, the intersections numbers are non-negative $\kappa_{abc}\geq 0$, 
	and the K\"ahler cone condition ensures that D4-brane charges are also non-negative $p^a\geq 0$.}
In particular, this ensures that $(p^3)>0$ where we introduced the notation $(xyz)=\kappa_{abc}x^a y^b z^c$.
In this subsection we present the main steps of the derivation
and, if the reader is not interested in this, one can safely skip it.

In fact, for the purpose of obtaining the modular constraints, it is not necessary to work
with the full metric on $\cM$ which is a highly complicated object.
Instead, one can consider a certain function on $\cM$, known as {\it contact potential} $e^\phi$ 
\cite{Alexandrov:2008nk}, which is related to the four-dimensional string coupling and 
must be a (non-holomorphic) modular form of weight $(-\hf,-\hf)$ for S-duality to be realized consistently 
with the quaternion-K\"ahler property of the metric \cite{Alexandrov:2008gh}. 
Furthermore, in the large volume limit, the part of the contact potential affected by D3-instantons in the type IIB picture
can be expressed through an even more fundamental function $\cG$, 
dubbed in \cite{Alexandrov:2018lgp} as {\it instanton generating potential},
\be
(e^\phi )_{\rm D3}
=\frac{\tau_2}{2}\Re\left(\cD_{-\frac{3}{2}}\cG \right)
+\frac{1}{32\pi^2}\,\kappa_{abc}t^c\p_{\tc_a}\cG\p_{\tc_b}\overline{\cG},
\label{Phitwo-tcF}
\ee
where $\tau_2$ is the inverse of the ten-dimensional string coupling and
the imaginary part of the axio-dilaton field $\tau=\tau_1+\I\tau_2$,
$\tc_a$ is the RR-axion coupled to D3-branes, and $\cD_{w}$ is
the Maass raising operator mapping modular forms of weight $(w,\bw)$ to modular forms of weight $(w+2,\bw)$.
Since S-duality transforms K\"ahler moduli as modular forms, $t^a\mapsto |c\tau+d|t^a$, and leaves $\p_{\tc_a}$ invariant,
the relation \eqref{Phitwo-tcF} shows that one gets the required transformation for the contact potential
if and only if $\cG$ transforms as a modular form of weight $(-\frac32\, ,\, \hf)$.
Thus, one should just calculate this function in terms of the MSW invariants and find under which conditions 
it has these transformation properties.

The instanton generating potential has a simple expression in terms of the 
Darboux coordinates on the twistor space denoted by $\cX_\gamma$.
These coordinates are the main object of the twistorial construction of the metric on $\cM$
because, once they are found, there is a straightforward, albeit non-trivial procedure to get the metric \cite{Alexandrov:2008nk}.
They are functions of all moduli and an additional coordinate $\zsf$ on the $\CP$-fiber of the twistor bundle over $\cM$,
which are determined as solutions of the following TBA-like equation\footnote{The integral equation \eqref{expcX}
is the large volume limit of a more general equation that holds for all charges $\gamma$ and appears to be identical
to the equation put forward in \cite{Gaiotto:2008cd} which describes an instanton corrected hyperk\"ahler target space
of a three-dimensional sigma model obtained by a circle compactification
of a four-dimensional $\cN=2$ gauge theory. The identification between D-instantons in string theory on a CY and 
instantons in $\cN=2$ gauge theory on a circle suggested by the coincidence of the equations encoding them
has its origin in the QK/HK correspondence \cite{Haydys,Alexandrov:2011ac} establishing a relation between 
the two types of quaternionic manifolds.}
\be
\cX_\gamma(\zsf)=\cXcl_\gamma(\zsf) \, \exp\[\sum_{\gamma'\in\Gamma_+}\bOm(\gamma')
\int_{\ell_{\gamma'}}\de \zsf'\, K_{\gamma\gamma'}(\zsf,\zsf')\,\cX_{\gamma'}(\zsf')\],
\label{expcX}
\ee
where the dependence on the moduli, not shown explicitly, is hidden in the functions $\cXcl_\gamma$ 
(and $K_{\gamma\gamma'}$).
Here $\Gamma_+$ is the lattice of charges with $p^0=0$ and $p^a$ belonging to the K\"ahler cone,
$\ell_\gamma$ is a contour known as the BPS ray which in the large volume approximation in the $\zsf$-plane
can be seen as a line passing through the saddle point $\zsf_\gamma=-\I (q_at^a+(pbt))/(pt^2)$,
$K_{\gamma\gamma'}(\zsf,\zsf')$ is an integration kernel given by
\be
K_{\gamma_1\gamma_2}(\zsf_1,\zsf_2)
=\frac{1}{2\pi}\((tp_1p_2)+\frac{\I\langle\gamma_1,\gamma_2\rangle}{\zsf_1-\zsf_2}\),
\label{defkerK}
\ee
and $\cXcl_\gamma$ is the classical limit of the Darboux coordinates which can be written as
\be
\cXcl_\gamma(\zsf) =e^{ - 2\pi\I \hq_0\tau}\, e^{-\pi\I \tau(q+b)^2+2\pi\I c^a q_a-2\pi\tau_2(pt^2)(\zsf-\zsf_\gamma)^2+\dots },
\label{clactinst}
\ee
where $\hq_0$ is the invariant charge \eqref{def-hq0}, $c^a$ is the RR-axion coupled to D1-branes,
and we dropped some terms irrelevant for our discussion.
For what follows it is important to note that the dependence on $\hq_0$ factorizes, while the dependence on
the charge $q_a$ is Gaussian with the quadratic form $q^2=\kappa^{ab}q_a q_b$ 
determined by $p^a$ via $\kappa_{ab}=\kappa_{abc}p^b$ and having appeared already in \S\ref{subsec-genfun}.

Returning to the function $\cG$, it is given by a double integral of the quantum corrected Darboux coordinates
\be
\cG= \frac{1}{4\pi^2}\sum_{\gamma\in\Gamma_+}\bOm(\gamma)\int\limits_{\ell_{\gamma}} \de \zsf\, \cX_{\gamma}(\zsf)
-\frac{1}{8\pi^2} \sum_{\gamma_1,\gamma_2\in\Gamma_+}\bOm(\gamma_1)\bOm(\gamma_2)
\int\limits_{\ell_{\gamma_1}}\de \zsf_1
\int\limits_{\ell_{\gamma_2}} \de \zsf_2
\, K_{\gamma_1\gamma_2}(\zsf_1,\zsf_2)\,\cX_{\gamma_1}(\zsf_1)\cX_{\gamma_2}(\zsf_2).
\label{defcF2}
\ee
To express it in terms of the MSW invariants in a form suitable for extracting their modular properties,
one should follow several steps.
\begin{enumerate} 
\item 
First, we need to compute the Darboux coordinates $\cX_\gamma(\zsf)$ to be substituted into \eqref{defcF2}.
Unfortunately, the TBA-like equation \eqref{expcX} cannot be solved in a closed form.
Nevertheless, it can always be solved by iterations: first plug in $\cXcl_\gamma$ into the r.h.s.
to get $\cX_\gamma$ up to the first order, then plug in the result to get the second order, and so on.
This procedure produces an asymptotic expansion which can be seen as an expansion in the number of instantons 
or, equivalently, in powers of the BPS indices $\bOm(\gamma)$.
The resulting perturbative solution can be expressed as a sum over {\it rooted trees} with vertices labeled by charges.
Below we will find many more different trees, so this is our first step into a ``forest".

\item 
The next step is to substitute the perturbative solution into the instanton generating potential
and to re-expand it in powers of $\bOm(\gamma)$. The result can again be written as a sum over trees, 
but this time these are {\it unrooted labeled trees}:
\be
\cG=\frac{1}{4\pi^2}\sum_{n=1}^\infty\[\prod_{i=1}^{n} \frac{1}{n!}
\sum_{\gamma_i\in \Gamma_+}\bOm(\gamma_i)\int_{\ell_{\gamma_i}}\de \zsf_i\, \cXcl_{\gamma_i}(\zsf_i) \]
\sum_{\cT\in\, \IT_n^\ell} \prod_{e\in E_{\cT}}
K_{\gamma_{s(e)},\gamma_{t(e)}}(\zsf_{s(e)},\zsf_{t(e)}),
\label{treeF}
\ee
where $n$ is the number of vertices, the last product goes over all edges of a tree $\cT$, and 
$s(e)$, $t(e)$ denote the source and target vertex of an edge $e$. 

\item 
The expansion \eqref{treeF} is not yet exactly what we want because it is expressed through 
the rational DT invariants $\bOm(\gamma)$, while we are looking for an expression in terms of the MSW invariants 
$\OmMSW(\gamma)$. The difference between them is due to the fact explained in \S\ref{subsec-DT} 
that the DT invariants are not actually constant, but depend on the moduli.
Fortunately, it is possible to express $\bOm(\gamma;z)$ through $\OmMSW(\gamma)$ using 
the {\it split attractor flow conjecture} which allows to count contributions of all bound states
to an index in terms of indices evaluated at their attractor points \cite{Denef:2001xn,Denef:2007vg}.
The result is represented as a sum over all possible consecutive splits of bound states into their constituents
and can be conveniently written as
\be
\bOm(\gamma,z) =
\sum_{\sum_{i=1}^n \gamma_i=\gamma}
\gtr(\{\gamma_i\}, z)\,
\prod_{i=1}^n \bOmMSW(\gamma_i) , 
\label{Omsumtreelv}
\ee
where the sum runs over ordered decompositions of charge $\gamma$ into elements of $\Gamma_+$ and 
the weight $\gtr(\{\gamma_i\}, z)$ is known as the {\it tree index}. 
It is given by a sum over yet another type of trees known as {\it attractor flow trees},
which are binary rooted trees with $n$ leaves labeled by $\gamma_i$ and other vertices labeled 
by charges equal to the sum of the charges of their children.
The contribution of each tree is a simple combination of factors given by the Dirac product of charges \eqref{defDirac} 
and certain sign functions responsible for the piece-wise constant moduli dependence of $\bOm(\gamma,z)$. 
In \cite{Alexandrov:2018iao} an alternative representation for the tree index has been found
that provided important hints for the construction explained below.

\item 
After one substitutes \eqref{Omsumtreelv} into \eqref{treeF}, one can make two observations.
First, since for charges with vanishing $p^0$ the tree index is independent of their $q_0$ components,
the only dependence on these components is in $\cXcl_{\gamma_i}$ through the factor $e^{-2\pi\I \hq_{i,0}\tau}$
(see \eqref{clactinst}) and in the MSW invariants. 
Therefore, the sum over $q_{0,i}$ gives rise precisely to the generating functions $h_{p_i,\mu_i}(\tau)$ \eqref{defhDT}
where $\mu_i$ are the residue classes appearing in the decomposition \eqref{decomp-spfl} of charges $q_{i,a}$.
Second, due to the spectral flow invariance, the MSW invariants and hence their generating functions 
are independent of the parameters $\eps_i$ also appearing in the decomposition \eqref{decomp-spfl}.
As a result, the sum over these parameters produces some theta series with non-trivial kernels $\Phi^{\rm tot}_n$ 
constructed from the integrals appearing in \eqref{treeF} and the tree indices arising from DT invariants.
Combining everything together, one arrives at the following representation
\be
\cG=\sum_{n=1}^\infty \[\prod_{i=1}^{n}\sum_{p_i,\mu_i}h_{p_i,\mu_i}\]
\vth_{\bfp,\bfmu}\bigl(\bfLam_\bfp,\Phi^{{\rm tot}}_n\bigr),
\label{treeFh-fl}
\ee
where we used boldface letters to denote tuples of $n$ variables like $\bfp=(p_1,\dots,p_n)$.
The theta series $\vth_{\bfp,\bfmu}$ is of type \eqref{gentheta} defined by the lattice 
$\bfLam_\bfp=\oplus_{i=1}^n\Lambda_{p_i}$ with the quadratic form
\be
\bfk^2= \sum_{i=1}^n\kappa_{i,ab}k_i^a k_i^b, 
\qquad
\kappa_{i,ab}=\kappa_{abc}p_i^c.
\label{defQlr-k2}
\ee
As follows from the Hodge index theorem, for $p^a$ corresponding to an ample divisor,
the associated quadratic form $\kappa_{ab}$ has signature $(1,b_2-1)$. Since $\cD_{p_i}$ are also ample,
$\vth_{\bfp,\bfmu}$ is an indefinite theta series of signature $(n,(b_2-1)n)$.
Its convergence is ensured by the integrals in \eqref{treeF} entering the kernel 
which can be shown to decay exponentially along the ``dangerous" directions of the lattice.

\item 
Since $\cG$ must be a modular form, the representation \eqref{treeFh-fl} implies that the modular properties of 
$h_{p,\mu}$ are determined by the modular properties of the theta series: 
if $\vth_{\bfp,\bfmu}\bigl(\bfLam_\bfp,\Phi^{{\rm tot}}_n\bigr)$ are all modular, the generating functions are also modular;
if not --- $h_{p,\mu}$ must have a modular anomaly to cancel the anomaly of the theta series.
The easiest way to check the modularity of $\vth_{\bfp,\bfmu}\bigl(\bfLam_\bfp,\Phi^{{\rm tot}}_n\bigr)$
is to verify whether its kernel $\Phi^{{\rm tot}}_n$ satisfies the differential equation \eqref{Vigdif}.
It turns out that all the non-trivial integrals pass through this equation, whereas the sign functions
coming from the tree indices spoil it. 
Thus, it is the existence of bound states and the corresponding wall-crossing 
that are responsible for the appearance of a modular anomaly, 
exactly as in the story about immortal dyons in $\cN=4$ compactifications \cite{Dabholkar:2012nd}.

\item 
Once the origin of the anomaly in each term with fixed $n$ has been identified, one can try to ``improve"
the expansion \eqref{treeFh-fl} by reshuffling it. Namely, we can look for the modular completion of the theta series and 
ask whether this completion can be achieved by ``redefining" the generating functions $h_{p,\mu}$.
This is an extremely non-trivial problem because the theta series depend on {\it all} scalar fields 
of the effective theory (playing the role of coordinates on the moduli space $\cM$), while $h_{p,\mu}$
are functions of only the axio-dilaton $\tau$. Nevertheless, the above idea can be realized! 
It leads to a new representation
\be
\cG=\sum_{n=1}^\infty \[\prod_{i=1}^{n}\sum_{p_i,\mu_i}\whh_{p_i,\mu_i}\]
\vth_{\bfp,\bfmu}\bigl(\bfLam_\bfp,\whPhi^{{\rm tot}}_n\bigr),
\label{treeFh-compl}
\ee
where the new kernels $\whPhi^{{\rm tot}}_n$ are such that the corresponding theta series are modular.
This implies that the new functions $\whh_{p,\mu}$, which are now non-holomorphic, should also be modular
of weight $(-\hf b_2-1,0)$.
Thus, they can be considered as modular completions of the original generating series and 
their form encodes the modular anomaly of $h_{p,\mu}$.

\end{enumerate}

\subsection{Equation for the modular completion}
\label{subsec-eqanom}

The explicit form of the modular completion  $\whh_{p,\mu}$, derived in \cite{Alexandrov:2018lgp}
and later corrected and simplified in \cite{Alexandrov:2024jnu},
is the main result containing all information about the modular behavior of the generating series $h_{p,\mu}$.
Its schematic form has been already presented in the Introduction in \eqref{exp-whh-new}.
Here we will explain it in detail. 

If $r$ is the degree of reducibility of divisor $\cD_p$, then the completion is given by
\be
\whh_{p,\mu}(\tau,\btau)=h_{p,\mu}(\tau)+ \sum_{n=2}^{r}\sum_{\sum_{i=1}^n p_i=p}
\sum_{\bfmu}
\rmRi{\bfp}_{\mu,\bfmu}(\tau, \btau)
\prod_{i=1}^n h_{p_i,\mu_i}(\tau),
\label{exp-whh}
\ee
where the coefficients $\rmRi{\bfp}_{\mu,\bfmu}$ can be written as a sum over D2-brane charges $q_{i,a}$
with fixed residue classes $\mu_{i,a}$ and a fixed total sum:
\be
\rmRi{\bfp}_{\mu,\bfmu}(\tau, \btau)=
\sum_{\sum_{i=1}^n q_i=\mu+\hf p \atop q_i\in \Lambda_{p_i}+\mu_i+\hf  p_i} 
\Sym\Bigl\{(-1)^{\sum_{i<j} \gamma_{ij} }\scR_n(\bfhgam;\tau_2)\Bigr\}\, e^{\pi\I \tau Q_n(\bfhgam)}.
\label{defRn}
\ee
Here $\bfhgam$ is the $n$-tuple of reduced charge vectors $\hgam_i=(p_i^a,q_{i,a})$, 
$\Sym$ denotes symmetrization (with weight $1/n!$) with respect to charges $\hgam_i$, 
$\gamma_{ij}=\langle\hgam_i,\hgam_j\rangle$,
and $Q_n(\bfhgam)$ is a the quadratic form on $\bfLam_\bfp/\Lambda_p$
\be
Q_n(\bfhgam)= \kappa^{ab}q_a q_b-\sum_{i=1}^n\kappa_i^{ab}q_{i,a} q_{i,b} \, .
\label{defQlr}
\ee
It is clear that $\rmRi{\bfp}_{\mu,\bfmu}$ can be seen as indefinite theta series on $\bfLam_\bfp/\Lambda_p$.
All non-trivialities are hidden in the functions $\scR_n(\bfhgam;\tau_2)$ which play the role of 
kernels of these theta series.
Their definition involves two types of trees and proceeds in two steps.

At the first step, we consider the set $\IT_n^\ell$ of {\it unrooted labeled trees}, as in \eqref{treeF},  
with $n$ vertices decorated by charges from the set $\bfhgam=(\hgam_1,\dots,\hgam_n)$.
Given a tree $\cT\in \IT_{n}^\ell$, we denote the set of its edges by $E_{\cT}$, the set of vertices by $V_{\cT}$, 
the source and target vertex\footnote{The orientation 
	of edges on a given tree can be chosen arbitrarily, the final result does not depend on this choice.} 
of an edge $e$ by $s(e)$ and $t(e)$, respectively,
and the two disconnected trees obtained from $\cT$ by removing the edge $e$ by $\cT_e^s$ and $\cT_e^t$.
Furthermore, to each edge we assign the vector 
\be
\bfv_e=\sum_{i\in V_{\cT_e^s}}\sum_{j\in V_{\cT_e^t}}\bfv_{ij},
\label{defue}
\ee
where $\bfv_{ij}$ are $nb_2$-dimensional vectors with the following components
\be
(\bfv_{ij})_k^a=\delta_{ki} p_j^a-\delta_{kj} p_i^a.
\label{defvij}
\ee
Using these notations, we define functions of $\tau_2$ parametrized by $n$ reduced charges $\hgam_i=(p_i^a,q_{i,a})$
and constructed from the generalized error functions \eqref{generrPhiME}.
To this end, we introduce 
\be
\EPhi_n(\bfx)=
\frac{1}{n!}\sum_{\cT\in\, \IT_n^\ell}
\[\prod_{e\in E_\cT} \cD(\bfv_{s(e) t(e)},\bfy)\]
\Phi^E_{n-1}(\{ \bfv_e\};\bfx)\Bigr|_{\bfy=\bfx},
\label{rescEn}
\ee
where
\be
\cD(\bfv,\bfy)=\bfv\cdot\(\bfy+\frac{1}{2\pi}\,\p_\bfx\)
\label{defcDif}
\ee
and the dot in \eqref{defcDif} denotes the bilinear form
\be
\bfx\cdot\bfy=\sum_{i=1}^n \kappa_{i,ab}x_i^a y_i^b,
\label{biform}
\ee
which is also used to define the generalized error functions $\Phi^E_n$ \eqref{generrPhiME}.
It is useful to note that with respect to this bilinear form one has $\bfv_{ij}\cdot\bfq=\gamma_{ij}$
where $\bfq=\bigl(\kappa_1^{ab} q_{1,b} ,\dots ,\kappa_n^{ab}q_{n,b}\bigr)$.
In terms of the vector $\bfq$, our functions are given by
\be
\Ev_n(\bfhgam;\tau_2)=
\frac{\EPhi_n(\sqrt{2\tau_2}\, \bfq)}{(\sqrt{2\tau_2})^{n-1}}\, .
\label{rescEnPhi}
\ee

An important fact is that these functions have a canonical decomposition
\be
\Ev_n(\bfhgam;\tau_2)=\Ef_n(\bfhgam)+\Ep_n(\bfhgam;\tau_2),
\label{twocEs}
\ee
where the first term $\Ef_n$ does not depend on $\tau_2$,
whereas the second term $\Ep_n$ is exponentially suppressed as $\tau_2\to\infty$ keeping
the charges $\hgam_i$ fixed. In \cite{Alexandrov:2024jnu} it was shown that 
\be 
\Ef_n(\bfhgam)\equiv  \lim_{\tau_2\to\infty}\Ev_n(\bfhgam;\tau_2)
= \frac{1}{n!}
\sum_{\cT\in\, \IT_n^\ell} S_\cT(\bfhgam)\prod_{e\in E_{\cT}}\gamma_{s(e) t(e)},
\label{newexprcEf}
\ee
where $S_\cT(\bfhgam)$ is a product of sign functions of the following combinations of Dirac products 
\be 
\Gamma_e=\sum_{i\in V_{\cT_e^s}}\sum_{j\in V_{\cT_e^t}}\gamma_{ij}=\bfv_e\cdot \bfq.
\label{defGammae}
\ee
However, it cannot be written as a simple product of $\sgn(\Gamma_e)$ because, when an even number of $\Gamma_e$'s vanish, 
it is not actually zero, but equals a rational number.
This means that the correct formula can be written as
\be
S_\cT(\bfhgam)=
\sum_{\cJ\subseteq E_\cT}e_{\cT_\cJ}
\,\prod_{e\in \cJ}\delta_{\Gamma_e}
\prod_{e\in E_\cT\setminus \cJ} \sgn (\Gamma_e),
\label{defST}
\ee
where $\cT_\cJ$ denotes the tree obtained from $\cT$ by contracting the edges $e\in E_\cT\backslash \cJ$
and $e_{\cT}$ are the above mentioned rational numbers.
They depend only on the topology of $\cT$, vanish for trees with an even number of vertices, 
and can be computed with the help of a recursive formula.
This formula involves yet another rational numbers $a_\cT$ for which there is their own recursive formula:
\be
a_\cT=\frac{1}{n_\cT}\sum_{\ver\in V_\cT} (-1)^{n_\ver^+} \prod_{s=1}^{n_\ver} a_{\cT_s(\ver)},
\label{res-aT}
\ee
where $n_\cT$ is the number of vertices, $n_\ver$ is the valency of the vertex $\ver$,
$n_\ver^+$ is the number of incoming edges at the vertex, and
$\cT_s(\ver)$ are the trees obtained from $\cT$ by removing the vertex.
Having computed $a_\cT$, one can obtain $e_\cT$ from
\be 
e_\cT=-\sum_{m=1}^{n_\cT-1}\sum_{\smash{\mathop{\cup}\limits_{k=1}^m\cT_k \simeq\cT }}\,
e_{\cT/\{\cT_k\}}\prod_{k=1}^m a_{\cT_k}.
\label{res-eT}
\ee 
where the second sum runs over all decompositions of $\cT$ into a set of 
non-intersecting subtrees
and $\cT/\{\cT_k\}$ denotes the tree obtained from $\cT$ by collapsing each subtree $\cT_k$ to a single vertex.
Both recursions are initiated by the values $e_\bullet=a_\bullet=1$ for a single vertex tree.
The values of both $e_\cT$ and $a_\cT$ for trees with $n_\cT\leq 7$ can be found in \cite[Ap.B]{Alexandrov:2024jnu}.

\begin{figure}
	\isPreprints{\centering}{} 
	\includegraphics[width=5.7cm]{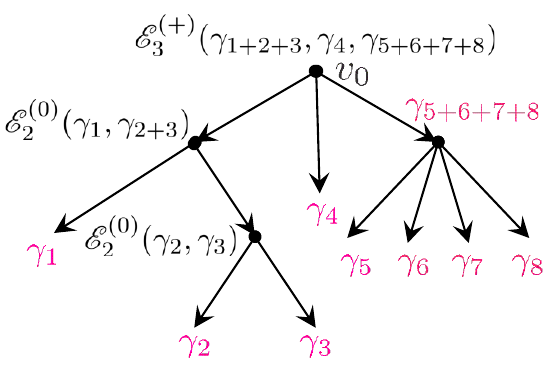}
	\vspace{-0.5cm}
	\caption{An example of Schr\"oder tree contributing to $R_8$. 
		Near each vertex we showed the corresponding factor
		using the shorthand notation $\gamma_{i+j}=\gamma_i+\gamma_j$.\label{fig-WRtree}}
\end{figure}   

At the second step, we introduce a new type of trees known as {\it Schr\"oder trees}.
They are defined as rooted planar trees such that all vertices $\vert\in V_T$ 
(the set of vertices of $T$ excluding the leaves) have $k_\vert\geq 2$ children.
The set of such trees with $n$ leaves will be denoted by $\IT_n^{\rm S}$. Besides, we take
$n_T$ to be the number of elements in $V_T$ and $\vert_0$ to denote the root vertex.
The vertices of $T$ are labeled by charges in a way similar to attractor flow trees: 
the leaves carry charges $\hgam_i$, whereas the charges assigned to other vertices
are given recursively by
the sum of charges of their children, $\hgam_\vert\in\sum_{\vert'\in\Ch(\vert)}\hgam_{v'}$.
Then, given a Schr\"oder tree $T$,
we set $\Ev_{\vert}\equiv \Ev_{k_\vert}(\{\hgam_{\vert'}\})$ (and similarly for $\Ef_{\vert}, \Ep_{\vert}$)
where $\vert'\in \Ch(\vert)$ runs over the $k_\vert$ children of the vertex $\vert$ (see Fig. \ref{fig-WRtree}). 
Using these notations, we can finally write a formula for the coefficients $\scR_n$:
\be
\scR_n(\bfhgam;\tau_2)= \frac{1}{2^{n-1}}\sum_{T\in\IT_n^{\rm S}}(-1)^{n_T-1} 
\Ep_{\vert_0}\prod_{\vert\in V_T\setminus{\{\vert_0\}}}\Ef_{\vert}.
\label{solRn}
\ee

\begin{figure}
	\isPreprints{\centering}{} 
	\includegraphics[width=13.7cm]{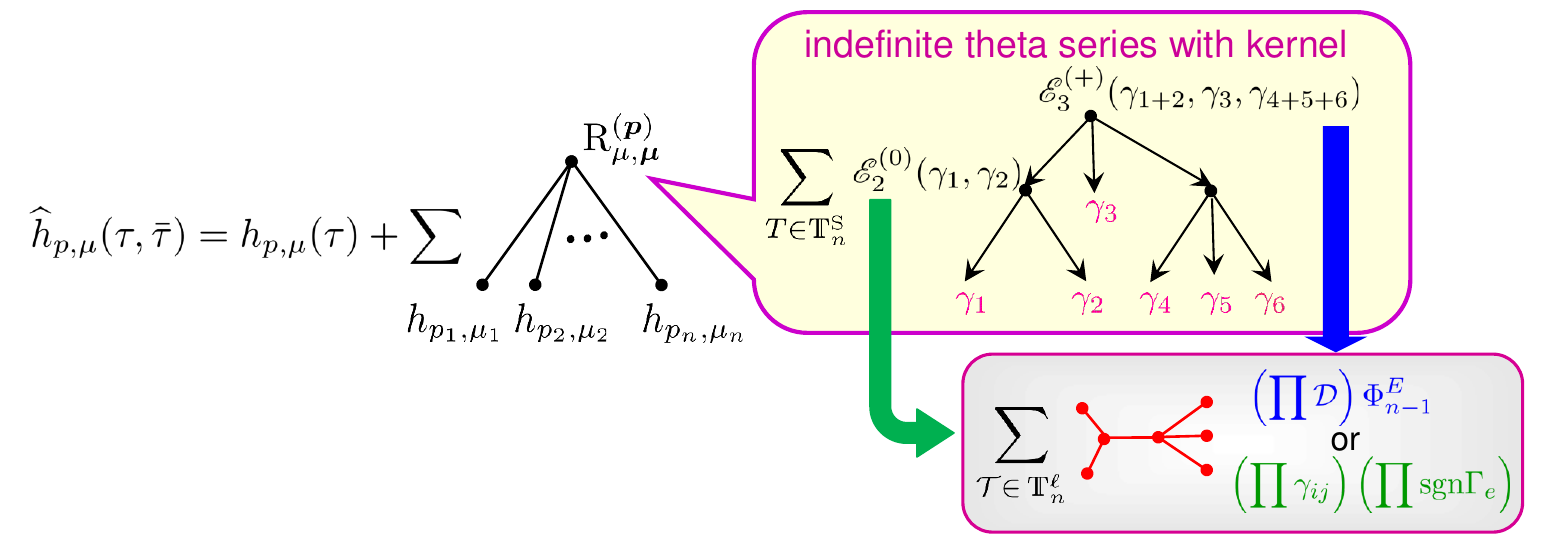}
	\vspace{-0.1cm}
	\caption{A schematic representation of the modular completion \eqref{exp-whh}. 
		The first sum over decompositions of the D4-brane charge $p^a$ can be seen as a sum over rooted trees of depth one 
		with generating series $h_{p_i,\mu_i}$ assigned to the leaves and the indefinite theta series $\rmRi{\bfp}_{\mu,\bfmu}$
		assigned to the root. The kernel of the theta series is a sum over Schr\"oder trees
		and the functions assigned to their vertices are sums over unrooted labeled trees.\label{fig-anomaly}}
\end{figure}

This completes the definition of the objects appearing in the expression \eqref{exp-whh}
of the modular completion.
Although the above construction appears to be complicated 
(a simpler version will be presented in \S\ref{subsec-refine} after incorporating an additional refinement parameter), 
its mathematical structure is transparent:
Eq. \eqref{exp-whh} expresses the completion as a sum over all possible bound states
with a given D4-brane charge, Eq. \eqref{defRn} represents the coefficients of this expansion as indefinite theta series on 
$\bfLam_\bfp/\Lambda_p$, while Eq. \eqref{solRn} builds up kernels of the theta series 
as a sum over Schr\"oder trees, which can be seen to encode different decay channels of bound states,
weighted by (derivatives of)
the generalized error functions and their limit at large $\tau_2$ where they reduce to a combination of sign functions
(see Fig. \ref{fig-anomaly}).
As explained in \S\ref{sec-indef}, both building blocks are very natural in the context of 
indefinite theta series.

Furthermore, since indefinite theta series are the classic examples of higher depth mock modular forms, 
it is natural to expect that the generating functions $h_{p,\mu}$ belong to the same class. 
In fact, one can explicitly compute the non-holomorphic derivative of the completion encoding the shadow.
The result reads as 
\be
\p_{\bar\tau}\whh_{p,\mu}(\tau,\btau)= \sum_{n=2}^r
\sum_{\sum_{i=1}^n \hgam_i=\hgam}(-1)^{\sum_{i<j} \gamma_{ij} }
\cJ_n(\bfhgam,\tau_2)
\, e^{\pi\I \tau Q_n(\bfhgam)}
\prod_{i=1}^n \whh_{p_i,\mu_i}(\tau,\btau),
\label{exp-derwh}
\ee
where $\hgam=(p^a,\mu_a+\hf\kappa_{ab}p^b)$ and 
\be
\cJ_n(\bfhgam,\tau_2)= \frac{\I}{2^n}\sum_{T\in\IT_n^{\rm S}}(-1)^{n_T-1} 
\p_{\tau_2}\cE_{v_0}\prod_{v\in V_T\setminus{\{v_0\}}}\cE_{v}.
\label{solJn}
\ee
It is expressed through modular forms, the completions $\whh_{p_i,\mu_i}$ and an indefinite theta series with kernel
$\cJ_n$ satisfying the Vign\'eras equation \eqref{Vigdif}, consistently with the fact that 
$\p_{\bar\tau}\whh_{p,\mu}$ is a modular form of weight $(-\hf b_2-1,2)$.
Since the shadow is expressed through the same types of objects as the completion, 
but the derivative decreases the rank of the generalized error functions by 1,
the result \eqref{solJn} confirms that $h_{p,\mu}$ are mock modular forms of depth $r-1$.
In particular, for $r=1$ the second term in \eqref{exp-whh} is absent and the generating series are ordinary modular forms,
consistently with the previous results \cite{Gaiotto:2006wm,deBoer:2006vg,Denef:2007vg}.
The first non-trivial case appears at $r=2$ and exhibits the mixed mock modularity.

Independently of $r$, the multiplier system of $h_{p,\mu}$ is obtained as the inverse of the multiplier system of 
$\vartheta_{p,\mu}\bigl(\bfLam_p,\whPhi^{{\rm tot}}_1\bigr)$ (see \eqref{treeFh-compl})
and is given by 
\be
\begin{split}
	M_{\mu\nu}(T)=&\, e^{\pi\I\(\mu+\frac{p}{2}\)^2+\tfrac{\pi\I}{12}\, c_{2,a}p^a}\,\delta_{\mu\nu},
	\\
	M_{\mu\nu}(S)=&\, \frac{(-1)^{\chi(\cO_{\cD_p})}}{\sqrt{|\det\kappa_{ab}|}}\, e^{(b_2-2)\frac{\pi\I}{4}}\,
	e^{-2\pi\I \mu\cdot\nu}\,,
\end{split}
\label{Multsys-hp}
\ee
where $\mu\cdot\nu=\kappa^{ab}\mu_a \nu_b$ and
$\chi(\cO_{\cD_p})=\frac12(b_2^+(\cD_p)+1)$ is the arithmetic genus of the divisor expressed in terms of 
the D4-brane charge as
\be 
\label{defL0}
\chi(\cO_{\cD_p}) = \frac16\, (p^3)+ \frac{1}{12}\, c_{2,a} p^a.
\ee

\subsubsection{Collinear charges}
\label{subsubsec-collinear}

Before we proceed further, it is worth to consider a special case that has many interesting applications.
Let $p_0^a$ corresponds to an irreducible divisor. Since in \eqref{exp-whh} one sums over
decompositions of $\cD_p$ only in ample divisors, for $p^a=rp_0^a$ only collinear charges $p_i^a=r_ip_0^a$
can appear and the sum is equivalent to the sum over decompositions $r=\sum_i r_i$.
It turns out that when all charges are collinear, the holomorphic anomaly \eqref{exp-derwh} 
of the completion enormously simplifies. In \cite{Alexandrov:2019rth} it was proven that
in this case $\cJ_n=0$ for all $n>2$. Thus, the only term that survives is the one with $n=2$!
It can be computed explicitly and the resulting anomaly equation takes the following form
\be
\p_{\bar\tau}\whh_{rp_0,\mu}=\frac{\sqrt{(p_0^3)}}{8\pi\I(2\tau_2)^{3/2}}
\sum_{r_1+r_2=r\atop q_{1}+q_{2}=\mu+\frac{1}{2}r^2 p_0}(-1)^{\gamma_{12}}
\,\sqrt{rr_1r_2}\,
e^{-\frac{2\pi\tau_2 \gamma_{12}^2}{ rr_1r_2(p_0^3)} }
\, e^{\pi\I \tau Q_2(\hgam_i,\hgam_2)}
\whh_{r_1p_0,\mu_1}\whh_{r_2p_0,\mu_2},
\label{exp-derwh-lim}
\ee
where $\gamma_{12}=p_0^a(r_2q_{1,a}-r_1 q_{2,a})$.
Note that this result does {\it not} imply that a similar cancellation happens in the expression for 
the completion \eqref{exp-whh}, and $h_{rp_0,\mu}$ are still mock modular forms of depth $r-1$.

The holomorphic anomaly \eqref{exp-derwh-lim} can be further simplified by introducing a partition function,
which can be identified with the modified elliptic genus \cite{Maldacena:1999bp} of the SCFT mentioned in the Introduction.
It is defined by
\be
\whcZ_r =\sqrt{\frac{(p_0^3)}{r}}\, \sum_{\mu}\whh_{rp_0,\mu}\,\overline{\vth^{\rm (S)}_{rp_0,\mu}(\tau,v)},
\label{pfZBN-compl}
\ee
where $\vth^{\rm (S)}_{p,\mu}$ is a Jacobi extension of the Siegel theta series \eqref{Siegel-ts}
on the lattice $\Lambda_{p}$
\be 
\vth^{\rm (S)}_{p,\mu}(\tau,v)=\sum_{q\in \Lambda_{p}+\mu+\frac{1}{2}p}
\sigma_\gamma\,  e^{-2\pi\tau_2 q_p^2+\pi\I\btau q^2+2\pi\I q_a v^a}
\label{def-vthSp}
\ee 
and we denoted $q^2=\kappa^{ab}q_a q_b$ and $q_p=q_ap^a/\sqrt{(p^3)}$ playing the role
of the projection on the positive definite sublattice (denoted by + in \eqref{Siegel-ts}),
which in this case is one-dimensional. 
The factor $\sigma_\gamma$ is just a sign factor satisfying 
$\sigma_{\gamma_1}\sigma_{\gamma_2}=(-1)^{\gamma_{12}}\sigma_{\gamma_1+\gamma_2}$
which is known as quadratic refinement.
Since the Siegel theta series is a modular form of weight $(\hf,\hf\,(b_2-1))$, the partition function $\whcZ_r$
is also modular with weight $(-\frac32\, ,\, \hf)$. Furthermore, the theta series almost cancels the multiplier system of 
$\whh_{rp_0,\mu}$, so that one remains with $(\Mi{\eta})^{rp_0^a c_{2,a}}$ where $\Mi{\eta}$ is the multiplier system
\eqref{Meta} of the Dedekind eta function.\footnote{It can be canceled by multiplying the partition function 
by $e^{2\pi\I rp_0^a \tc_a}$ where $\tc_a$ is the RR-field appearing in \eqref{Phitwo-tcF}.}
Given \eqref{exp-derwh-lim}, it is straightforward to check that \cite{Alexandrov:2019rth}
\be
\label{holanomZp}
\overline{\mathcal{D}}\whcZ_r=\frac{\sqrt{ 2\tau_2}}{32\pi\I}\sum_{r_1+r_2=r} r_1r_2\, \whcZ_{r_1}\whcZ_{r_2},
\ee
where
\be
\overline{\mathcal{D}}=\tau_2^2\( \partial_{\bar \tau}
-\frac{\I}{4\pi}\, \p_{v_+}^2 \)
\ee
is designed to commute with the Siegel theta series and, acting on the completion of a mock modular form,
decreases its holomorphic weight by 2.
Such a holomorphic anomaly equation has been found, for example, for the completion of the elliptic genus of the $\hf K3$ surface \cite{Minahan:1998vr}. The result \eqref{holanomZp} shows that this is a much more general and universal phenomenon
taking place as soon as the relevant magnetic charges are all collinear.

\section{Extensions}
\label{sec-ext}

The expression for the modular completion presented in the previous section has been derived in a concrete setup:
D4-brane wrapping an ample divisor of a compact CY threefold in compactified type IIA string theory.
However, it turns out that some conditions, like the ampleness of the divisor, can be relaxed and the whole construction
can be extended either to more general settings or to include additional parameters.
In this section, we describe three such extensions.

\subsection{Degenerations}
\label{subsec-degen}

First, we drop the assumption that the divisor wrapped by D4-brane is ample and replace it by a weaker condition that
it should be {\it effective}.\footnote{In the basis of the K\"ahler cone used above the components of the magnetic charge
	corresponding to an effective divisor can be negative. However, since effective divisors form a cone, 
	there is a basis where $p^a\geq 0$. The price to pay is that the intersection numbers can be negative in this basis.}
In particular, we are interested in the situation when the charge $p^a$ satisfies
\be 
(p^3)=\kappa_{abc}p^a p^b p^c=0,
\label{degcase}
\ee 
which was not allowed in the original construction.
Nevertheless, as we will see now, one can still make sense of it.
However, one should distinguish between two very different cases.

\subsubsection{Degenerate quadratic form}
\label{subsubsec-degqf}

The first case we need to consider is when the quadratic form $\kappa_{ab}$ is degenerate.
Namely, it has at least one zero eigenvalue.
In \cite{Alexandrov:2020qpb} it was argued that in such situation the above construction of the modular completion 
still holds provided the lattice used to sum over D2-brane charges is restricted to 
the non-degenerate part of the full original lattice.
More precisely, let $\lambda_s^a$ be a set of null eigenvectors, i.e. $\kappa_{ab}\lambda_s^a=0$.
Then we take 
\be
\Lambda_p^*=\{ q_a\in \IZ  + \frac12 \,\kappa_{ab}p^b \ :\  \lambda_s^a q_a=0\}.
\label{Lam-p}
\ee
Furthermore, one can introduce the inverse quadratic form $\kappa^{ab}$ known as Moore–Penrose or pseudoinverse of $\kappa_{ab}$. 
It is defined by the conditions
i) $\rank(\kappa^{ab})=\rank(\kappa_{ab})$, ii) $\kappa^{ac}\kappa_{cb}=\delta^a_b-\sum_{s,t} e^{st} \lambda_s^a\lambda_{t,b}$
where $e^{st}$ is the inverse of $e_{st}=\lambda_s^a\lambda_{t,a}$.
Using these definitions in the equations of \S\ref{subsec-eqanom} leads to a well-defined modular completion.
The only change to be made is to replace $b_2$ by $\rank(\Lambda_p):=\rank(\kappa_{ab})$. 
In particular, the weight of the generating functions is now given by 
\be
w(p)=-\hf\, \rank(\Lambda_p)-1.
\label{weightgen}
\ee

If $p^a$ is one of the null eigenvectors, one gets an even stronger condition than \eqref{degcase}, 
\be 
\kappa_{abc} p^b p^c=0.
\label{degpp}
\ee 
Let us assume for simplicity that $p^a$ is a multiple of $p_0^a$ representing an irreducible effective divisor. 
It turns out that in this case the modular anomaly completely disappears!
Indeed, it is trivial to see that all reduced charges $\hgam_i=(p_i^a,q_{i,a})$ corresponding to the bound state constituents
have $p_i^a=r_ip_0^a$ and $q_{i,a}$ orthogonal to $p_0^a$ 
and therefore satisfy $\langle\gamma_i,\gamma_j\rangle=0$.
As a result, the modular anomaly simply does not arise in this case because all its sources
such as multi-instanton contributions to the instanton generating potential or bound state contributions to DT invariants
are weighted by the Dirac products $\langle\gamma_i,\gamma_j\rangle$ and hence vanish.
Thus, the generating functions $h_{p,\mu}$ with $p^a=rp_0^a$ satisfying \eqref{degpp} must be modular forms.\footnote{It is likely
that this conclusion continues to hold even for more general effective divisors satisfying \eqref{degpp}.
To prove this, one should show that $\langle\gamma_i,\gamma_j\rangle=0$ for all possible decay products.}
This is the case relevant, for example, for vertical divisors of K3-fibered CYs 
\cite{Gholampour:2013hfa,Diaconescu:2015koa,Bouchard:2016lfg,Doran:2024kcb}.

If however $p^a$ is not a null eigenvector of the quadratic form that it defines, the reasoning
leading to the vanishing of all relevant Dirac products $\langle\gamma_i,\gamma_j\rangle$
does not hold anymore and the generating functions $h_{p,\mu}$ 
can still have very non-trivial mock modular properties
encoded by their completions \eqref{exp-whh}. 
This case will be relevant below in \S\ref{subsec-nonxompCY}.

\subsubsection{Non-degenerate quadratic form}

In the second case, $(p^3)=0$ but the quadratic form is non-degenerate.
This case has not been analyzed before in the literature, therefore we consider it here in detail.
Let us again assume that $p^a=rp_0^a$ where $p_0^a$ is an irreducible effective divisor 
satisfying $(p_0^3)=0$, so that all charges are collinear as in \S\ref{subsubsec-collinear}.
Then the holomorphic anomaly of the completion is greatly simplified and should be given by \eqref{exp-derwh-lim}.
However, this formula is not directly applicable to our case. 
On one hand, due to $(p_0^3)=0$, the first exponential factor vanishes 
so that one might think that the anomaly disappears and the generating functions become usual modular forms
as in the previous case. On the other hand, for charges satisfying $\gamma_{12}=0$, this reasoning fails 
because the exponential does not vanish anymore.
It is tempting to say that, even for these charges, the vanishing is still ensured by the overall factor $\sqrt{(p_0^3)}$.
But the problem is that the remaining theta series is actually divergent: since $p_0^a$ is a null vector,
the component of D2-brane charges along this vector does not contribute to the quadratic form $Q_2(\hgam_i,\hgam_2)$.
As a result, we arrive at ambiguity $0\times \infty$.

In appendix \ref{ap-null} we show how this ambiguity can be resolved by taking the magnetic charge 
slightly off the null direction and then removing the regularization. The result of this analysis 
is the following holomorphic anomaly equation
\be
\p_{\btau}\,\whh_{rp_0,\mu} (\tau,\btau)
= \frac{\tau_2^{-2}}{16\pi \I}\sum_{r_1+r_2=r} \frh^{(r_1,r_2)}_{p_0,\mu}(\tau,\btau), 
\label{anom-null}
\ee
where
\be 
\frh^{(r_1,r_2)}_{p_0,\mu}(\tau,\btau)=
r_0
\sum_{\mu_1,\mu_2} \delta_{\Delta\mu\in r_0\Lambda_0}\,
\whh_{r_1p_0,\mu_1}\,\whh_{r_2p_0,\mu_2}
\sum_{\Asf=0}^{n_g-1}\delta^{(\xi\rr_{12})}_{p_0\cdot (\mupr_{12}+\rr_{12}\gluegl_\Asf)}
\vth^{\perp}_{\mu_{12}^\perp+\rr_{12}\gluegp_\Asf},
\label{def-frh}
\ee
is a modular form of weight $(-\hf b_2-3,0)$, 
$\delta^{(n)}_x$ is the mod-$n$ Kronecker delta defined by
\be
\label{defdelta}
\delta^{(n)}_x=\left\{ \begin{array}{ll}
	1\  & \mbox{if } x=0\!\!\!\mod n,
	\\
	0\ & \mbox{otherwise},
\end{array}\right.
\ee
while other notations can be found in appendix \ref{ap-null}.
A remarkable feature of the anomaly equation \eqref{anom-null} is that the non-holomorphic dependence,
besides the one due to the functions $\whh_{r_ip_0,\mu_i}$, is completely captured by the overall factor of $\tau_2^{-2}$. 
This implies that the generating series $h_{rp_0,\mu}$ are actually {\it quasi-modular} forms which can be constructed 
as polynomials in the Eisenstein series $E_2(\tau)$ (see Ex. \ref{ex-E2}) with coefficients given by usual modular forms.
For example, for $r=2$, $\frh^{(1,1)}_{p_0,\mu}$ is holomorphic because $\whh_{p_0,\mu}=h_{p_0,\mu}$
and the non-holomorphic dependence of the completion can be captured by $\whE_2=E_2-\frac{3}{\pi\tau_2}$.
This is equivalent to the statement that
\be
h_{2p_0,\mu} (\tau)=h^{(0)}_{2p_0,\mu} (\tau)-\frac{E_2(\tau)}{24}\,\frh^{(1,1)}_{p_0,\mu}(\tau) ,
\label{anomalynull}
\ee
where $h^{(0)}_{2p_0,\mu}$ is a holomorphic modular form.
A description in terms of quasi-modular forms is known to hold, for example, for elliptically fibered CYs
\cite{Klemm:2012sx,Cota:2017aal,Cota:2019cjx,Lee:2020gvu} and their geometric data turn out to perfectly fit 
the framework presented here and implied by the condition \eqref{degcase} \cite{PStorus}.

\subsection{Non-compact Calabi-Yau}
\label{subsec-nonxompCY}

Next, we show what happens if one takes a non-compact CY threefold.
This extension is important because, on one hand, it can serve as a test-ground for 
the results presented in \S\ref{sec-anomaly} since, in contrast to the compact case,
there are various powerful techniques to compute BPS indices and other topological invariants for non-compact CYs
(see, e.g., \cite{Douglas:2000qw,Nishinaka:2013mba,Banerjee:2019apt,Beaujard:2020sgs,DelMonte:2021ytz,LeFloch:2024cwl}),
while, on the other hand, it can still tell us something new.
This is particularly important from the physical viewpoint because the non-compact case typically provides
a geometric realization of supersymmetric gauge theories which are of great interest \cite{Katz:1996fh,Katz:1997eq}.

A non-compact CY can be obtained as the {\it local limit} of a compact CY where
one zooms in on the region near some singularity in the moduli space corresponding to shrinking of one or several cycles. 
Let us recall a description of this limit from \cite{Alexandrov:2017mgi} which puts it in the same framework
that was used above. As we will see, the local limit fits the degenerate case 
considered in \S\ref{subsubsec-degqf}.

Instead of specifying either a set of shrinking 4-cycles or 2-cycles,
let us start from a set of $\nA$ linearly independent vectors $\vtA^a$ belonging to the closure of the K\"ahler cone of $\CY$,
where the index $A$ labels different vectors and takes $\nA$ values.
Given these vectors, we define a set of matrices
\be
\kappa_{A,ab}=\kappa_{abc}\vtA^c.
\label{matA}
\ee
We assume that the vectors $\vtA^a$ are chosen so that the matrices $\kappa_A$ have a non-trivial common
kernel of dimension $\nI$, which in particular implies that all $\vtA^a$ must belong to the boundary of the K\"ahler cone. 
We denote a basis of this kernel by $\vtI^a$. Obviously, these vectors satisfy
\be
\kappa_{A,ab}\,\vtI^b=0
\label{deftI}
\ee
for any $A$ and $I$. We also assume that the two sets, $\vtA^a$ and $\vtI^a$, are linearly independent
and complete them to a basis in $H_2(\CY,\IR)$ by providing an additional set of 
$b_2-\nA-\nI\equiv\nX$ vectors $\vtX^a$.
This allows to expand the K\"ahler moduli in the new basis
\be
t^a=\vtA^a\, \hatt^A+\vtX^a\, \hatt^X+\vtI^a\, \hatt^I\equiv \vt^a_b\, \hatt^b,
\label{changet}
\ee
where we combined three indices $A$, $X$ and $I$ into one index $b$.
Then the local limit is defined by taking the moduli $\hatt^A$ to scale to infinity, whereas keeping $\hatt^X$ and $\hatt^I$
finite.\footnote{In the dual gauge theory, $\hatt^I$ become dynamical Coulomb branch moduli, $\hatt^X$
turn into physical parameters such as masses and the gauge coupling, and $\hatt^A$ drop out from the theory.} 
It is important that this definition does not depend on the choice of $\vtX^a$
because changing $\vtX^a$ in \eqref{changet} can at most shift $\hatt^A$ and $\hatt^I$ by a combination of $\hatt^X$,
which does not affect the split between growing and finite variables.

By computing the volumes of divisors and curves in the rotated basis, 
$\hcD_a=v^b_a \cD_b$ and $\hcC^a=(v^{-1})^a_b \cC^b$, it is easy to see that in the local limit defined above the volumes of
$\hcD_A$, $\hcD_X$ and $\hcC^A$ grow, while the volumes of $\hcD_I$, $\hcC^X$ and $\hcC^I$ remain finite.
(Or, if one divides the K\"ahler moduli by some physical scale which grows as $\hatt^A$ in the local limit,
$\hcD_A$, $\hcD_X$ and $\hcC^A$ stay finite, while $\hcD_I$, $\hcC^X$ and $\hcC^I$ shrink.
Thus, the definition of the local limit is equivalent to specifying either the set of shrinking divisors $\hcD_I$
or the set of shrinking curves $\hcC^X,\hcC^I$.)
Therefore, only the D4-branes wrapping $\hcD_I$ and the D2-branes wrapping $\hcC^X,\hcC^I$ survive in the limit.
In other words, we have access only to the magnetic charges $p^a$ that are linear combinations of $\vtI^a$.
The associated quadratic forms $\kappa_{I,ab}=\kappa_{abc}\vtI^c$ are degenerate because $\vtA^a$ are their null eigenvectors 
due to \eqref{deftI}. Hence, we fall into the case described in the previous subsection where 
$\rank(\kappa_{ab})<b_2$ but $p^a$ is not a null eigenvector. In our case $\rank(\kappa_{ab})\leq b_2-\nA$.
Note that the curves that could be wrapped by D2-branes satisfy 
the orthogonality relation 
\be 
\hcC^X,\hcC^I\cap \hcD_{A}=0.
\label{orthCD}
\ee 
It provides a physical interpretation of the reduction of the charge lattice \eqref{Lam-p} 
given in this case by the condition $\vtI^a q_a=0$ which is equivalent to \eqref{orthCD}.

Since $p^a$ is not a null eigenvector of the quadratic form which it defines, 
in general, the expression \eqref{exp-whh} for modular completions is not simplified.
The only change induced by the local limit is 
the reduction of charge lattices: magnetic charges should be linear combinations of $\vtI^a$ 
and electric charges should be orthogonal to all null eigenvectors of the associate quadratic forms,
which also affects the modular weight determined now by the rank of the electric charge lattice as in \eqref{weightgen}.

{\Example \label{ex-elliptic}
	Elliptically fibered CY \cite{Alexandrov:2019rth}.

\rm
Let us consider a smooth elliptic fibration $\pi\,:\,\CY\to S$ with a single section $\sigma$ 
over a compact, smooth almost Fano base $S$.
For all smooth elliptic fibrations a basis of $H^{1,1}(\CY)$ generating the K\"ahler cone is given by
$\{\omega_e,\pi^* \omega_{\alpha}\}$, $\alpha=1,\dots,h^{1,1}(\base)$,
where
\be
\omega_e=\sigma+\pi^* c_1(\base)
\label{section}
\ee
and $\omega_\alpha$ are the generators of the K\"ahler cone on the base.
We denote the corresponding basis of dual divisors by $\{\cD_e,\cD_\alpha\}$. 
The divisor $\cD_e$ is dual to the elliptic fiber curve $\cE$
in the sense that it does not intersect any curve in $\base$ and obeys $\cD_e\cap \cE=1$.
In this basis the triple intersection numbers of $\CY$ can be shown to be
\be
\kappa_{\alpha\beta\gamma}=0,
\qquad
\kappa_{e\,\alpha\beta}=C_{\alpha\beta},
\qquad
\kappa_{ee\,\alpha}=C_{\alpha\beta}c_1^\beta,
\qquad
\kappa_{eee}=C_{\alpha\beta}c_1^\alpha c_1^\beta,
\label{kappa-ell}
\ee
where
\be
C_{\alpha\beta}=\int_\base \omega_\alpha\wedge \omega_\beta,
\qquad 
c_1(\base)=c_1^\alpha\, \omega_\alpha,
\ee
while the components of the second Chern class $c_2(T\CY)$ are
\be
\label{c2loc}
c_{2,e} = 11 C_{\alpha\beta}c_1^\alpha c_1^\beta+ \chi(S),
\qquad
c_{2,\alpha} = 12 \, C_{\alpha\beta} c_1^{\beta}.
\ee

A crucial property of the intersection numbers \eqref{kappa-ell} is that the matrix $\kappa_{eab}$ is degenerate, 
i.e. its determinant vanishes.
This suggests that the vector $\vt_1^a=\delta^a_e$, playing the role of $\vtA^a$ in \eqref{matA}, 
defines a non-trivial local limit. 
The kernel of $\kappa_{eab}$ is one-dimensional and described by the vector
\be
p_0^a=(1,-c_1^\alpha),
\label{vecvI}
\ee
playing the role of $\vtI^a$ in \eqref{deftI}.
The corresponding shrinking divisor $\hcD_0=p_0^a \cD_a$ is nothing but  
the base of the elliptic fibration
\be
\hcD_0=\cD_e-c_1^\alpha \cD_\alpha=\base.
\ee
It is not an ample divisor, as some of the coefficients of the charge vector \eqref{vecvI} are negative, but it is effective. 
We observe that in the local limit defined by the vector $\vt_1^a$, 
one obtains a non-compact CY given by the total space $\mbox{Tot}(K_S)$ of the canonical bundle over the surface $S$.
In this limit, all magnetic charges should be multiples of \eqref{vecvI}, $p^a=rp_0^a$.
The associated quadratic forms are given by 
\be
\kappa_{ab} = \kappa_{abc} p^c = r\(\begin{array}{cc}
	0 & 0
	\\
	0 & C_{\alpha\beta}
\end{array}\).
\label{kappaEll}
\ee
They are all degenerate along the fiber direction described by $\vt_1^a$, 
in agreement with the fact that the divisor $S$ is not ample.
Since $\rank(\kappa_{ab})=b_2(S)$, one finds that the generating functions $h_{p,\mu}$
are higher depth mock modular forms of weight $-\hf\,b_2(S)-1$.
Note that it is different from the weight in the compact case because $b_2(S)\ne b_2(\CY)$.
Since all magnetic charges are collinear, the results presented in \S\ref{subsubsec-collinear},
including the holomorphic anomaly equation \eqref{holanomZp} for the partition function,
apply to this example.
\vsp}

As mentioned above, string theory on non-compact CYs can often be reinterpreted as 
a supersymmetric gauge theory. As a result, BPS indices can also acquire a new interpretation.
In particular, in the above example of the non-compact CY given by $\mbox{Tot}(K_S)$,
the DT invariants counting D4-branes wrapped $r$ times around the surface $S$ are expected to coincide
\cite{Minahan:1998vr,Alim:2010cf,gholampour2017localized} with
the Vafa-Witten invariants with gauge group $U(r)$ on $S$ \cite{Vafa:1994tf}.
In \S\ref{subsec-VW} we show how the formula for the completion $\whh_{p,\mu}$ allows 
to find the generating series of these VW invariants for arbitrary rank $r$ for various rational surfaces.

\subsection{Refinement}
\label{subsec-refine}

Our third extension is the inclusion of a refinement parameter. At a formal level, it is done by simply replacing 
the sign factor $(-1)^{2J_3}$ in the definition of the BPS index, where $J_3$ generates rotations around a fixed axis in $\IR^3$, 
by the factor $(-y)^{2J_3}$, where $y=e^{2\pi\I z}$ is a new (in general complex) parameter.
Physically, the refinement corresponds to switching on the $\Omega$-background \cite{Moore:1998et,Nekrasov:2002qd},
while mathematically, it gives access to Betti numbers of moduli spaces of stable objects, 
in contrast to the unrefined DT invariants computing only their Euler characteristic.
More precisely, the {\it refined BPS indices} can be written as Poincar\'e polynomials
\be
\label{defOmref}
\Omega(\gamma,y) = 
\sum_{p=0}^{2d} (-y)^{p-d}\, b_p(\cM_\gamma),
\ee
where $\cM_\gamma$ is the moduli space of coherent sheaves of charge $\gamma$
and $d$ is its complex dimension.
As in \eqref{def-bOm}, we also introduce their rational counterparts \cite{Manschot:2010qz}
\be
\label{defbOm}
\bOm(\gamma,y) = \sum_{m|\gamma} \frac{y-1/y}{m(y^m-1/y^m)}\, \Omega(\gamma/m,y^m) ,
\ee
which simplify the well-known wall-crossing relations satisfied by the refined indices 
\cite{ks,Manschot:2010qz,Alexandrov:2018iao}.
We use them to define the generating functions
\be 
\hr_{p,\mu}(\tau,z) = \sum_{\hat q_0 \leq \hat q_0^{\rm max}}
\frac{\bOm_{p,\mu}(\hq_0,y)}{y-y^{-1}}\,e^{-2\pi\I \hq_0 \tau },
\label{defhDTr}
\ee  
where, as in \S\ref{subsec-genfun}, we restricted our attention to D4-D2-D0 bound states, 
specified the moduli to be at the large volume attractor point, and used the spectral flow invariance 
to reduce the dependence on charges. An important new feature of \eqref{defhDTr} is the presence of the denominator 
which generates a singularity in the unrefined limit $y\to 1$.
While its inclusion looks artificial, it turns out to be indispensable for $\hr_{p,\mu}$ to have nice modular properties.

In \cite{Alexandrov:2019rth} it has been shown that the construction of the modular completion $\whh_{p,\mu}$
presented in \S\ref{subsec-eqanom} and encoding the modular anomaly of the generating series $h_{p,\mu}(\tau)$
of the unrefined BPS indices, has a natural generalization to the refined case.
According to this more general construction, the generating series $\hr_{p,\mu}(\tau,z)$ \eqref{defhDTr} 
of the refined BPS indices are {\it higher depth mock Jacobi forms} 
where the role of the elliptic argument is played by the refinement parameter $z$. 
Thus, the refinement parameter must transform under $SL(2,\IZ)$ as $z\mapsto z/(c\tau+d)$ together with the modular parameter $\tau$.

A formula for the refined completion takes exactly the same form as \eqref{exp-whh},
\be
\whhr_{p,\mu}(\tau,\btau,z)=\hr_{p,\mu}(\tau,z)+ \sum_{n=2}^{r}\sum_{\sum_{i=1}^n p_i=p}
\sum_{\bfmu}
\rmRirf{\bfp}_{\mu,\bfmu}(\tau, \btau,z)
\prod_{i=1}^n \hr_{p_i,\mu_i}(\tau,z),
\label{exp-whhr}
\ee
but with the coefficients given now by 
\be
\label{Rirf-to-rmRrf}
\rmRirf{\bfp}_{\mu,\bfmu}(\tau,\btau,z)
=
\sum_{\sum_{i=1}^n q_i=\mu+\hf p \atop q_i\in \Lambda_{p_i}+\mu_i+\hf  p_i}  
\Sym \Bigl\{ (-y)^{\sum_{i<j} \gamma_{ij}}\, \scRrf_n(\bfhgam;\tau_2,\beta)
\Bigr\} \,e^{\pi\I \tau Q_n(\bfhgam)},
\ee
where we set $z=\alpha-\tau\beta$ with $\alpha,\beta\in \IR$.
The main difference here, besides the appearance of a power of $y$, 
lies in the form of the functions $\scRrf_n$. They turn out to be much simpler than their unrefined 
version $\scR_n$.
In particular, while $\scR_n$ involve a sum over two types of trees weighted 
by generalized error functions and their derivatives,
$\scRrf_n$ are defined using only one type of trees (or even without them at all!) and no derivatives.
More precisely, the difference is hidden in the functions $\Er_n(\bfhgam;\tau_2,\beta)$,
a refined analogue of $\Ev_n(\bfhgam;\tau_2)$ \eqref{rescEnPhi}.
Although they depend on the additional parameter $\beta$, they are actually much simpler than $\Ev_n$ 
because in their definition there is not any sum over trees.
They simply coincide with the generalized error functions evaluated at appropriate variables (cf. \eqref{rescEn} and \eqref{rescEnPhi}):
\be
\Er_n(\bfhgam;\tau_2,\beta)= \Phi^E_{n-1}\(\{ \bfv_{\ell}\};\sqrt{2\tau_2}\,(\bfq+\beta\bftet )\),
\label{Erefsim}
\ee
where
\be
\bfv_\ell= \sum_{i=1}^\ell\sum_{j=\ell+1}^n\bfv_{ij},
\qquad
\bftet = \sum_{i<j} \bfv_{ij},
\label{def-bfvk}
\ee
while other notations are the same as in \S\ref{subsec-eqanom}.
Note that $\bfv_\ell$ can be thought of as the vectors $\bfv_e$ \eqref{defue} assigned to edges of
the simplest unrooted linear tree $\cT_{\rm lin}=
\bullet\!\mbox{---}\!\bullet\!\mbox{--}\cdots \mbox{--}\!\bullet\!\mbox{---}\!\bullet\,$.
The functions $\Er_n$ have a canonical decomposition similar to \eqref{twocEs},
$\Er_n=\Efrf_n+\Eprf_n$, where\footnote{$\Efrf_n$ is related, but in general not equal to 
the large $\tau_2$ limit of $\Er_n$ evaluated at $\beta=0$. The relation involves a symmetrization
and can be written as 
$$
\Sym\Bigl\{(-y)^{\sum_{i<j} \gamma_{ij}}\Efrf_n(\bfhgam) \Bigr\}
=\Sym\Bigl\{(-y)^{\sum_{i<j} \gamma_{ij}}\lim_{\tau_2\to\infty}\Er_n(\bfhgam;\tau_2,0)\Bigr\}.
$$
\label{foot-limy}}
\be 
\Efrf_n(\bfhgam)= S_{\cT_{\rm lin}}(\bfhgam).
\label{Efref}
\ee
Here $S_\cT$ is defined in \eqref{defST} and for the linear tree the coefficients $e_\cT$ can be computed
to be $e_{\cT_{\rm lin}}\equiv e_{n-1}=\frac1n\,\delta^{(2)}_{n-1}$ with $n$ being the number of vertices.
Finally, the formula for $\scRrf_n$ looks exactly as \eqref{solRn}: 
\be
\scRrf_n\(\bfhgam;\tau_2,\beta\) = \frac{1}{2^{n-1}}\sum_{T\in\IT_n^{\rm S}}(-1)^{n_T-1} 
\Eprf_{v_0} \prod_{v\in V_T \backslash \{v_0\}}\Efrf_v.
\label{refsolRn}
\ee

The claim is that the functions \eqref{exp-whhr} transform as vector valued Jacobi forms of 
weight and index given by
\be 
w=-\hf\, b_2,
\qquad
m(p)=-\chi(\cO_{\cD_p}),
\label{index-Hr}
\ee 
where $\chi(\cO_{\cD_p})$ is the arithmetic genus \eqref{defL0}, 
and with the same multiplier system \eqref{Multsys-hp} as in the unrefined case.
Furthermore, in the unrefined limit, after multiplication by $y-y^{-1}$ to cancel this factor 
in \eqref{defhDTr}, the refined completion reduces to \eqref{exp-whh}. Namely,
\be
\whh_{p,\mu}(\tau,\btau)=
\lim_{y\to 1}\[ (y-y^{-1})\, \whhr_{p,\mu}(\tau,\btau,z)\].
\label{unreflim-h}
\ee
This fact follows from a similar very non-trivial property of the coefficients $\rmRirf{\bfp}_{\mu,\bfmu}$:
\be
\rmRi{\bfp}_{\mu,\bfmu}(\tau,\btau)=
\lim_{y\to 1} \[(y-y^{-1})^{1-n} \,\rmRirf{\bfp}_{\mu,\bfmu}(\tau,\btau,z)\],
\label{unreflim-Rn}
\ee
which ensures the cancellation of the poles of the refined generating functions appearing on the r.h.s. of \eqref{exp-whhr}.
It also explains the appearance of the derivatives of the generalized error functions in the unrefined construction
(see \eqref{rescEn}) as a consequence of applying the L'H\^opital's rule to evaluate the limit in \eqref{unreflim-Rn}.
Besides, it is worth mentioning that the results presented in \S\ref{subsubsec-collinear}
have originally been proven in the refined case and then followed by taking the limit \eqref{unreflim-h} \cite{Alexandrov:2019rth}.

In fact, it was recently noticed \cite{Pioline:2025xgf} that in the collinear case as in \S\ref{subsubsec-collinear}, for $n=3$ and 4,  
the sum over Schr\"oder trees in \eqref{refsolRn},
upon substitution of the expression \eqref{expPhiE-mod} of the generalized error functions $\Phi^E_n$
in terms of their complementary counterparts $\hPhi^M_n$ \eqref{def-hPhiM}, 
results in a huge cancellation leaving a single term supplemented 
only by contributions involving Kronecker deltas of the type appearing in \eqref{defST}.\footnote{Strictly speaking, 
$\hPhi^M_n$ are not defined at the loci of their discontinuities where the Kronecker deltas are supported.
Here we define them at these loci through the relation \eqref{expPhiE-mod} to $\Phi^E_n$, 
which are smooth and do not have this problem.} 
This observation suggests the following\footnote{This conjecture has been proven in \cite{AC}.}

\begin{conj} \label{conj-simple-ref}
	In the case of collinear magnetic charges as in \S\ref{subsubsec-collinear},
	the functions $\scRrf_n$ determining the coefficients \eqref{Rirf-to-rmRrf} are given by
	\be
	\scRrf_n\(\bfhgam;\tau_2,\beta\) =
	\frac{1}{2^{n-1}}\sum_{\cJ\subseteq \Zv_{n-1}}
	\hPhi^M_{|\cJ|}\(\{ \bfv_{\ell}\}_{\ell\in\cJ};\sqrt{2\tau_2}\,(\bfq+\beta\bftet ),\sqrt{2\tau_2}\,\beta\bftet\) 
	\prod_{k=1}^{m_{\cJ}}
	b_{|\cI_k|+1}\,\delta_{\cI_k\perp\cJ},
	\label{refsolRn-simple}
	\ee	
	where $\cI_k$ are subsets of $\Zv_{n-1}\setminus\cJ$ separated by the elements of $\cJ$ in $\Zv_{n-1}$,
	$m_{\cJ}$ is their number,
	and\footnote{The coefficients $b_n$
	have been introduced in \cite{Alexandrov:2018lgp} and are inverse to the coefficients $e_{n}$ introduced below \eqref{Efref}
	in the sense that $b_n$ and $e_{n-1}$ are the Taylor series coefficients of the mutually inverse functions,
	$\tanh(x)$ and $\mbox{arctanh}(x)$.}
	\be 
	b_{n-1}=\frac{2^n(2^n-1)}{n!}\, B_n,
	\qquad 
	\delta_{\cI\perp\cJ}=\prod_{i\in \cI}\delta_{\bfv_{i\perp\cJ}\cdot\bfq},
	\label{defdel}
	\ee 
	with $B_n$ being the Bernoulli number and $\bfv_{i\perp \cJ}$ denoting the projection of $\bfv_i$ 
	orthogonal to the subspace spanned by $\{\bfv_j\}_{j\in\cJ}$.
\end{conj}

This conjecture further simplifies the anomaly by expressing each term in \eqref{exp-whhr} as 
an iterated integral of depth $n-1$.
Since the anomaly in the unrefined case can be obtained by taking the limit \eqref{unreflim-Rn} of 
the refined coefficients, for collinear charges, the construction presented in \S\ref{subsec-eqanom} is also expected to 
have a simpler version in terms of the complementary generalized error functions $\hPhi^M_n$.

Finally, it was argued in \cite{Alexandrov:2019rth} that, if the quadratic form is degenerate, 
as it happens in the case of an elliptically fibered CY or its local limit (Ex. \ref{ex-elliptic}),
this affects not only the weight of the refined generating functions
but also their index. The previous expressions \eqref{index-Hr} are then replaced by
\be 
w(p)=-\hf\, \rank(\Lambda_p),
\qquad
m(p)=-\chi(\cO_{\cD_p})-\lambda_a p^a, 
\label{index-Hr0}
\ee 
where the last term should be an integer. Its precise value in the generic case and its origin remain unclear. 
In \S\ref{subsec-VW}, this term will be needed to get the correct value of the index of the generating functions of refined
VW invariants.

We observe that the refined BPS indices appear to possess very similar modular properties to the unrefined ones 
and even simplify the description of the corresponding modular anomaly.
However, in the case of a compact CY threefold, there is a problem: it is not clear whether 
the refined BPS indices can actually be well-defined. 
More precisely, there seems to be no natural deformation invariant way of defining them.
In physics terms, this means that their naive definition is not protected by supersymmetry
and they may change under the variation of hypermultiplet moduli.
In a drastic contrast, in the case of non-compact CYs, it is possible to define protected refined indices 
due to the existence of a certain $\IC^\times$ action carried by the moduli space of semi-stable sheaves,
corresponding to an additional $SU(2)$ R-symmetry in the dual supersymmetric gauge theories.
This is achieved by changing the factor $(-y)^{2J_3}$ by $(-1)^{2J_3} y^{2(J_3+I)}$ 
where $I$ is the generator of the additional symmetry \cite{Gaiotto:2010be}.\footnote{Recently, in \cite{Huang:2025xkc}
the refined BPS indices were defined for elliptically fibered CY threefolds through the gauge theories emerging 
in various local limits.\label{foot-refine}}

Thus, even if the compact case remains problematic, there are two possible ways to use the construction presented in this subsection:
\begin{itemize}
	\item 
	as a description of modular properties of the generating functions of refined BPS indices on non-compact CYs
	(or in any other case, like in footnote \ref{foot-refine}, where these indices can be well-defined);
	\item 
	as a useful trick to compute modular completions in the unrefined case.
\end{itemize}
Below we explain an additional structure associated with the refined construction which 
suggests that it is not just a mere trick, but reveals something fundamental even in the compact case.

\subsubsection{Non-commutative structure}

An important difference between the unrefined and refined constructions is that the former was derived
(following the steps sketched in \S\ref{subsec-deriv}), whereas the latter was simply guessed.
One of the reasons for this was explained above: this is the lack of a satisfactory definition of refined BPS indices
for compact CYs and, as a result, the lack of understanding of the implications on them of string dualities.
Even the fate of S-duality under the refinement is not clear.  
  
Nevertheless, in \cite{Alexandrov:2019rth} it was shown that it is possible to revert the logic 
explained in \S\ref{subsec-deriv} and, starting from \eqref{exp-whhr}, to reconstruct a function
on the moduli space $\cM$ whose modularity is equivalent to the expression for $\whhr_{p,\mu}$. 
This function $\cGr$ is supposed to be a refined analogue of the instanton generating potential $\cG$ \eqref{defcF2}.
Although its geometric meaning remains unclear, its existence hints that S-duality is preserved by the refinement.

Moreover, it strongly suggests that the refinement makes the moduli space $\cM$ {\it non-commutative}! 
On one hand, this is somewhat expected from previous studies of refined indices \cite{Gaiotto:2010be,Cecotti:2010fi}.
On the other hand, it is remarkable that the non-commutativity appears as an indispensable ingredient
in the definition of $\cGr$. 
It manifests through the following {\it modular invariant} star product defined on functions on the moduli space:
\be
f \star g =  f \exp\[ \frac{1}{2\pi\I}\( \overleftarrow{\Dv}_{\!\!a}\overrightarrow{\p}_{\!\tc_a}-
\overleftarrow{\p}_{\!\tc_a} \overrightarrow{\Dv}_{\!\! a}\) \] g,
\label{starproduct-alt}
\ee
where
\be
\Dv_a=\alpha\p_{c^a}+\beta\p_{b^a}=z\p_{v^a}+\bz\p_{\bv^a}
\ee
with $v^a=c^a-\tau b^a$. The modular invariance is due to the fact that $z$ and $v^a$ both transform as
elliptic variables, whereas $\p_{\tc_a}$ is modular invariant.
One can show that with respect to this star product the classical Darboux coordinates \eqref{clactinst}
satisfy
\be
\cXcl_{\gamma_1}(\zsf_1)\star\cXcl_{\gamma_2}(\zsf_2)=
y^{\gamma_{12}+\frac{\beta}{2}\, (p_1p_2 (p_1+p_2))}
(y\by)^{-\I(p_1p_2t)(\zsf_1-\zsf_2)}\cXcl_{\gamma_1}(\zsf_1)\cXcl_{\gamma_2}(\zsf_2).
\label{starXX}
\ee
This non-commutativity relation might look unusual, but this is because we allowed the refinement parameter
$z$ to have a non-vanishing imaginary part. If it vanishes, i.e. $\beta=0$ so that $y$ is a pure phase, 
the only remaining factor in \eqref{starXX} is $y^{\gamma_{12}}$, as in \cite{Gaiotto:2010be,Cecotti:2010fi}.

Using the star product \eqref{starproduct-alt}, we define {refined Darboux coordinates}
as solutions of the following integral equation (cf. \eqref{expcX})
\be
\cXr_{\gamma}=\cXcl_{\gamma}\star\(1+\cJr_\gamma\),
\qquad
\cJr_\gamma(\zsf)=\sum_{\gamma'\in \Gamma_+} \bOm(\gamma,y)\,
\int_{\ell_{\gamma'}}\de \zsf'\, \Kr_{\gamma\gamma'}(\zsf,\zsf')\,\cXr_{\gamma'}(\zsf'),
\label{inteqH-star}
\ee  
where the integration kernel is given by
\be
\Kr_{\gamma_1\gamma_2}(\zsf_1,\zsf_2)
=\frac{\I}{2\pi}\, \frac{y^{-\beta m(p)}}{y-y^{-1}}\, \frac{1}{\zsf_1-\zsf_2}
\label{defkerKr}
\ee
and $m(p)$ is the index \eqref{index-Hr}.
Then the refined instanton generating potential can be shown to have the following integral representation 
\be
\cGr=\frac{1}{4\pi^2}\,\frac{y^{-\beta m(p)}}{y-y^{-1}}
\sum_{\gamma\in \Gamma_+}  \bOm(\gamma,y) \int_{\ell_{\gamma}}\de \zsf\, \cXr_\gamma(\zsf).
\label{nonpert-cGr}
\ee
It is a Jacobi form of weight $(-\hf\, ,\, \hf)$ and index 0 provided $\whhr_{p,\mu}$ \eqref{exp-whhr} 
have the modular properties specified in \eqref{index-Hr}. 
Comparing to \eqref{defcF2}, one again observes that the refined version is simpler as it is given by a single integral 
in contrast to the unrefined version involving also a double integral.

One can wonder why the integral equation \eqref{inteqH-star} has so unusual form which has nothing to do with 
TBA-like equations. In fact, it does not have the unrefined limit due to the pole in the kernel \eqref{defkerKr},
just as there is no such limit for the refined Darboux coordinates $\cXr_\gamma$.
Nevertheless, the limit exists for $(y-y^{-1})\cGr$ where it reproduces $\cG$. 
The situation was clarified recently in \cite{ABrefine}. It turns out that one can introduce another version 
of refined Darboux coordinates defined through $\cXr_\gamma$ by

\be 
\hcXr_\gamma
=\(1+\cJr_\gamma \)_\star^{-1}\star \cXr_\gamma 
= \(1+\cJr_\gamma\)_\star^{-1}\star \cXcl_\gamma \star \(1+\cJr_\gamma\),
\label{hcXranz}
\ee 
where the star index means that $(1+x)^{-1}_\star=\sum_{n=1}^\infty(-1)^n x\star \cdots \star x $.
One can show that the new coordinates do have the unrefined limit where they reduce to $\cX_\gamma$.
Besides, they satisfy the standard refined wall-crossing relations \cite{ks} and,
for $\zsf_1=\zsf_2$, the commutation relations \eqref{starXX}.
Thus, these are $\hcXr_\gamma$, rather than $\cXr_\gamma$, that should be considered as a refined version of $\cX_\gamma$
and as a solution of the quantum Riemann-Hilbert problem introduced in \cite{Barbieri:2019yya}
and studied in \cite{Bridgeland:2020zjh,Alexandrov:2021wxu,Chuang:2022uey}.

Since $\cX_\gamma$ are the Darboux coordinates on the twistor space over $\cM$, one can view 
$\hcXr_\gamma$ as Darboux coordinates on a quantization of this twistor space.
Although a theory of such non-commutative spaces has not been developed yet, 
a small step towards it has been performed in \cite{Alexandrov:2023wdj}
where it was shown how one can define a {\it quantized contact structure}.
Thus, one can hope that the above construction opens a door into a potentially rich and still poorly explored topic
of quantum twistor spaces.

\section{Solution of the modular anomaly}
\label{sec-solution}

The expression for the modular completion \eqref{exp-whh} (or its refined version \eqref{exp-whhr})
can be seen as an iterative system of modular anomaly equations on the generating functions $h_{p,\mu}$. 
Of course, these equations cannot fix $h_{p,\mu}$ uniquely but only up to addition of a holomorphic modular form,
which is not ``seen" by the modular anomaly. To fix the modular ambiguity, one should provide an additional information.
For example, as explained in \S\ref{sec-modular}, this can be information about the coefficients of the polar terms of $h_{p,\mu}$.  

Thus, the equations \eqref{exp-whh} can be used to find the generating functions, but one should follow a two step strategy:
\begin{enumerate}
	\item 
	find {\it any} mock modular form $\han_{p,\mu}$ having the modular anomaly
	described by \eqref{exp-whh};
	\item 
	represent
	\be
	h_{p,\mu}=\han_{p,\mu}+\hh_{p,\mu},
	\label{han}
	\ee
	where $\hh_{r,\mu}$ is a modular form, and find the second term by computing the polar terms.
\end{enumerate}
Of course, for $p^a$ corresponding to irreducible divisors, the first step is not necessary as the generating function
is not anomalous.

The computation of polar terms is, in general, a highly non-trivial problem.
We show how it can be systematically approached in \S\ref{subsec-topinv} in the case of one-parameter CY threefolds, 
i.e. when $b_2=1$.
However, it seems impossible to give any general formulas for the polar coefficients, so that they have to be computed 
example by example. And even in the one-parameter case, the results remain quite limited. 
In contrast, as we will see, it is possible to solve the modular anomaly equations \eqref{exp-whh},
at least in the same one-parameter case, in full generality.
Given this situation, we postpone the second step of the above procedure and concentrate in this section on the first one.

\subsection{Anomalous coefficients}
\label{subsec-anomcoef}

An immediate problem arising when one tries to solve \eqref{exp-whh} to get $\han_{p,\mu}$
is that the r.h.s. of this equation depends on the {\it full} generating functions $h_{p_i,\mu_i}$
of the constituents and hence on all functions $\hh_{p_i,\mu_i}$ that remain unknown 
because we decided to fix them at a later stage.
In such situation, the best we can do is to find $\han_{p,\mu}$ in a form parametrized by $\hh_{p_i,\mu_i}$.
It is clear that the dependence on these functions must be polynomial and in each monomial the charges $p_i^a$ must 
sum up to $p^a$. This brings us to the following ansatz
\be
\han_{p,\mu}(\tau)= \sum_{n=2}^\Nr\sum_{\sum_{i=1}^n p_i=p}
\sum_{\bfmu}
\gi{\bfp}_{\mu,\bfmu}(\tau)
\prod_{i=1}^n \hh_{p_i,\mu_i}(\tau).   
\label{genansatz}
\ee
Note that one can write a similar formula for the full generating function $h_{p,\mu}$ 
if one starts the sum from $n=1$ and sets by definition $\gi{p}_{\mu,\mu'}=\delta_{\mu,\mu'}$.
The other coefficients $\gi{\bfp}_{\mu,\bfmu}$ with $n\geq 2$, where $n$ is the length of the tuples $\bfp$ and $\bfmu$, 
are the functions to be found. To do this, we need to know the constraints that they satisfy and that can be derived 
by plugging the ansatz \eqref{genansatz} into \eqref{exp-whh}.
In \cite{Alexandrov:2024wla}, the following result has been proven\footnote{In fact, in \cite{Alexandrov:2024wla}
Theorem \ref{thm-ancoef} has been proven only in the one-modulus case, but it is easy to see that exactly 
the same proof applies to the generic case.}

{\Theorem\label{thm-ancoef}
	Let $\hh_{p,\mu}$ be a set of holomorphic modular forms.
	Then $h_{p,\mu}$ is a depth $r-1$ modular form whose completion has the form \eqref{exp-whh}
	provided $\gi{\bfp}_{\mu,\bfmu}$ are depth $n-1$ mock modular forms 
	whose completions satisfy
	\be
	\whgi{\bfp}_{\mu,\bfmu}=
	\Sym\Bigg\{\sum_{m=1}^n 
	\sum_{\sum_{k=1}^m n_k=n}
	\sum_{\bfnu}\rmRi{\bfs}_{\mu,\bfnu}
	\prod_{k=1}^m \gi{\frp_k}_{\nu_k,\frm_k}
	\Bigg\},
	\label{compl-gi}
	\ee
	where
	\be
	j_k=\sum_{l=1}^{k-1} n_l,
	\qquad
	\Ms_k^a=\sum_{i=1}^{n_k} p_{j_k+i}^a,
	\qquad
	\begin{array}{c}
		\frp_k=(p_{j_k+1}^a,\dots, p_{j_{k+1}}^a),
		\\
		\frm_k=(\mu_{j_k+1},\dots,\mu_{j_{k+1}}).
		\vphantom{\sum\limits^{a}}
	\end{array}
	\label{split-rs}
	\ee
}

\begin{figure}
	\isPreprints{\centering}{} 
	\includegraphics[width=9cm]{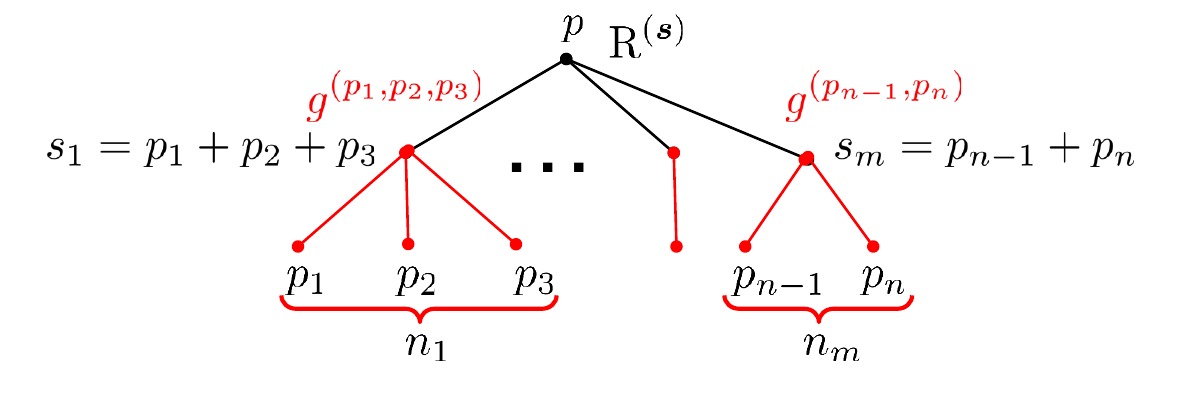}
	\vspace{-0.5cm}
	\caption{A representation of contributions to the r.h.s. of \eqref{compl-gi} 
		in terms of planar rooted trees with all leaves having depth 2.\label{fig-g-trees}}
\end{figure}

Note that while the sets $\bfp$ and $\bfmu$ have $n$ elements, 
the sets $\bfs$ and $\bfnu$ have only $m\le n$ elements. 
To comprehend the structure of the equation \eqref{compl-gi}, it might be 
useful to notice the fact that the sum on its r.h.s. is equivalent to the sum over
planar rooted trees of uniform leaf-depth 2 with leaves labeled by charges $p_i^a$ and other vertices
labeled by the sum of the charges of their children. Using this labeling, we assign 
the function $\rmRi{\bfs}_{\mu,\bfnu}$ to the root vertex and 
the coefficients $\gi{\frp_k}_{\nu_k,\frm_k}$ to the vertices of depth 1 with arguments determined by the charges 
of their children (see Fig. \ref{fig-g-trees} and cf. Fig. \ref{fig-anomaly}).
Then the contribution of a tree is given by the product of the weights of its vertices.

Theorem \ref{thm-ancoef} reformulates the problem of finding $\han_{p,\mu}$ as the problem of finding 
the functions $\gi{\bfp}_{\mu,\bfmu}$, which were called {\it anomalous coefficients}.
It states that they are also higher depth mock modular forms satisfying an iterative system of anomaly equations.
Then why is it better than the previous one? The difference between \eqref{exp-whh} and \eqref{compl-gi},
is that, solving the latter, there is no need to fix the modular ambiguity in $\gi{\bfp}_{\mu,\bfmu}$.
As was emphasized above, {\it any} solution will suit our purposes because a difference between two solutions 
can always be absorbed into a redefinition of the modular ambiguities $\hh_{p,\mu}$ 
in the formula for the generating functions (\eqref{han} combined with \eqref{genansatz}).
It is important however to take into account that a choice of solution for $n$ charges affects all 
anomaly equations for $\gi{\bfp}_{\mu,\bfmu}$ with $n'>n$ charges.
One should remember this when one deals with different systems of solutions to avoid potential inconsistencies.

\subsection{One-modulus case}

The problem of finding the anomalous coefficients $\gi{\bfp}_{\mu,\bfmu}$ has been addressed 
in the case of compact CY threefolds with one K\"ahler modulus \cite{Alexandrov:2024wla}.
Since in this case the D4-brane charge $p^1$ is equal to the degree of reducibility of the corresponding divisor $r$,
we will use the latter variable to denote magnetic charges.
Besides, it is convenient to make a redefinition which allows us to absorb some annoying sign factors.
Namely, let us introduce redefined generating functions
\be
\tlh_{r,\mu}(\tau) =
(-1)^{(\Nr-1)\mu }h_{r,\mu - \frac{\kappa \Nr(\Nr-1)}{2}}(\tau).
\label{def-tgi}
\ee
This leads to two simplifications: 
i) the shift of $\mu$ replaces the last term in the spectral flow decomposition \eqref{decomp-spfl},
which in the one-parameter case reads as $\hf \kappa r^2$, by a term linear in $r$,
\be
\label{defmu-shift}
q =  \kappa r \eps+\mu + \frac12\, \kappa r;
\ee
ii) the sign factor in \eqref{def-tgi} cancels the sign factor in \eqref{defRn}.
In particular, due to the first simplification, the condition on the sum over $q_i$ in \eqref{defRn} takes
the following simple form
\be
\kappa\sum_{i=1}^n r_i\eps_i=\Delta\mu, 
\qquad
\Delta\mu=\mu-\sum_{i=1}^n\mu_i.
\label{constr-epsi}
\ee
Then we introduce redefined anomalous coefficients,
for which, to avoid cluttering, we will use the same notation $\gi{\bfr}_{\mu,\bfmu}$ as before the redefinition. 
We define them by the same ansatz \eqref{genansatz}, but with the functions $\han_{p,\mu}$ and $\hh_{p,\mu}$
replaced by their redefined versions. Equivalently, they should satisfy \eqref{compl-gi} with $\rmRi{\bfr}_{\mu,\bfnu}$
replaced by 
\be
\trmRi{\bfr}_{\mu,\bfmu}(\tau, \btau)=
\sum_{\sum_{i=1}^n q_i=\mu+\kappa r/2 \atop q_i\in \kappa r_i \IZ+\mu_i+\kappa r_i/2} 
\Sym\Bigl\{\scR_n(\bfhgam;\tau_2)\Bigr\}\, e^{\pi\I \tau Q_n(\bfhgam)}.
\label{redefRn}
\ee

\subsection{Partial solutions}

If one restricts to $n=2$, there is a very simple way to solve for $\gi{\Nr_1,\Nr_2}_{\mu,\mu_1,\mu_2}$.
From the holomorphic anomaly equation \eqref{anom-collin} specified to $p_0=1$, it follows that one can take
\be
\gi{\Nr_1,\Nr_2}_{\mu,\mu_1,\mu_2}(\tau)=\Nr_0\delta^{(\kappa\Nr_0)}_{\Delta\mu}
\Gi{\kappa_{12}}_{\mu_{12}}(\tau),
\label{sol-n=2}
\ee
where $\Nr_0=\gcd(\Nr_1,\Nr_2)$, 
$\kappa_{12}=\kappa \rr_{12}/2$, $\rr_{12}$ and\footnote{In the definition \eqref{def-param12} of $\mu_{12}$, 
one should drop the term proportional to $p_0$ which disappears due to the redefinition \eqref{def-tgi}.}
$\mu_{12}$ are parameters introduced in \S\ref{ap-null}, 
$\mu_{12}$ runs over $2\kappa_{12}$ values,
and $\Gi{\kappa}_{\mu}$ is a vector valued mock modular form 
of weight 3/2 with the shadow proportional to the following unary holomorphic theta series
\be
\ths{\kappa}_\mu(\tau)
=\sum_{k\in 2\kappa\IZ+\mu}\q^{\frac{k^2}{4\kappa} }.
\label{deftheta-kap}
\ee
Thus, $\Gi{\kappa}_{\mu}$ is not a mixed but ordinary mock modular form, 
and the problem reduces to its reconstruction given its shadow.
Remarkably, exactly this problem was solved in \cite{Dabholkar:2012nd} in the context of $\cN=4$
string compactifications, with an additional condition that the mock modular form should have the slowest possible 
asymptotic growth of its Fourier coefficients. 
Such functions have been called mock modular forms of optimal growth.
Although we do not impose any restrictions on the asymptotic growth, we can take $\Gi{\kappa}_{\mu}$ 
to be the solution found in \cite{Dabholkar:2012nd} because, as was discussed above,
any solution is equally suitable for us. All other solutions would differ just by a pure modular form.

The mock modular forms of optimal growth are determined by a single parameter $\kappa$ and 
constructed by acting by certain Hecke-like operators on a set of ``seed"
mock modular forms $\cGi{d}_\mu$ which have to be introduced for each square-free integer $d$ 
with an even number of prime factors, such as 1, 6, 10, 14, 15, etc.
In particular, for $d=1$ the seed function is given by the generating series of Hurwitz class numbers (see Ex. \ref{ex-Hurwitz}):
\be
\cGi{1}_\mu(\tau)=H_\mu(\tau).
\label{cG-H}
\ee
This implies that for all $\kappa_{12}$ given by a power of a prime number, the mock modular form $\Gi{\kappa_{12}}_{\mu_{12}}$
is generated by the Hurwitz class numbers. For a detailed description of 
the solution for $\Gi{\kappa_{12}}_{\mu_{12}}$ in terms of the mock modular forms of optimal growth
we refer to \cite{Alexandrov:2024wla}. 
 
Another class of anomalous coefficients that can be found almost for free appears for CYs 
with the intersection number $\kappa=1$.
It comprises the anomalous coefficients with all charges $r_i$ equal to one.
A crucial simplification in this case is that one can drop all indices $\mu_i$ 
because they take only $\kappa r_i=1$ value. Due to this, the corresponding anomalous coefficients
can be denoted simply as $g_{n,\mu}\equiv\gi{1,\dots,1}_{\mu}$.
It turns out that they can be identified with the normalized generating series of $U(n)$ VW invariants 
on $\IP^2$ (see \S\ref{subsec-VW})
\be
\vwgi{n,\mu}(\tau) = \eta^{3n}(\tau)\,h^{\IP^2}_{n,\mu}(\tau),
\label{def-VWnorm}
\ee
where we used the fact that $h^{\IP^2}_1=\eta^{-3}$.
More precisely, one can take \cite{Alexandrov:2024wla}
\be
g_{n,\mu}=3^{1-n}\vwgi{n,\mu}(\tau).
\qquad
\label{rel-tggP2-n}
\ee
Note that for $n=2$ this choice is consistent with the solution given by the mock modular form of optimal growth 
which coincides with \eqref{cG-H}.

Unfortunately, neither of the above solutions seems to be generalizable to other cases.
Therefore, below we present a different construction which is more complicated, but works for generic charges and parameters.
It produces anomalous coefficients different from the ones introduced in this subsection.
Therefore, as explained at the end of \S\ref{subsec-anomcoef}, 
if one wants to go beyond $n=2$ or the very special case $\kappa=r_i=1$,
even for $n=2$ one should use the solution constructed in the next subsection and not here.

\subsection{General solution}
\label{subsec-gensol-ac}

A solution for the anomalous coefficients that works for any $n$ and any charges $r_i$ 
can be constructed in terms of indefinite theta series. 
This is a very natural approach given that the functions $\rmRi{\bfr}_{\mu,\bfnu}$, 
as well as their redefined version $\trmRi{\bfr}_{\mu,\bfmu}$,
determining the modular anomaly of $\gi{\bfr}_{\mu,\bfmu}$ are themselves such indefinite theta series.\footnote{More precisely, 
	they are defined by the quadratic form \eqref{defQlr} which is {\it negative definite} 
    in the one-modulus case. As we will see, this fact leads to additional complications.\label{foot-signlat}}
Although we could just present the final result found in \cite{Alexandrov:2024wla} together with its necessary ingredients,
we prefer first to explain where it comes from. Otherwise its rather non-trivial form would look
completely mysterious to the reader.

\subsubsection{Strategy}

The anomalous coefficients must be holomorphic functions.
Therefore, if we express them in terms of indefinite theta series, as was explained in \S\ref{sec-indef}, 
the kernels of these theta series must be combinations of sign functions. A typical example of such kernel 
is provided in Theorem \ref{th-conv} and is determined by two sets of vectors of dimension of the lattice.
On the other hand, after substitution into \eqref{compl-gi}, the theta series must recombine into a modular form.
According to the recipe \eqref{replace-alln}, this means that each product of sign functions 
should be effectively replaced by the corresponding generalized error function.
However, already for $n=2$ it is easy to see that this is impossible. Indeed, in this case the anomaly equation 
to be solved takes the simple form
\be
\whgi{\Nr_1,\Nr_2}_{\mu,\mu_1,\mu_2}(\tau,\btau)
=\gi{\Nr_1,\Nr_2}_{\mu,\mu_1,\mu_2}(\tau)+\trmRi{\Nr_1,\Nr_2}_{\mu, \mu_1, \mu_2}(\tau, \btau).
\label{whh2}
\ee
The function $\trmRi{\Nr_1,\Nr_2}_{\mu, \mu_1, \mu_2}$ is built of a single complementary error function,
whereas the kernel of an indefinite theta series, to make it convergent, 
should be a linear combination of at least two sign functions.
Thus, one sign function can be recombined with $\trmRi{\Nr_1,\Nr_2}_{\mu, \mu_1, \mu_2}$ to produce an error function,
while the second sign remains "uncompleted".
Fortunately, this problem can be solved by choosing the vector defining the second sign function to be {\it null}
with respect to the relevant quadratic form. Then, due to the property \eqref{Phinull}, 
the completion is not required. This solution generalizes to any $n$: choosing the kernel to be of the type \eqref{kerconverge},
one set of vectors will be determined by the vectors appearing in the definition of 
$\trmRi{\Nr_1,\Nr_2}_{\mu, \mu_1, \mu_2}$, while the second set should consist of null vectors. 

However, null vectors give rise to other problems. The first one is related to convergence
because null vectors spoil the conditions of Theorem \ref{th-conv}. On the other hand, as discussed below 
Theorem \ref{th-conv}, they can still be included provided i) they belong to the relevant lattice, 
ii) the theta series includes a non-vanishing elliptic parameter. 
In our story, the latter can be associated with the refinement parameter $z$
(see \S\ref{subsec-refine}). Thus, we must switch on the refinement if we want to use indefinite theta series!
Of course, in the end one should take the unrefined limit which can be non-singular only if 
the indefinite theta series are combined with some other types of Jacobi forms.
In fact, since we are interested only in the behavior near $z=0$, the elliptic property \eqref{Jacobi-ell}
of Jacobi forms is not essential and can be abandoned, so that it is sufficient to require
that all relevant functions are Jacobi-like forms.

The second problem with null vectors is that they simply do not exist in our lattice just because 
it is negative definite (see footnote \ref{foot-signlat}).
This problem can be solved by a well-known trick in the theory of mock modular forms (see, e.g., \cite{Zwegers-thesis})
which is to effectively extend the lattice by multiplying, for example, with Jacobi theta functions \eqref{free-theta-1}.
Though this trick works, it gives rise to a serious technical complication. 
It turns out that for a solution on the extended lattice to be reducible to a solution on the original lattice,
it should have zero at $z=0$ of order given by the difference of the dimensions of the two lattices.
This property is very difficult to achieve.
Fortunately, this issue can be avoided by introducing multiple refinement parameters combined into a vector 
$\zbbm=(z\bftet,\vec z)$ of dimension of the extended lattice, 
such that $(0,\vec z)$ is orthogonal to all null vectors used in the construction.
The latter condition ensures the decoupling of the auxiliary part of the lattice.
As a result, the indefinite theta series we have to deal with are multi-variable (mock) Jacobi forms as in \eqref{gentheta}.
Once they are constructed and combined with proper Jacobi-like forms to ensure the existence of the unrefined limit,
they have to be reduced to a solution of the original problem.

\begin{figure}
	\isPreprints{\centering}{} 
	\includegraphics[width=10cm]{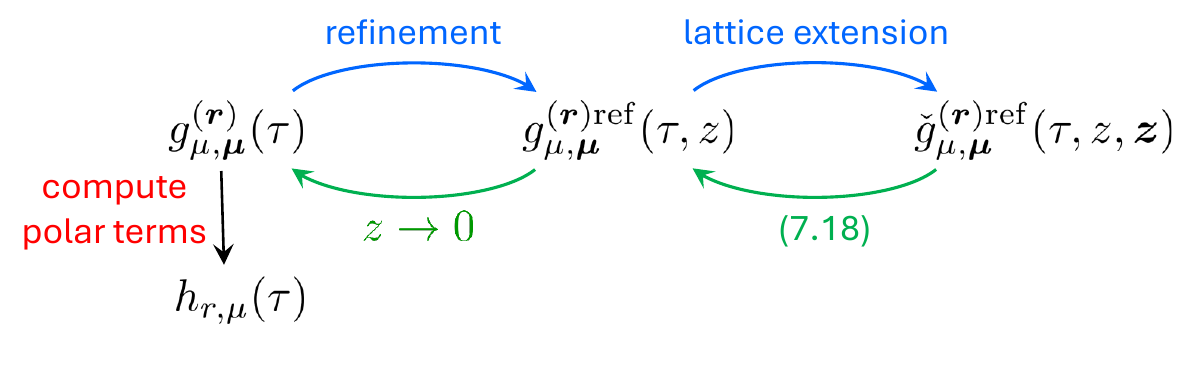}
	\vspace{-0.5cm}
	\caption{Construction of the anomalous coefficients through the refinement and lattice extension
		and their relation to the generating functions of BPS indices.\label{fig-strategy}}
\end{figure}

Thus, the solution presented below is constructed by performing the following steps:
\begin{enumerate}
	\item
	First, one introduces the refinement and looks for vector valued mock Jacobi-like forms $\girf{\bfr}_{\mu,\bfmu}(\tau,z)$
	of depth $n-1$, weight $\hf(n-1)$, index
	\be 
	m_{\bfr}=-\frac{\kappa}{6}\,\biggl(\Nr^3-\sum_{i=1}^n\Nr_i^3\biggr),
	\label{index-mr}
	\ee
	and the multiplier system specified in \cite[Eq.(B.4)]{{Alexandrov:2024wla}},
	satisfying the analogue of the equations \eqref{compl-gi} with $\rmRi{\bfr}_{\mu,\bfnu}$
	replaced by $\trmRirf{\bfr}_{\mu,\bfmu}$, the redefined version of \eqref{Rirf-to-rmRrf},
	and having a zero of order $n-1$ at $z=0$ to ensure the unrefined limit (see \eqref{lim-ancoef} below).
	
	\item 
	Next, one extends the charge lattice so that it possesses a set of null vectors suitable for solving the anomaly equation 
	and associates with the lattice extension a vector of additional refinement parameters
	satisfying certain orthogonality properties with the null vectors.
	
	To this end, let us define $\eps=\delta_{\kappa-1}$, $d_r=4^\eps \kappa r$, $d_\bfr=\sum_{i=1}^n d_{r_i}$,
	and introduce $d_r$-dimensional vectors $\frt^{(\Nr)}$ such that their components 
	are all non-vanishing integers and sum to zero, $\sum_{\alpha=1}^{d_\Nr}\frt^{(\Nr)}_\alpha=0$.
	Of course, there are plenty of possible choices of such vectors and 
	the following construction does not depend on their concrete form.
	Given these data, one looks for vector valued multi-variable mock Jacobi-like forms $\chgirf{\bfr}_{\mu,\bfmu}(\tau,z,\bfz) $
	depending on $n+1$ refinement parameters $(z,\bfz)$ where $\bfz=(z_1,\dots,z_n)$ and 
	satisfying a new anomaly equation:
	\be
	\whchgirf{\bfr}_{\mu,\bfmu}(\tau,z,\bfz) 
	=
	\Sym \Bigg\{
	\sum_{m=1}^n \sum_{\sum_{k=1}^m n_k=n} \sum_{\bfnu}
	\trmRirf{\bfs}_{\mu,\bfnu}(\tau,z)
	\prod_{k=1}^m \chgirf{\frr_k}_{\nu_k,\frm_k} (\tau,z,\frz_k),
	\Bigg\},
	\label{extmodan-manyz}
	\ee
	where $\frz_k=(z_{j_k+1},\dots,z_{j_{k+1}})$. Although it looks identical to the previous equation 
	on $\girf{\bfr}_{\mu,\bfmu}(\tau,z)$, it is supplemented by a new normalization condition for $n=1$:
	\be 
	\chgirf{\Nr}_{\mu,\mu'}(\tau,z,z')=\delta_{\mu,\mu'}\, 
	\prod_{\alpha=1}^{d_\Nr} \theta_1(\tau,\frt^{(\Nr)}_\alpha z').
	\label{newnorm-g}
	\ee 
	The crucial property of \eqref{extmodan-manyz} is that its solutions
	that are regular at $\bfz=0$ give rise to the functions $\girf{\bfr}_{\mu,\bfmu}(\tau,z)$
	introduced at the previous step. The relation between the two sets of functions is given by
	\be
	\girf{\bfr}_{\mu,\bfmu}(\tau,z)= 
	\frac{1}{\(-2\pi\eta^3(\tau)\)^{d_\bfr}} 
	\(\prod_{i=1}^{n}\frac{\cD^{(d_{\Nr_i})}_{\hf(\frt^{(\Nr_i)})^2}(z_i)}
	{d_{\Nr_i}!\prod_{\alpha=1}^{d_{\Nr_i}}\frt^{(\Nr_i)}_\alpha}\) 
	\chgirf{\bfr}_{\mu,\bfmu}|_{\bfz=0},
	\label{recover-gref}
	\ee
	where the differential operators $\cD_m^{(n)}$ are defined in \eqref{defcDmn}.
	
	It is the presence of the additional factors of the Jacobi theta function in \eqref{newnorm-g} that leads 
	to an effective extension of the lattice defining the theta series 
	that capture the coefficients on the r.h.s. of \eqref{extmodan-manyz}. 
	While the original lattice, which can be read off, e.g., 
	from \eqref{redefRn}, is given by
	\be
	\bfLami{\bfr}=\left\{\bfk\in \IZ^n \ :\ \sum_{i=1}^n \Nr_i k_i=0 \right\}
	\label{def-bfLam}
	\ee
	with the bilinear form
	\be
	\bfx\cdot\bfy=\kappa\sum_{i=1}^n \Nr_i x_i y_i, 
	\label{bf-r}
	\ee
	its extended version turns out to be
	\be
	\bbLami{\bfr}=\bfLami{\bfr}\oplus \IZ^{d_\bfr}
	\label{extlatNr}
	\ee
	and carries the bilinear form
	\be
	\xbbm\ast\ybbm=\sum_{i=1}^n \(\kappa \Nr_i x_i y_i-\sum_{\alpha=1}^{d_{\Nr_i}} x_{i,\alpha} y_{i,\alpha}\),
	\label{bb-r}
	\ee
	where $\xbbm=\{x_i,x_{i,\alpha}\}$ with $i=1,\dots ,n$ and $\alpha=1,\dots,d_{r_i}$.
	Since the signature of \eqref{bb-r} is $(n-1,d_\bfr)$, $\bbLami{\bfr}$ has many null vectors.
	In the following we will use two sets of vectors belonging to $\bbLami{\bfr}$ with the second set consisting of null vectors.
	Both sets are extensions of the vectors $\bfv_{ij}\in \bfLami{\bfr}$,
	defined as in \eqref{defvij}
	\be 
	(\bfv_{ij})_k=\delta_{ki}\Nr_j-\delta_{kj}\Nr_i,
	\label{def-bfvij}
	\ee 
	and given by 
	\be 
	\begin{split}
		(\vbbm_{ij})_k=&\,(\bfv_{ij})_k,
		\qquad\ \
		(\vbbm_{ij})_{k,\alpha}= 0,
		\\
		(\wbbm_{ij})_k=&\, 2^\eps(\bfv_{ij})_k,
		\qquad
		(\wbbm_{ij})_{k,\alpha}= (\bfv_{ij})_k,
	\end{split}
	\label{def-bfcvij}
	\ee 
	where the factor of $2^\eps$ compensates the factor of $4^\eps$ appearing in $d_r$
	and ensures that $\wbbm_{ij}^2=0$. We will also use their normalized versions
	$\hvbbm_{ij}=\vbbm_{ij}/\Nr_{ij}$ and $\hwbbm_{ij}=\wbbm_{ij}/\Nr_{ij}$
	where $\Nr_{ij}=\gcd(\Nr_i,\Nr_j)$.
	
	Note also that the theta series appearing on the r.h.s. of \eqref{extmodan-manyz} depend on
	the following vector of refinement parameters
	\be 
	\zbbm= (\bftet^{(\bfr)} z;-\frt^{(\Nr_1)}z_1;\dots;-\frt^{(\Nr_n)}z_n),
	\qquad
	\bftet^{(\bfr)}=\sum_{i<j}\bfv_{ij}.
	\label{vecz}
	\ee		
    For all null vectors, $\zbbm\ast\wbbm_{ij}$ is proportional to $z$ and independent of $\bfz$.
	
	\item 
	Then one solves the refined system of anomaly equations on the extended lattice \eqref{extmodan-manyz}
	using the null vectors introduced in \eqref{def-bfcvij}.
	
	\item 
	After that, one reduces the solution to the original lattice using the relation \eqref{recover-gref}.
	
	\item 
	Finally, one evaluates the unrefined limit by means of 
	\be
	\gi{\bfr}_{\mu,\bfmu}(\tau)=
	\lim_{y\xrightarrow{}1} (y-y^{-1})^{1-n} \girf{\bfr}_{\mu,\bfmu}(\tau,z).
	\label{lim-ancoef}
	\ee  
\end{enumerate}
This procedure is schematically presented in Fig. \ref{fig-strategy}.

\subsubsection{Lattices, glue vectors and discriminant groups}
	
Before we present the result of the procedure outlined above, 
let us spell out a few properties of the relevant lattices.

First, let us consider a sublattice generated by the vectors \eqref{def-bfcvij}.
More precisely, we denote by $\bbLami{\bfr}_{||}$ a sublattice of $\bbLami{\bfr}$ given by
the span with integer coefficients of the normalized vectors $\hvbbm_{ij},\hwbbm_{ij}$, $1\leq i<j\leq n$.
It is easy to see that
\be 
\label{defLampar}
\bbLami{\bfr}_{||}=\bfLami{\bfr}\oplus \tbfLami{\bfr},
\ee
where $\bfLami{\bfr}=\Span\{\hvbbm_{ij}\}$, while $\tbfLami{\bfr}=\Span\{\hubbm_{ij}\}$ with $\hubbm_{ij}=\hwbbm_{ij}-2^\epsilon\hvbbm_{ij}$ is isomorphic to $\bfLami{\bfr}$ with quadratic form rescaled by $-4^\eps$.
The orthogonal completion of $\bbLami{\bfr}_{||}$ in $\bbLami{\bfr}$, denoted by $\bbLami{\bfr}_{\perp}$,
is a direct sum of $n+1$ lattices
\be 
\bbLami{\bfr}_{\perp}=\IZ\oplus \IA_{d_{\Nr_1}-1}\oplus \cdots\oplus \IA_{d_{\Nr_n}-1},
\label{factor-perp}
\ee 
where $\IA_{N}$ is the root lattice of the corresponding Lie algebra. 
The different factors in \eqref{factor-perp} are generated by the vectors 
\be
\begin{split}
	(\ebbm_0)_k=&\,0,
	\qquad
	(\ebbm_0)_{k,\beta}=1,
	\\
	(\ebbm_{i,\alpha})_k=&\,0,
	\qquad
	(\ebbm_{i,\alpha})_{k,\beta} =\delta_{ik}(\delta_{\alpha+1,\beta}-\delta_{\alpha\beta}).
\end{split}
\label{def-basise}
\ee
Importantly, the full lattice $\bbLami{\bfr}$ is {\it not} a direct sum of $\bbLami{\bfr}_{||}$
and $\bbLami{\bfr}_{\perp}$, but requires
the introduction of glue vectors (see \S\ref{ap-fact}). Namely, one has 
\be 
\bbLami{\bfr}=\bigcup\limits_{\Asf=0}^{n_g-1} \Bigl[\(\bbLami{\bfr}_{||}+\glueg^{||}_\Asf\)
\oplus \(\bbLami{\bfr}_{\perp}+\glueg^{\perp}_\Asf\)\Bigr],
\label{lat-glue-our}
\ee 
where the glue vectors are labeled by the set $\Asf=\{\asf_0,\asf_1,\dots,\asf_n\}$ with 
the indices taking values in the following ranges:
$\asf_0= 0,\dots,\Nr/\Nr_0-1$ and $\asf_i= 0,\dots,\di{i}-1$.
Explicitly, they can be chosen as 
\be
\glueg_\Asf = \asf_0\glueg_0 + \sum_{i=1}^{n} \glueg_{i,\asf_i},
\label{gluegA}
\ee 
where 
\be
\begin{split}
	(\glueg_{i,\asf})_k=&\,0,
	\qquad
	(\glueg_{i,\asf})_{k,\alpha} =\delta_{ik}\sum_{\beta=1}^\asf \delta_{\alpha\beta},
	\qquad
	\glueg_0=\sum_{i=1}^n\rho_i \glueg_{i,\di{i}},
\end{split}
\label{def-g}
\ee
and $\rho_i$ are fixed integers satisfying 
$\sum_{i=1}^n \rho_i\Nr_i=\Nr_0\equiv \gcd(\Nr_1,\dots,\Nr_n)$.

The factorization \eqref{lat-glue-our} plays an important role in solving the anomaly equations 
for the anomalous coefficients because it allows to disentangle the refinement parameters $z$ and $\bfz$: 
they appear only in the theta series defined by $\bbLami{\bfr}_{||}$ and $\bbLami{\bfr}_{\perp}$, respectively.
This in turn leads to two simplifications. First, the theta series based on $\bbLami{\bfr}_{\perp}$
decouples from the problem of ensuring the existence of zero of order $n-1$ at $z=0$ needed for the unrefined limit.
Second, due to the additional factorization property \eqref{factor-perp}, 
the refinement parameter $z_i$ appears only in the theta series defined by the corresponding $\IA_{N}$ lattice.
As a result, applying \eqref{recover-gref} to recover $\girf{\bfr}_{\mu,\bfmu}(\tau,z)$, 
each differential operator acts on one theta series only.

Next, we note that there are two ways to represent the elements of 
the discriminant group $\bfD^{(\bfr)}=(\bfLami{\bfr})^*/\bfLami{\bfr}$.
On one hand, they can be parametrized by the $(n+1)$-tuple $(\mu,\bfmu)$ taking values in 
$\IZ_{\kappa r} \otimes \prod_{i=1}^n \IZ_{\kappa\Nr_i}$, satisfying the condition $\Delta\mu\in \kappa r_0\IZ$
and subject to the identification $(\mu,\bfmu)\simeq (\mu+r/r_0,\bfmu+\bfr/r_0)$.
This is the same tuple that labels the components of the anomalous coefficients.
Moreover, the condition on $\Delta\mu$ already appeared in our equations. For example, 
in \eqref{sol-n=2} it is imposed by the Kronecker delta and, in general,  
it is a direct consequence of \eqref{constr-epsi}. 
On the other hand, the elements of $\bfD^{(\bfr)}$ can be represented by a rational $n$-dimensional vector $\bfhmu$
with components satisfying $\sum_i r_i\hmu_i=0$.
In terms of the previous parametrization, they are given by
\be 
\hmu_i=\frac{\mu_i}{\kappa\Nr_i}-\frac{\mu}{\kappa\Nr}+\frac{\rho_i\Delta\mu}{\kappa\Nr_0} \, .
\label{def-hmu}
\ee 
where $\rho_i$ are as in \eqref{def-g}.
The discriminant group of $\tbfLami{\bfr}$ has exactly the same description with $\kappa$ replaced by $4^\eps\kappa$.
We will denote $\bfhtmu$ the vector \eqref{def-hmu} defined by $(\tmu,\bftmu)$ after this replacement.
It is clear that the residue classes of $\bbLami{\bfr}_{||}$ can then be seen either as 
tuples $(\mu,\bfmu;\tmu,\bftmu)$ or as bi-vectors $(\bfhmu,\bfhtmu)$.
In particular, the residue class given by the projection of the glue vector \eqref{gluegA} on $\bbLami{\bfr}_{||}$ 
is represented by the tuple with
\be 
\mu=\bfmu=0,
\qquad
\tmu(\Asf)=4^\eps \kappa\Nr_0\asf_0 +\sum_{i=1}^n\asf_i,
\qquad 
\tmu_i(\Asf)=\asf_i.
\label{tildemus}
\ee

\subsubsection{The result}

Let us introduce the following objects:
\begin{itemize}
	\item 
	a vector valued Jacobi-like form\footnote{For $n=1$, it is a trivial scalar function $\phi^{(r)}=1$.}
	\be  
	\phi^{(\bfr)}_{(\bfhmu,\bfhtmu)}(\tau,z)=\frac{\Sym\{ \cbfr \}}{ z^{n-1}}\,e^{-\frac{\pi^2}{3}\,m_{\bfr} E_2(\tau)z^2}
	\prod_{i=1}^n \delta^{(1)}_{\hmu_i-2^\eps \htmu_i}\, ,
	\label{solfullphi}
	\ee
	where $(\bfhmu,\bfhtmu)\in \ID^{(\bfr)}_{||}$ is a residue class of $\bbLami{\bfr}_{||}$ in the bi-vector representation, 
	$m_{\bfr}$ is the index \eqref{index-mr}, and 
	\be  
	\cbfr
	=\frac{\Nr_0}{(2^{\eps}\pi\I \kappa)^{n-1}\Nr}
	\prod\limits_{k=1}^{n-1}\(\sum\limits_{i=1}^k\Nr_i\sum\limits_{j=n-k+1}^n\Nr_j\)^{-1}\, ;
	\label{constphi}
	\ee 
	
	\item 
	a function $\Fvi{\bfr}$ playing the role of a kernel of indefinite theta series on $\bbLami{\bfr}_{||}$
	\be
		\Fvi{\bfr}(\xbbm)=
		\sum_{\cJ\subseteq\Zv_{n-1}} e_{|\cJ|} \delta_\cJ
		\prod_{\ell \in \Zv_{n-1} \backslash \cJ}
		\Bigl(\sign (\xbbm_\bbbeta\ast\vbbm_\ell)-\sign ( \xbbm\ast\wbbm_{\ell,\ell+1}) \Bigr),
    \label{kern-manyz}
	\ee
	where $\xbbm_\bbbeta=\xbbm+\sqrt{2\tau_2}\,\bbbeta$,
	$\Zv_{n}=\{1,\dots,n\}$, 
	\be
	\label{def-genthm}
	e_m=\left\{\begin{array}{ll}
		0 & \mbox{\rm if $m$ is odd},
		\\
		\frac{1}{m+1}\ & \mbox{\rm if $m$ is even},
	\end{array}\right.
	\qquad
	\delta_\cJ=\prod_{\ell\in\cJ}\delta_{\xbbm_\bbbeta\ast\vbbm_\ell},
	\quad\mbox{\rm and}\quad
	\vbbm_\ell=\sum_{i=1}^{\ell}\sum_{j=\ell+1}^{n} \vbbm_{ij} ;
	\ee 
	
	\item 
	a function associated with the lattice $\bbLami{\bfr}_{||}$ 
	\be
	\vthls{\bfr}_{(\bfhmu,\bfhtmu)}(\tau, z)
	=
	\sum_{m=1}^{n} \sum_{\sum_{k=1}^m n_k = n} 
	\sum_{\bfnu,\bftnu} \vth_{\mu,\bfnu;\tmu,\bftnu} (\tau, \zbbm_{||}^{(\bfs)};\bbLami{\bfs}_{||}, \Fvi{\bfs},0) 
	\prod_{k=1}^{m} \phi^{(\frr_k)}_{\nu_k,\frm_k;\tnu_k,\tfrm_k}(\tau,z),
	\label{thet-phi-thm}
	\ee
	where $\bfs$, $\frr_k$ and $\frm_k$ are as in \eqref{split-rs} restricted to the one-modulus case,
	$\tfrm_k$ is defined from $\bftmu$ similarly to $\frm_k$, 
	$\zbbm_{||}^{(\bfs)}=(\bftet^{(\bfs)},0)$ denotes the projection of $\zbbm$ \eqref{vecz} to $\bbLami{\bfs}_{||}$ 
	which can be seen as a sublattice of $\bbLami{\bfr}_{||}$,
	$\vth_\bbmu$ is the generic theta series \eqref{gentheta},
	and on the r.h.s. we used the representation of $\ID^{(\bfr)}_{||}$ in terms of $(2n+2)$-tuples;

	\item 
	the theta series associated with the lattices appearing in the decomposition \eqref{factor-perp}
	\bea
	\label{3thetaZ}
	\vth^{(d)}_{\nu}(\tau) &=& 
	\sum_{\ell\in \IZ +\frac{\nu}{d}+\hf} 
	(-1)^{d\ell} \, \q^{\frac{d}{2}\,\ell^2} ,
	\\
	\vthA{N}_\asf(\tau,z;\frt) &=&
	\(\prod_{\alpha=1}^{N-1}\sum_{\ell_\alpha\in \IZ +\frac{\alpha\asf}{N}} \)
	\q^{\sum\limits_{\alpha=1}^{N-1}\( \ell_{\alpha}^2-\ell_\alpha\ell_{\alpha+1} \)}
	y^{\sum\limits_{\alpha=1}^{N-1 } \(\frt_{\alpha+1}-\frt_\alpha \)\ell_{\alpha}} ,
	\label{3thetas}
	\eea
	where we used the convention $\ell_N=0$,
	and the result of the action on $\vthA{N}_\asf$ of the differential operator in \eqref{recover-gref}
	\be
	\cD\vthA{N}_{\asf}(\tau;\frt) =
	\frac{\cD^{(N)}_{\frt^2/2}\vthA{N}_{\asf}(\tau,z;\frt)\bigr|_{z_i=0}}
	{N!\(\prod_{\alpha=1}^{N}\frt_\alpha\)\(-2\pi\eta^3(\tau)\)^{N}}\,.
	\label{def-cDvth}
	\ee
	
\end{itemize}
In terms of these quantities, a solution for the refined anomalous coefficients was found to be
\be
\girf{\bfr}_{\mu,\bfmu}(\tau,z) =  \frac{\delta^{(\kappa\Nr_0)}_{\Delta\mu} }{2^{n-1}}\, \sum_{\Asf}
\Sym\bigl\{\vthls{\bfr}_{(\bfhmu,\bfhtmu(\Asf))}(\tau, z) \bigr\}
\vth^{(d_\Nr)}_{\tmu(\Asf)}(\tau) 
\prod_{i=1}^{n} \cD\vthA{d_{\Nr_i}}_{\asf_i}(\tau;\frt^{(\Nr_i)}),
\label{refgr}
\ee
where $\bfhtmu(\Asf)$ denotes the residue class corresponding to
the tuple $(\tmu(\Asf),\bftmu(\Asf))$ defined in \eqref{tildemus}.

\begin{figure}
	\isPreprints{\centering}{} 
	\includegraphics[width=10cm]{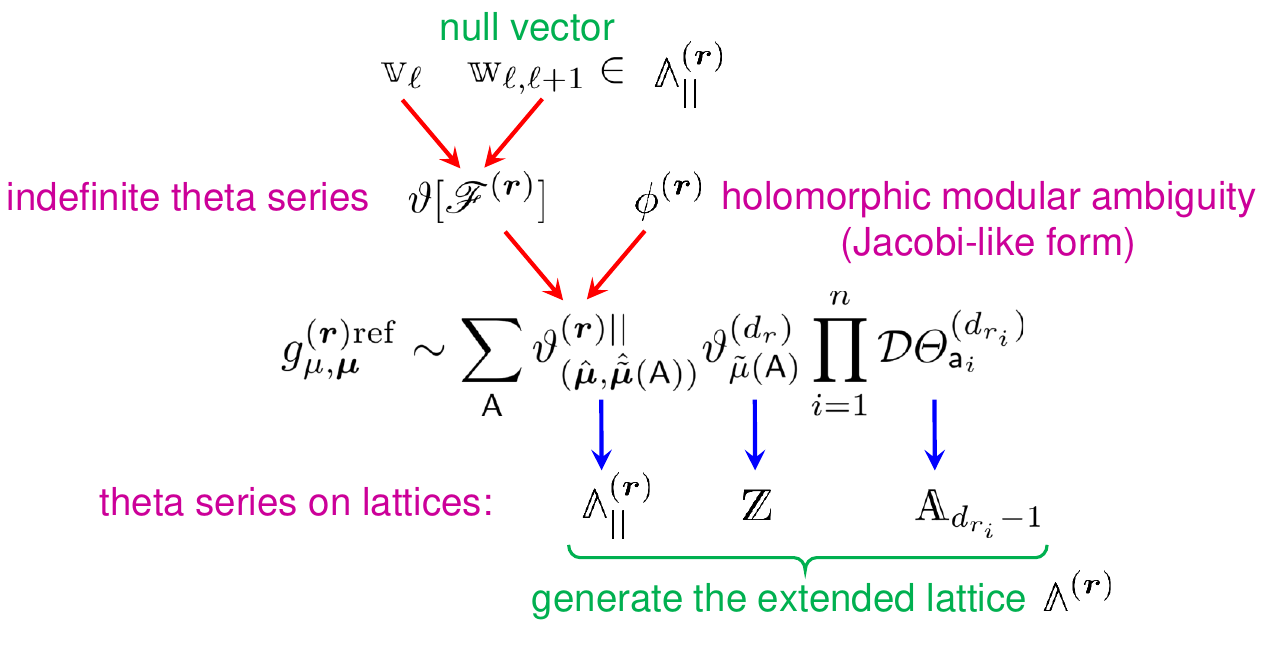}
	\vspace{-0.2cm}
	\caption{A schematic representation of the solution \eqref{refgr} for the refined anomalous coefficients
		and its ingredients.\label{fig-solution}}
\end{figure}

The most non-trivial ingredient of this construction is the function $\vthls{\bfr}_{(\bfhmu,\bfhtmu)}$.
It recombines indefinite theta series defined by the kernels $\Fvi{\bfr}(\xbbm)$ and the Jacobi-like forms  
$\phi^{(\bfr)}_{(\bfhmu,\bfhtmu)}$ in such a way that it has a zero of order $n-1$ at $z=0$.\footnote{Although this fact
was not proven in all generality, it has been extensively checked both analytically and numerically.}
As a result, $\girf{\bfr}_{\mu,\bfmu}$ have a well-defined unrefined limit that should be computed 
using \eqref{lim-ancoef} to extract the anomalous coefficients $\gi{\bfr}_{\mu,\bfmu}$ we are interested in. 
Unfortunately, this turns out to be the most non-trivial step and 
so far it has been accomplished analytically only for $n=2$ and 3.
Explicit expressions for $\gi{\Nr_1,\Nr_2}_{\mu, \mu_1, \mu_2}$ and $\gi{\Nr_1,\Nr_2,\Nr_3}_{\mu, \mu_1, \mu_2,\mu_3}$
for generic charges as well as their Fourier series for a few small charges and some choice of the vectors $\frt^{(\Nr)}$
can be found in \cite{Alexandrov:2024wla}.
In Fig. \ref{fig-solution}, we showed schematically the main ingredients of the solution \eqref{refgr}.

\section{Applications}
\label{sec-application}

In this section we present three very different applications of the results described in the previous sections.
The first is an explicit evaluation of various topological invariants of compact CY threefolds,
the second is a solution of Vafa-Witten theory on rational surfaces, 
and the third is a surprising extension of the above formalism 
to string compactifications preserving more than 8 supercharges which allows not only to reproduce many known results,
but also to obtain something new.

\subsection{DT, PT, GV and topological strings}
\label{subsec-topinv}

\subsubsection{Polar terms from wall-crossing}
\label{subsubsec-wall}

The results explained in \S\ref{sec-solution} reduce the problem of finding the generating series of rank 0 DT invariants
to the problem of computing their polar terms. First attempts to do this have been undertaken already quite some time ago
for several one-modulus CYs and unit D4-brane charge, in which case the anomaly is absent and the two-step procedure
advocated in the beginning of the previous section reduces to the second step.
In \cite{Gaiotto:2006wm,Gaiotto:2007cd} the polar terms were calculated by a direct count of D-branes in the geometry and 
in \cite{Collinucci:2008ht,VanHerck:2009ww} through a representation via
the attractor flow trees \cite{Denef:2001xn,Denef:2007vg} in terms of bound states of D6 and $\aD$-branes.
However, at this point it is not clear how to generalize these results to other cases 
by applying the same methods. Therefore, to make progress in the computation of polar terms,
one should look for alternative approaches.

An interesting possibility, which often works in the non-compact case, is to find a chamber
in the moduli space where the BPS spectrum is simple enough to be computed exactly and then perform wall-crossing 
to the large volume chamber. Unfortunately, for compact CY threefolds there is no such chamber.
Nevertheless, an interesting phenomenon happens if one goes off the physical slice in the space of 
stability conditions defining the generalized DT invariants as explained in \S\ref{subsec-DT}.

\begin{figure}
	\isPreprints{\centering}{} 
	\includegraphics[width=12cm]{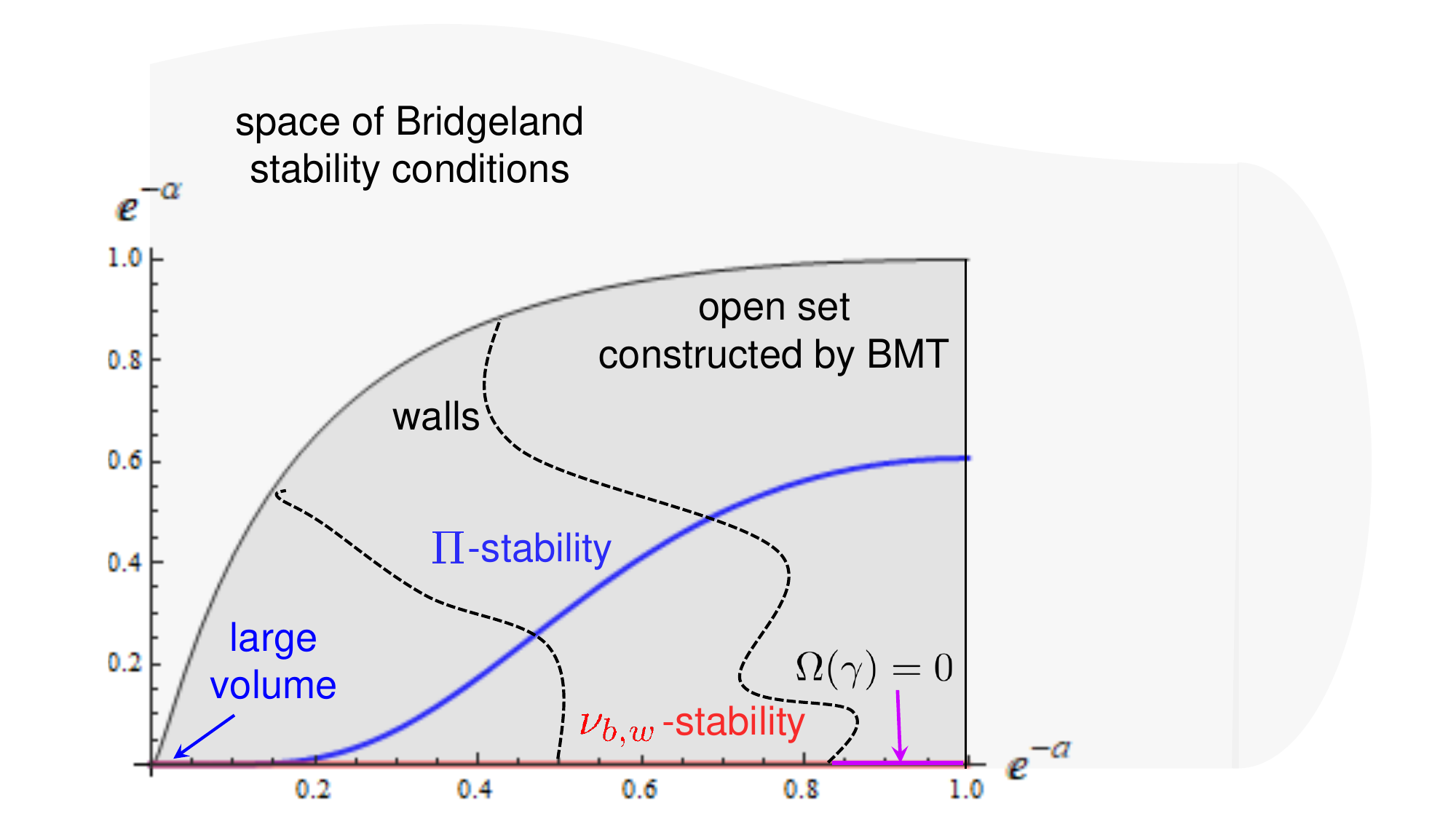}
	\vspace{-0.0cm}
	\caption{Section of the space of Bridgeland stability conditions by the plane $\beta=0$, $b=\mbox{const}$, 
		drawn in coordinates $(e^{-a},e^{-\alpha})$ and the set 
		constructed in \cite{bayer2011bridgeland,bayer2016space}. 
		The red line is the slice of weak $\nu_{b,w}$-stability conditions with the central charge \eqref{Ztilt} and 
		the blue line represents the physical slice of $\Pi$-stability conditions parametrized  by 
		the complexified K\"ahler moduli of $\CY$. The large volume limit corresponds to the region near the origin 
		where the two slices approach each other.\label{fig-stabBr}}
\end{figure}

Let us again restrict to CY threefolds with one K\"ahler modulus.
In this case the space of stability conditions, modulo the action of a symmetry group,
can be parametrized by four real variables $(a,b,\alpha,\beta)$
and an open set of such stability conditions has been rigorously constructed in \cite{bayer2011bridgeland,bayer2016space}.
The subset corresponding to the physical $\Pi$-stability is a real-codimension two slice of this open set parametrized
by the complexified K\"ahler modulus $z=b+\I t$. 
Its explicit parametrization is determined by the prepotential $F(X)$ and can be found
in \cite[Eq.(2.51)]{Alexandrov:2023zjb}.
On the other hand, the boundary of the open set corresponding to $\alpha=\infty$ and $\beta=0$ 
defines another set of (weak) stability conditions playing a crucial role in our story (see Fig. \ref{fig-stabBr}).
It is called $\nu_{b,w}$-stability\footnote{Here $w$ refers the parameter $w=\hf(a^2+b^2)$ which
simplifies the wall-crossing analysis.} 
and defined by the central charge
\be
\label{Ztilt}
Z_{b,a}(\gamma) =-a \ch_2^b   +  \frac12\, a^3 \ch_0^b  + \I \, a^2\ch_1^b \, ,
\ee
which we wrote in terms of shifted Chern classes that can be compactly written as
\be
\ch^b_k (E)=\int_{\CY} e^{-b \omega} \omega^{3-k}\ch(E),
\ee 
where $\omega$ is the generator of $H^2(\CY,\IZ)$.
The central charge \eqref{Ztilt} can be obtained from the physical central charge 
by dropping all quantum corrections to the classical prepotential \eqref{Fcl},
omitting the contributions proportional to the D0-brane charge and, after substituting \eqref{ch-Ch}, 
to the second Chern class $c_2(T\CY)$, and finally setting $t=a\sqrt3$. 
Importantly, in the large volume region, the two notions of stability coincide 
which means that the DT invariants defined by them are also the same.

The advantage of the $\nu_{b,w}$-stability conditions is twofold. 
First, all stable objects with respect to the $\nu_{b,w}$-stability were conjectured to satisfy
the following {\it BMT inequality} \cite{bayer2011bridgeland,bayer2016space}
\be
\label{BMTorig}
\ch_3^b \leq \frac{a^2}{6}\, \ch_1^b. 
\ee
It implies that for a given charge there can be a region in the moduli space at small $a$ 
where the corresponding DT invariant must vanish. This suggests that the same DT invariant 
evaluated in the large volume chamber can be calculated by starting from the chamber where 
it vanishes and then performing wall-crossing towards the large volume.
The second advantage of the $\nu_{b,w}$-stability is its independence of the D0-brane charge which
significantly simplifies the wall-crossing analysis and makes the proposed strategy doable. 
This idea has been pursued in a recent series of 
works~\cite{Toda:2011aa,Feyzbakhsh:2020wvm,Feyzbakhsh:2021rcv,Feyzbakhsh:2021nds},
and led to explicit formulas \cite{Feyzbakhsh:2022ydn}
relating rank 0 DT invariants, counting D4-D2-D0 bound states,
to rank $\pm 1$ invariants, counting D6-D2-D0 bound states with one unit of D6 or $\aD$-brane charge.
These formulas have been further improved in \cite{Alexandrov:2023zjb,Alexandrov:2023ltz}
and made possible to formulate a systematic procedure for computing the polar and other terms of the generating series 
$h_{p,\mu}$.

There are actually two types of formulas that can be used to compute rank 0 DT invariants.
The first one is in the spirit of \cite{Collinucci:2008ht,VanHerck:2009ww} and expresses
rank 0 DT invariants as a sum of products of rank 1 and rank $-1$ DT invariants, given by the standard DT and PT invariants.
Physically, this precisely corresponds to a sum over D6-$\aD$ bounds states.  
The formula reads
\be
\bOm_{r,\mu}(\hq_0) =\tors^2  \sum_{r_i,Q_i,n_i} 
(-1)^{\gamma_{12}}\,\gamma_{12} \,
\PT(Q_1,n_1) \, \DT(Q_2,n_2)\,,
\label{conseqTh1}
\ee
where $\DT(Q,n)$ and $\PT(Q,n)$ denote the standard invariants
which depend just on two charges due to the spectral flow invariance, 
$\gamma_{12}=r(Q_1+Q_2)+n_1+n_2-\chi(\cO_{\cD_r})$ with $\chi(\cO_{\cD_r})$ defined in \eqref{defL0}, 
and $\tors$ is the order of the torsion part of the second cohomology group 
which in general has the form $H^2(\CY,\IZ)=\IZ^{b_2}\oplus\IZ_{\tors}$.
Up to the torsion factor, each term in the sum can be recognized as a contribution of  
the primitive wall-crossing formula \eqref{prim-wc}. 
The factor $\tors$ appears because the rank $\pm 1$ generalized DT invariants, corresponding to
the BPS indices on the r.h.s. of \eqref{prim-wc}, coincide with $\DT(Q,n)$ and $\PT(Q,n)$ 
only up to this factor \cite{Toda:2011aa}.
Its presence has recently been confirmed in \cite{McGovern:2024kno}.
To fully specify the formula, one also needs to provide the range of summation for all six variables in \eqref{conseqTh1}.
We refer to \cite[\S4.1]{Alexandrov:2023zjb} for these details. 
Finally, the derivation of the formula implies that it is expected to be valid only in the range
\be
\label{condFeyz}
0 \leq \frac{\chi(\cD_r)}{24} -\hq_0 < \frac{\kappa r}{12} \min\(\frac{r^2}{2}-\frac18\, ,\, r-\hf\),
\ee
where $\chi(\cD_r)$ is the topological Euler characteristic of the divisor $\cD_r$ 
\be
\chi(\cD_r) = \kappa r^3+c_{2}r.
\label{defchiD}
\ee
The first inequality in \eqref{condFeyz} is just the Bogomolov-Gieseker bound \eqref{qmax}, while 
the second is the condition of the existence of the chamber violating the BMT inequality \eqref{BMTorig}, 
which we can call {\it empty chamber} due to the absence there of stable objects of the given charge.

Unfortunately, the condition \eqref{condFeyz} is so restrictive that for $r=1$ the formula \eqref{conseqTh1} 
can only apply, at best, to the most polar term in each component of the modular vector $h_{1,\mu}$. 
In particular, for $\mu=0$ it is valid only for 
$\hq_0=\frac{\chi(\cD_r)}{24}$. In this case only the term with $Q_i=n_i=0$ contributes to \eqref{conseqTh1}, which gives
\be
\label{eqFeyzus}
\bOm_{r,0}\left( \frac{\chi(\cD_r)}{24} \right) = \tors^2 (-1)^{1+\chi(\cO_{\cD_r})} \chi(\cO_{\cD_r})\,.
\ee
In practice, however, it was observed that the formula \eqref{conseqTh1} predicts the correct polar terms in many examples with $r=1$, 
provided one restricts the sum only to $Q_1=n_1=0$. Using $\PT(0,0)=1$, one arrives at the naive ansatz 
for polar coefficients suggested in \cite[(5.20)]{Alexandrov:2022pgd}:
\be
\label{naive}
\bOm_{r,\mu}(\hq_0) = \tors^2 (-1)^{r \mu+n + \chi(\cO_{\cD_r})} (r \mu + n-\chi(\cO_{\cD_r}))\, \DT(\mu,n)\,,
\ee
where
\be
\label{nfromhq0}
n = \frac{\chi(\cD_r)}{24} -\frac{\mu^2}{2\kappa r} - \frac{r \mu}{2} - \hq_0\in\IZ \,.
\ee
The fact that it holds in many cases indicates that it should be possible to extend the range of validity of 
the formula \eqref{conseqTh1}. On the other hand, it remains still unclear what determines when it works and when it fails.

The second formula instead expresses a PT invariant $\PT(Q,m)$ in terms of invariants 
$\PT(Q',m')$ with $Q'<Q$ and rank 0 DT invariants $\Omega_{r,\mu}(\hq_0)$.\footnote{This formula was derived assuming 
	the triviality of the torsion, i.e. $\tors=1$.}
In its simplest version it reads
\be
\label{thmS11}
\PT(Q,m) =   \sum_{Q', m'} (-1)^{\gamma_{12}}
\gamma_{12} \, \PT(Q',m') \, \Omega_{1,Q-Q'}(\hq'_0)\,,
\ee 
where $\gamma_{12}=Q+Q'+m -m'-\chi(\cO_{\cD_1})$,
\be 
\hq'_0 = m'-m -\frac{1}{2\kappa}\(Q'-Q \)^2-\hf\, (Q+Q')+\frac{\chi(\cD)}{24}\,,
\label{hq0-Qm}
\ee 
and we refer to \cite[\S4.2]{Alexandrov:2023zjb} for the exact range of summation.
This formula holds provided $f_1(x)<\alpha$ where 
\be
\label{defxa}
x=\frac{Q}{\kappa}\,, 
\qquad 
\alpha=-\frac{3m}{2Q}\,,
\ee
and $f_1(x)$ is a piece-wise linear function given by $\hf x+1$ for $x\geq 3$.	
Similarly to \eqref{condFeyz}, this condition ensures the existence of the empty chamber,
but in addition it also guarantees that on the way from this chamber to the large volume region
only the walls corresponding to bound states with a single D4-brane are encountered.

An important property of \eqref{thmS11} is that the term with $(Q',m')=(0,0)$ 
for which $\PT(0,0)=1$ always contributes to the sum.
This fact allows to invert the relation and express $\Omega_{1,Q}(\hq_0)$ with $\hq_0$ given by \eqref{hq0-Qm}
through other invariants. In practice, however, this does not allow yet to get the rank 0 DT invariants for 
the charges of interest. Such charges typically spoil the condition $f_1(x)<\alpha$.
Fortunately, one can always use the spectral flow invariance \eqref{spectr-flow} to make D2-brane charge large enough 
so that the condition becomes satisfied. As a result, {\it any} rank 0 DT invariant with $r=1$ can be expressed
through PT invariants and other rank 0 DT invariants with smaller charges \cite[Eq.(4.19)]{Alexandrov:2023zjb}.
Importantly, however, the smaller D0-brane charge $q_0$ we want to consider, the larger the degree $Q$
for which PT invariants have to be calculated.

If one relaxes the condition $f_1(x)<\alpha$ in a way that still ensures the existence of the empty chamber, 
the formula \eqref{thmS11} gets modified by acquiring terms on the r.h.s. that involve bound states 
with the D4-brane charge $r>1$. One can show that $r$ is the maximal appearing D4-brane charge 
if $f_r(x)<\alpha$ where $f_r(x)=f_{r-1}(x)$ for $x\leq (r+1)^2$ and $f_r(x)=\frac{x}{r+1}+\frac{r+1}{2}$ in the range $x>(r+1)^2$.
The modified version of the formula has been computed so far only for $r=2$ \cite{Alexandrov:2023ltz}.
For $x>4$ and $\alpha\leq \frac38 x +\frac32$, it has the same feature as \eqref{thmS11} that the r.h.s. 
contains a term proportional to $\bOm_{2,Q}(\hq_0)$. Thus, using again the spectral flow invariance, 
it can be used to express the rank 0 DT invariants with $r=2$ through PT invariants 
and other rank 0 DT invariants with smaller charges.

\subsubsection{The role of topological strings}
\label{subsubsec-method}

Above we showed how rank 0 DT invariants can be expressed through rank 1 and $-1$ invariants.
How does this help find them? The point is that, for any CY threefold, there is a systematic procedure to compute
rank $\pm1$ invariants (which however has its own limitations to be discussed below).
This is done using the so called MNOP relation~\cite{gw-dt,gw-dt2,Pandharipande:2011jz}  
which allows to express $\DT(Q,m)$ and $\PT(Q,m)$
in terms of the Gopakumar-Vafa (GV) invariants $\GV^{(g)}_{Q'}$ with $Q'\leq Q$ and $g\leq g_{\rm max}(Q)$.
The latter are integer valued and count embedded curves $\cC$ of genus $g$ and degree $Q$, 
while from the physical viewpoint they count BPS particles in the 5d theory obtained by compactifying M-theory on $\CY$
\cite{Gopakumar:1998ii,Gopakumar:1998jq}. A crucial fact is that the GV invariants of genus $g$
determine the A-model topological string free energy $F^{(g)}$ of the same genus and hence can be deduced
from a solution of the topological string. For genus $g=0$, the free energy $F^{(0)}$ can be found by mirror symmetry
techniques, while for $g\geq 1$ the free energies are obtained by integrating the holomorphic anomaly equations
that they satisfy \cite{Bershadsky:1993ta} following the procedure known as ``direct integration"
\cite{Huang:2006hq,Grimm:2007tm}.

Thus, one proceeds through the following steps:
\begin{enumerate} 
	\item
	First, one solves the A-model topological string by the direct integration method and obtains its partition function
	\be
	\label{PsitopF}
	\Psi_{\rm top}(z,\lambda) = \exp\left( \sum_{g\geq 0} \lambda^{2g-2} F^{(g)}(z) \right).
	\ee
	
	\item 
	It is used to extract the GV invariants by applying the formula \cite{Gopakumar:1998ii,Gopakumar:1998jq}
	\be
	\label{eqn:gvex}
	\log \Psi_{\rm top}(z,\lambda)=\sum\limits_{g=0}^\infty\sum\limits_{k=1}^\infty
	\sum\limits_{Q=1}^\infty\frac{\GVg{g}}{k}\left(2\sin\frac{k\lambda}{2}\right)^{2g-2}e^{2\pi \I k Q z}\,.
	\ee
	
	\item 
	In principle, the previous step could be skipped because the MNOP formula directly relates the topological
	string partition function with the generating functions of DT and PT invariants
	\be
	\label{defZDT}
	\begin{split}
		Z_{DT} (y,\q):=&\, \sum_{Q=0}^\infty \sum_{m=m_{\rm min}(Q)}^\infty \DT(Q,m)\,
		y^Q\,  \q^{m},
		\\
		Z_{PT} (y,\q):=&\, \sum_{Q=0}^\infty \sum_{m=m_{\rm min}(Q)}^\infty\PT(Q,m)\,
		y^Q\,  \q^{m},
	\end{split}
	\ee
	which are well defined because the D0-brane charge $m$ is restricted from below by the Castelnuovo bound 
	\cite{Liu:2022agh,Alexandrov:2023zjb}
	\be
	\label{CastPT}
	m\geq m_{\rm min}(Q)=- \left\lfloor \frac{Q^2}{2\kappa} + \frac{Q}{2}\right\rfloor.
	\ee
	The formula reads 
	\be
	\label{PsitopPT}
	\Psi_{\rm top}(z,\lambda) =  M(-e^{\I\lambda})^{\hf\,\chi_{\scriptstyle{\CY}}}
	Z_{PT} \left(e^{2\pi\I z/\lambda},e^{\I\lambda} \right) , 
	\ee
	where $M(\q)=\prod_{k>0}(1-\q^k)^{-k}$ is the Mac-Mahon function and $\chi_\CY$ is the Euler characteristic of $\CY$.
	A similar formula for the generating function of DT invariants follows from 
	the DT/PT relation conjectured in \cite{pandharipande2009curve} and proven in \cite{toda2010curve,bridgeland2011hall}, 
	which has the following simple form
\be
Z_{DT} (y,\q)= M(-\q)^{\chi_{\scriptstyle\CY}} \, Z_{PT} (y,\q).
\label{eqn:DTPTrelation}
\ee

In practice, however, one computes the PT and DT invariants always by passing through the GV invariants 
and evaluating the generating series degree by degree.
Then, it is more convenient to use the plethystic form of the MNOP relation \cite{Pandharipande:2011jz}
\be
\label{PTGVpleth}
Z_{PT}(y,\q)= \PE\[
\sum_{Q>0} \sum_{g=0}^{\gmax(Q)} (-1)^{g+1} \GVg{g}
\(1-x\)^{2g-2} x^{(1-g)} y^Q\](-\q,y),
\ee
where $\PE$ denotes the plethystic exponential 
\be
\PE[f](x,y)=\exp\(\sum_{k=1}^\infty\frac{1}{k}\, f(x^k,y^k)\).
\ee
Note that the Castelnuovo bound \eqref{CastPT} combined with the MNOP formula 
implies a similar bound on the genus of GV invariants
\be
\label{defgmax}
g\leq g_{\rm max}(Q) =  \left\lfloor \frac{Q^2}{2\kappa} + \frac{Q}{2}\right\rfloor + 1.
\ee

\item 
At the final step, one applies the formulas from \S\ref{subsubsec-wall} to compute the rank 0 DT invariants
appearing in the generating series $h_{r,\mu}$.
Note that {\it a priory} this approach is {\it not} restricted to the polar terms 
and can be applied to compute any rank 0 DT invariant.
\end{enumerate}

Unfortunately, the described procedure has a fundamental limitation.
The problem is that, to determine the topological string free energies by the direct integration method, one should 
supplement the holomorphic anomaly equations with some boundary conditions fixing the holomorphic ambiguity.
Currently, the known conditions include the Castelnuovo bound \eqref{defgmax}, the conifold gap constraints, 
and the value of GV invariants for $(Q,g)=(n\kappa,1+\hf n(n+1) \kappa)$ with $n\in\IN$ which saturate  
the Castelnuovo bound. However, the number of these conditions grows slower with genus 
than the number of parameters to be fixed in the holomorphic ambiguity.
As a result, without further input, the direct integration method works only up to a certain genus $g_{\rm integ}$,
and hence the PT and DT invariants can be computed only up to a certain finite degree $Q_{\rm integ}$.
In turn, this imposes limitations on the number of terms in the generating series that can be computed by this method. 
This is why the computation of polar terms remains a challenging problem for most compact CY threefolds.

\subsubsection{Results}

There are two classes of compact CY threefolds that have been analyzed so far by the method explained 
in \ref{subsubsec-method}. In both cases the generating functions of rank 0 DT invariants have been expressed through 
an over-complete basis of vector valued weakly holomorphic modular forms constructed from 
unary theta series with quadratic form $\kappa r$,
Dedekind eta function $\eta(\tau)$, Serre derivative $D$ acting as $\q\partial_{\q}-\frac{w}{12}E_2$ 
on modular forms of weight $w$,
and Eisenstein series $E_4(\tau)$ and $E_6(\tau)$. The theta series and the Dedekind eta function
allow to obtain the required multiplier system \eqref{Multsys-hp} and a number of polar terms,
while the Serre derivative and the Eisenstein series are needed to get the weight $-3/2$.

\paragraph{\it Hypergeometric threefolds}

The first set of CYs is given by one-parameter smooth complete intersections in weighted projective space,
which are known as the hypergeometric CY threefolds and include the famous quintic manifold. 
There are 13 of such CYs, but for 2 of them the current knowledge of GV invariants is insufficient to find 
the polar terms by the above method even for $r=1$. For the remaining 11 CYs listed in  
Table \ref{tab_cydata}, for $r=1$, all polar terms and hence the generating series $h_{1,\mu}$ 
have been found in \cite{Alexandrov:2023zjb}.
Furthermore, for 2 CYs, $X_8$ and $X_{10}$, in \cite{Alexandrov:2023ltz} 
the same has been done for $r=2$ where the generating series become mock modular forms
and one should use one of the solutions of the modular anomaly equation presented in \S\ref{sec-solution}.

It should be emphasized that, for most of the analyzed CYs, together with the polar terms, 
many non-polar ones have been computed and perfectly matched the coefficients obtained by 
the Fourier expansion of the modular forms uniquely fixed by the polar terms.
This provided a striking test of (mock) modularity as well as of various mathematical conjectures, 
such as the BMT inequality \eqref{BMTorig}, which underlie the analysis. 
Actually, even the fact that the resulting generating series (for $r=2$, after applying the inverse of \eqref{def-bOm})
produce integer valued invariants is a highly non-trivial check of their correctness. 

It is interesting that for most of the polar coefficients, and even for some non-polar ones, 
the correct value turns out to be given by the naive ansatz \eqref{naive}, which {\it a priori} has no reason to hold.
It fails only for 8 out of the 72 calculated polar terms. This fact suggests that there should be a way to correct the ansatz, 
which would open a possibility to compute the polar terms for higher D4-brane charges $r$ or other CY threefolds
because it requires much less data than the approach based on PT invariants and wall-crossing relations of type 
\eqref{thmS11} used in these calculations. However, it remains unclear which bound state contributions 
could account for the discrepancy between the ansatz and the correct values even in the eight mentioned cases.

\begin{table}
	$$
	\begin{array}{|l|c|c|c|c|c|c|c|c|}
		\hline
		\CY & \chi_{\CY} &\ \kappa\ &\ c_2\ & g_{\rm integ} & g_{\rm mod}^{(r)} 
		&\ g_{\rm avail} & Q_{\rm integ} & Q_{\rm avail} \\
		\hline
		X_5(1^5) & -200 & 5 & 50 & 53 & 69 & 64 & 22 & 26 \\
		X_6(1^4,2) & -204 & 3 & 42 & 48 & 66 & 48 & 15 & 17 \\
		X_8(1^4,4) & -296 & 2 & 44 & 60 & 84(112) & 64 & 15 & 17  \\
		X_{10}(1^3,2,5) & -288 & 1 & 34 & 50 & 70(95) & 71 & 11 & 13 \\
		X_{3,3}(1^6) & -144 & 9 & 54 & 29 & 33 & 33 & 20 & 21 \\
		X_{4,2}(1^6) & -176 & 8 & 56 & 50 & 64 & 64 & 28 & 31 \\
		X_{4,3}(1^5,2) & -156 & 6 & 48 & 20 & 24 & 24 & 14 & 15  \\
		X_{6,2}(1^5,3) & -256 & 4 & 52 & 63 & 78 & 49 & 17 & 20  \\
		X_{4,4}(1^4,2^2) & -144 & 4 & 40 & 26 & 34 & 34 & 14 & 16 \\
		X_{6,4}(1^3,2^2,3) & -156 & 2 & 32 & 14 & 17 & 17 & 7 & 8  \\
		X_{6,6}(1^2,2^2,3^2) & -120 & 1 & 22 & 18 & 21 & 24 & 6 & 7  \\
		\hline
	\end{array}
	$$
	\caption{Relevant data for the 11 hypergeometric CY threefolds for which $h_{1,\mu}$ has been found. 
		In the first column we use the notation $X_{d_1,\dots,d_k}(w_1^{m_1},\dots,w_p^{m_p})$
		to denote a complete intersection of multidegree $(d_1,\dots,d_k)$ in weighted projective space 
		$\IP^{k+3}(w_1,\dots,w_p)$ where $m_i$ is the number of repetitions of the weight $w_i$.
		The second to fourth columns indicate the Euler number of $\CY$, the self-intersection number $\kappa$,
		and the second Chern class $c_2$. 
		The column $g_{\rm integ}$ gives the maximal genus 
		for which GV invariants $\GVg{g}$ can be determined by the direct integration method 
		using only the usual boundary conditions. The column $g_{\rm mod}^{(r)}$ shows how this bound changes
		after adding information about the GV invariants predicted by the knowledge of $h_{r,\mu}$ for $r=1$ and 
		in brackets, for the two manifolds where these generating series are available, for $r=2$.
		The column $g_{\rm avail}$ indicates the genus up to which 
		complete tables of GV invariants are currently known. Finally, the columns $Q_{\rm integ}$ and $Q_{\rm avail}$
		provide the maximal degrees for DT and PT invariants attainable through the direct integration and
		available now due to the additional information about the rank 0 DT invariants.\label{tab_cydata}
		\vspace{-0.5cm}}
\end{table}

\paragraph{\it Quotients}

The second set of CYs consists of various quotients and has been studied in \cite{McGovern:2024kno}.
It includes 4 one-parameter threefolds ($X_5/\IZ_5$, $X_{3,3}/\IZ_3$, $(\mbox{Pfaffian in }\IP^6)/\IZ_7$
and a smooth double cover of determinantal quintic in $\IP^4$ quotient by $\IZ_5$)
and 5 two-parameter models given in the CICY notation \cite{Candelas:1987kf} by
\be 
\begin{split}
	\begin{matrix}
		\IP^2 \\ \IP^2
	\end{matrix}
	\[\begin{matrix} 3 \\ 3 \end{matrix}\]/\IZ_3, &
	\qquad
	\begin{matrix}
		\IP^2 \\ \IP^5
	\end{matrix}
	\[\begin{matrix} 1 & 1 & 1 & 0 \\ 1 & 1 & 1 & 3 \end{matrix}\]/\IZ_3,
	\qquad
	\begin{matrix}
		\IP^4 \\ \IP^4
	\end{matrix}
	\[\begin{matrix} 1 & 1 & 1 & 1 & 1 \\ 1 & 1 & 1 & 1 & 1 \end{matrix}\]/\IZ_5,
	\\
	&
	\begin{matrix}
		\IP^2 \\ \IP^2 \\ \IP^2
	\end{matrix}
	\[\begin{matrix} 1 & 1 & 1 \\ 1 & 1 & 1  \\ 1 & 1 & 1 \end{matrix}\]/\IZ_3,
	\qquad 
	\begin{matrix}
		\IP^2 \\ \IP^2 \\ \IP^5
	\end{matrix}
	\[\begin{matrix} 1 & 1 & 1 & 0 & 0 & 0 \\ 0 & 0 & 0 & 1 & 1 & 1 \\ 1 & 1 & 1 & 1 & 1 & 1 \end{matrix}\]/\IZ_3.
\end{split}
\ee 
We refer to \cite[Table 1]{McGovern:2024kno} for their topological data.
Because of the quotients, the intersection numbers and the second Chern classes of these manifolds 
are sufficiently small so that, for $r=1$, the generating series of rank 0 DT invariants 
have only a single polar term, except $X_{3,3}/\IZ_3$ which has two such terms.
Therefore, the whole generating series is determined just by one coefficient, which can found by using \eqref{eqFeyzus}
and requires the knowledge of GV invariants only at small genera. The case of $X_{3,3}/\IZ_3$ is a bit more complicated,
but can be treated using a combination of \eqref{eqFeyzus} and \eqref{thmS11}.

Note that all the quotients are non-simply connected manifolds and have a non-trivial torsion in the second cohomology group.
Thus, the results of \cite{McGovern:2024kno}  have allowed for the first time to test the torsion factor in 
\eqref{eqFeyzus}.

\subsubsection{Implications for topological strings}

The knowledge of the generating functions $h_{r,\mu}$ gives access to infinitely many rank 0 DT invariants.
This data can be used in the r.h.s. of the wall-crossing relations \eqref{thmS11} 
to compute PT invariants that were unknown before. They in turn can feed the MNOP relation 
to get new GV invariants. In other words, if a generating series $h_{r,\mu}$ has been successfully found,
one can invert the procedure of \S\ref{subsubsec-method} and calculate new sets of topological invariants.

\begin{figure}
	\isPreprints{\centering}{} 
	\includegraphics[width=12cm]{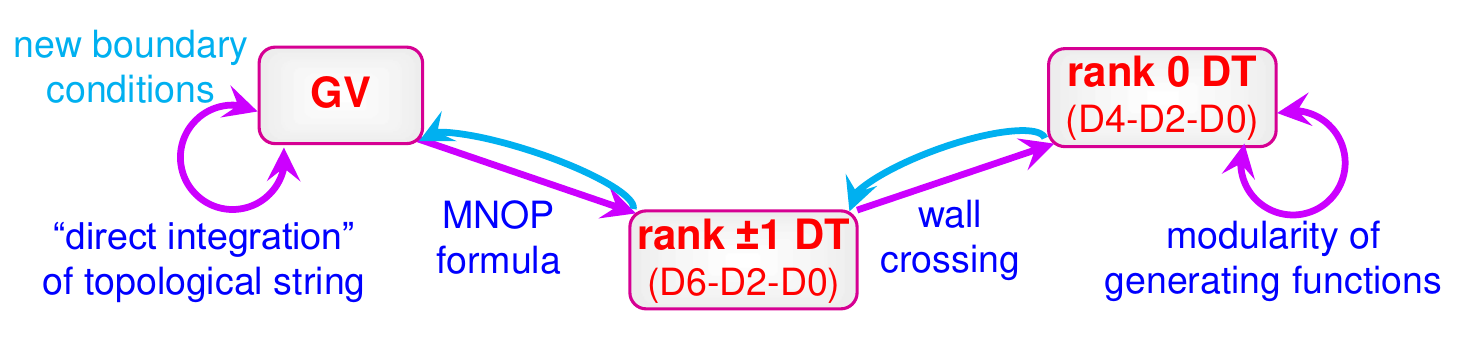}
	\vspace{-0.3cm}
	\caption{The procedure to obtain the rank 0 DT invariants through the direct integration of topological strings,
		MNOP formula and wall-crossing relations. The inverse arrows show that, once the generating functions 
		of the rank 0 DT invariants are found, the procedure can be inverted to get new boundary conditions
		for fixing the holomorphic ambiguity in the direct integration method.
		One can run this loop multiple times. \label{fig-method}}
\end{figure}

Importantly, the GV invariants obtained in this way can serve as new boundary conditions
to be used in the direct integration method for fixing the holomorphic ambiguity of the topological string. 
Thus, one can overcome the limitation of this method explained at the end of \S\ref{subsubsec-method} 
and go beyond the genus $g_{\rm integ}$. 
We summarize the whole procedure in Fig. \ref{fig-method}.

Unfortunately, if one knows only a finite number of the generating series $h_{r,\mu}$, 
one cannot go infinitely far, but only up to a certain new bound $g_{\rm mod}^{(r)}$
where $r$ is the maximal D4-brane charge for which $h_{r,\mu}$ is known.
In Table \ref{tab_cydata}, we provide $g_{\rm integ}$, $g_{\rm mod}^{(1)}$ (and $g_{\rm mod}^{(2)}$ for $X_8$ and $X_{10}$)
as well as the genus up to which the GV invariants have been currently calculated for the  11 hypergeometric CY threefolds 
analyzed in \cite{Alexandrov:2023zjb,Alexandrov:2023ltz}. We also give there the bound $Q_{\rm integ}$ 
on the degree of PT invariants and the degree $Q_{\rm avail}$ that has already been achieved.
The full tables of known GV, PT and DT invariants are available at the website
\cite{KlemmCYdata}.

Of course, once one obtained new boundary conditions for the direct integration, one can repeat the procedure
shown in Fig. \ref{fig-method} and one may hope that this should allow us to compute more polar terms 
and find more generating functions. In principle, one may think that, if one leaves aside computational problems
related to computer speed and memory, it might be possible to overcome all limitations and compute the GV invariants 
up to arbitrary genus. Unfortunately, this is not the case as the number of polar terms 
and hence the required genus grow with $r$ faster 
than the number of genera one can gain due to the new boundary conditions \cite{Alexandrov:2023ltz}.
Thus, if one wants to push the idea of the combined use of the holomorphic anomaly of topological string theory 
and the modular anomaly of the generating series of rank 0 DT invariants, one has to find more powerful
wall-crossing relations that would not be so demanding for the GV invariants.

\subsection{Vafa-Witten theory}
\label{subsec-VW}

Vafa-Witten theory is a topological field theory defined on any 4-manifold $S$, 
obtained as one of the three possible topological twists of $\cN=4$ SYM theory \cite{Vafa:1994tf}. 
We restrict to the case of the gauge group $U(r)$ and $S$ a complex Fano or weak Fano surface, 
which is equipped with a polarization $J\in H^2(S,\IR)$ such that $J \cdot c_1(S) > 0$.
Due to this condition and certain vanishing theorems, the functional integral localizes 
on solutions of hermitian Yang-Mills equations\footnote{In general, the partition function 
receives also contributions from what is known as the monopole branch. 
See \cite{Labastida:1999ij,gottsche2017virtual,Laarakker:2018isn,Thomas:2018lvm,Gottsche:2018meg,Gottsche:2021dye} 
for a progress in this direction.}
and the partition function is completely determined by the topological invariants, 
called VW invariants, given by the Euler numbers or, in the refined case, 
Poincar\'e polynomials similar to \eqref{defOmref} of the moduli spaces of instantons on the surface $S$.
Since the refined invariants contain more information and, as we saw in \S\ref{subsec-refine}, 
their description is actually simpler, in the following we will mostly concentrate on the refined case but omit
the index ``ref" to avoid cluttering.

S-duality of $\cN=4$ SYM implies that the partition function of VW theory should be a modular form.
Naively, it appears to be a Jacobi form with a theta expansion of the form \eqref{thetadecomp}
where the role of $h_\mu$ is played by the generating series of VW invariants, which we will denote $h^{S}_{r,\mu}$.
This would imply that these generating series are modular forms. 
However, when $b_2^+(S)=1$, the case of our interest, this expectation turns out to be naive 
because the generating series have a modular anomaly. In fact, they turn out to be examples of 
mock modular forms of depth $r-1$ or mock Jacobi forms in the presence of refinement.  
For instance, already in \cite{Vafa:1994tf}, on the basis of the previous mathematical results \cite{Klyachko:1991,Yoshioka:1994},
it was shown that the generating series of (unrefined) $SU(2)$ VW invariants on $\IP^2$ is given by the generating series 
of the Hurwitz class numbers (Ex. \ref{ex-Hurwitz}), which was one of the earliest examples of mock modularity in physics.
The modularity of the partition function is then supposed to be restored by taking into account non-holomorphic contributions 
from reducible connections \cite{Vafa:1994tf}. 
For the simplest case of $SU(2)$ theory on $\IP^2$, this restoration of modularity was demonstrated
in \cite{Dabholkar:2020fde}, where the required non-holomorphic contributions have been shown to
be generated by Q-exact terms due to boundaries of the moduli space, similarly to the holomorphic anomaly in
topological string theory \cite{Bershadsky:1993ta}.
Mathematically, this is nothing but the construction of the completion $\whh^{S}_{r,\mu}$ 
of the mock modular form $h^{S}_{r,\mu}$.
Thus, finding the modular completion is an important physical problem in the context of VW theory.

Until a few years ago, only very limited results existed about the modular completions of the generating series of VW invariants, 
not going beyond $r=2$ \cite{Vafa:1994tf,Bringmann:2010sd,Manschot:2011dj} and $r=3$ for $\IP^2$ \cite{Manschot:2017xcr}.
A breakthrough came from the introduction of the generalized error functions and the results presented
in \S\ref{sec-anomaly} and \S\ref{sec-ext}. 
The point is that when $\CY$ is the non-compact CY given by the total space ${\rm Tot}(K_S)$ of the canonical bundle 
over a projective surface $S$ with $b_1(S)=0$ and $b_2^+(S)=1$, as in Ex. \ref{ex-elliptic}, 
the BPS indices of $r$ D4-branes supported on $S$ are expected
to be equal to the VW invariants of $S$ for gauge group $U(r)$
\cite{Minahan:1998vr,Alim:2010cf,Gholampour:2013jfa,gholampour2017localized},
both at the unrefined and refined levels.
Physically, this expectation follows from the fact that the topologically twisted $\cN=4$ $U(r)$ SYM 
describes the world-volume dynamics of $r$ M5-branes wrapped on $S$ and dimensionally reduced 
along $S^1$ times the Euclidean time circle.
The large volume attractor chamber of $\CY$ then corresponds to the canonical chamber in 
the moduli space parametrized by $J$. It is defined as the chamber containing the point $J=c_1(S)$.
This is important to take into account 
since the VW invariants for $b_2^+(S)=1$ and $b_2(S)>1$ depend on the choice of polarization. 
Therefore, one can identify the generating series of (refined) VW invariants evaluated in the canonical chamber
with the generating series of (refined) D4-D2-D0 BPS indices or, more precisely,
with their redefined versions $\tlh_{r,\mu}$ \eqref{def-tgi} (or $\thr_{r,\mu}$ in the refined case).\footnote{One can wonder
why the redefined version of the generating functions introduced in the one-modulus case can be applied here.
The reason is that the only thing that is important for its definition is that the allowed D4-brane charges are all collinear, 
whereas the lattice of D2-brane charges can have any dimension. This does hold in our case where all D4-charges 
are multiplies of $p_0^a$ \eqref{vecvI} corresponding to the divisor $S$. To account for multiple dimensions
of the electric charge lattice in \eqref{def-tgi}, one should replace $\mu$ by $\mu_a p_0^a$ in the sign factor
and multiply the shift of $\mu$ in the index of the generating function by $p_0$.}
As a result, the expressions for the completions \eqref{exp-whh} and \eqref{exp-whhr} can be directly translated 
to the context of VW theory.
In \cite{Alexandrov:2019rth}, it was checked that this does reproduce the known completions in the case of $S=\IP^2$.

However, one can do better and use the modular anomaly equations to actually find the generating series of VW invariants,
similarly to how this problem was addressed in the compact case in \S\ref{sec-solution} and \S\ref{subsec-topinv}.
Before presenting the results in this direction, let us briefly review what was known before.
For $r=1$, the generating function has been known for any $S$ for a long time \cite{Gottsche:1990} and, when
$b_1(S)=0$, it is given in terms of the Jacobi theta and Dedekind eta functions
\be
\label{h10anySref}
h^S_{1,0}(\tau,z) = \frac{\I}{\theta_1(\tau,2z)\, \eta(\tau)^{b_2(S)-1}}.
\ee
For $r>1$, many explicit expressions already existed in the literature
\cite{Yoshioka:1994,Yoshioka:1995,Manschot:2010nc,Manschot:2011dj,Manschot:2011ym,Klemm:2012sx}.
Furthermore, for $\IP^2$ a closed formula has been found in \cite{Manschot:2014cca} 
in terms of generalized Appell-Lerch sums and, in principle, 
for any other Fano surface the generating functions
$h^S_{r,\mu}$ could be determined by applying a sequence of blow-ups and wall-crossing transitions to this result
(see \cite[Fig.1]{Beaujard:2020sgs} for a scheme of the blow-up relations between various Fano surfaces).
However, this procedure is complicated by the fact that one should pass through stack invariants, 
which are polynomial combinations of rational VW invariants having simpler transformation properties under wall-crossing.
Another general method to compute the VW invariants based on a relation to quivers 
and the tree index introduced in \cite{Alexandrov:2018iao} was proposed in \cite{Beaujard:2020sgs},
where many generating functions were given explicitly.
However, an explicit formula for the generating functions of arbitrary rank $r$ for $S$ other than $\IP^2$ remained unknown.

\subsubsection{Hirzebruch and del Pezzo} 
\label{subsubsec-HdP} 
 
This problem has been addressed in \cite{Alexandrov:2020bwg}
for the Hirzebruch surfaces $\IF_m$ with $0\leq m\leq 2$
and the del Pezzo surfaces $\IB_m$ with $1\leq m \leq 8$
by solving the modular anomaly equations for the generating functions of refined VW invariants 
following from \eqref{exp-whhr}.
The solution is very similar to the one described in \S\ref{subsec-gensol-ac} but significantly simpler
due to several reasons:
\begin{itemize}
	\item 
	Since we work with refined invariants from the very beginning, there is no need to introduce the refinement
	artificially and take the unrefined limit in the end. Besides, the generating functions must be mock Jacobi forms
	and not just Jacobi-like as in \S\ref{subsec-gensol-ac}. This is because here we are computing 
	the generating functions of Poincar\'e polynomials depending on the refinement parameter $z$ only through $y=e^{2\pi\I z}$,
	whereas there the refinement was just a trick to compute some auxiliary functions.
	\item 
	There is no need to do a lattice extension because $\Lambda_S=H^2(S,\IZ)$ is a unimodular lattice
	of signature $(1,b_2(S)-1)$ and for the Hirzebruch and del Pezzo surfaces $b_2(S)>1$. 
	In all relevant cases, $\Lambda_S$ has several null vectors (at least two),
	but only one of them appears in the construction of indefinite theta series.
	\item
	Finally, despite one can solve \eqref{exp-whhr}, as any anomaly equation, only up to a holomorphic modular ambiguity,
	the ambiguity is severely constrained by the requirements to be a Jacobi form with given modular properties and 
	to ensure the existence of the unrefined limit. As a result, after comparing with known results in the literature, 
	it is possible to suggest a universal ansatz for this ambiguity so that there is no need to fix it through 
	the computation of any polar terms.
\end{itemize}

To present the resulting generating functions, let us introduce several ingredients:
\begin{itemize} 
\item 
From Ex. \ref{ex-elliptic}, we know that the relevant magnetic charges are all collinear, $p^a=rp_0^a$,
with $p_0^a$ determined by the first Chern class of the surface \eqref{vecvI}.
The relevant lattice of electric charges is $\Lambda_r=r\Lambda_S$ with the quadratic form $rC_{\alpha\beta}$
where $C_{\alpha\beta}=\cD_\alpha \cap\cD_\beta$ is the intersection matrix on $S$ (see \eqref{kappaEll}), specified for 
Hirzebruch and del Pezzo surfaces in \S\ref{ap-surface}. This motivates to introduce the reduced charge vector
$\hgam=(r,q_\alpha)$ where the electric charge can be decomposed as (cf. \eqref{defmu-shift})
\be
\label{quant-q}
q_\alpha =  r \, C_{\alpha\beta} \epsilon^{\beta} + \mu_\alpha- \frac{r}{2}\, C_{\alpha\beta} c_1^\beta,
\qquad
\eps^\alpha\in \IZ.
\ee
One can also show that the quadratic form \eqref{defQlr} takes the form
\be
Q_n(\{\hgam_i\})= \frac{1}{r}\, q^2-\sum_{i=1}^n \frac{1}{r_i} q_{i}^2
=-\sum_{i<j}\frac{(r_i q_j - r_j q_i)^2}{rr_i r_j}\, ,
\label{defQlr-VW}
\ee
where $q^2=C^{\alpha\beta}q_\alpha q_\beta$ and $C^{\alpha\beta}$ is the inverse of $C_{\alpha\beta}$, and
for the charges satisfying $\sum_i q_i=q$ with $q$ fixed, its signature is $((n-1)(b_2(S)-1),n-1)$.

\item
We define the anti-symmetrized Dirac product of charges depending on a vector $v\in \Lambda_S$:
\be
\label{gam12}
\gamma_{ij}(v)=
v^{\alpha} (r_i q_{j,\alpha} -r_j q_{i,\alpha}).
\ee
Note that $\gamma_{ij}(c_1)$ coincides with the usual Dirac product of the reduced charge vectors $(r_ip_0^a,q_{i,a})$
relevant for the non-compact CY underlying this construction.

\item 	
For each surface $S$, we pick up a specific null vector $\nv\in\Lambda_S$.
For $S=\IF_m$ and $\IB_m$, in the basis described in \S\ref{ap-surface}, it is given by
\be
\nv(\IF_m)=[f],
\qquad
\nv(\IB_m)=\cD_1-\cD_2.
\label{nullv}
\ee

\item 
We define the theta series
\be
\vthA{r}_{\ell}(\tau,z)
= \sum_{\sum_{i=1}^r k_i=0 \atop k_i\in\IZ+{\ell}/{r}}
\q^{- \sum_{i<j} k_i k_j}\, y^{\sum_{i<j}(k_j-k_i)},
\label{defthN}
\ee
which transforms as a vector valued Jacobi form of weight $\hf(r-1)$ and index $\frac16 (r^3-r)$.
One can show that it is a specification of the $A_{r-1}$ theta series  $\vthA{r}_\ell(\tau,z;\frt)$ \eqref{3thetas}
for $\frt_\alpha=r+1-2\alpha$.
Combined with a power of the Dedekind eta function, it produces the blow-up functions 
\be
\label{defBNk}
B_{r,\ell}(\tau,z) = \frac{\vthA{r}_{\ell}(\tau,z)}{\eta(\tau)^r}\, ,
\ee
which relate the generating functions of stack invariants on manifolds related by the blow-up of an exceptional divisor
\cite{Yoshioka:1996,0961.14022,Li:1998nv}.
In turn, the generating functions of stack invariants evaluated at $J=\nv$ \eqref{nullv}
are given by \cite{Manschot:2011ym,Mozgovoy:2013zqx}
\be
H^S_{r,\mu}=\delta^{(r)}_{\nv\cdot\mu} \, H_r \prod_{\alpha=3}^{b_2(S)}B_{r,\mu^\alpha},
\qquad
H_r=\frac{\I (-1)^{r-1} \eta(\tau)^{2r-3}}
{\theta_1(\tau,2rz)\, \prod_{m=1}^{r-1} \theta_1(\tau,2mz)^2}\, ,
\label{gfHN}
\ee
where in the last factor, which is relevant only for del Pezzo surfaces, $\mu^\alpha$
are the components of the residue class $\mu$, i.e. $\mu=\mu^\alpha \cD_\alpha$, in the basis defined in \S\ref{ap-surface}. 
The functions $H^S_{r,\mu}$ are vector valued
Jacobi forms of weight $-\hf b_2(S)$ and index $-(\frac16(r^2-1)c_1^2+2)r$.
Note that these are the same weight and index that $h^{S}_{r,\mu}$ are expected to have \cite[Eq.(4.16)]{Alexandrov:2019rth}
and they agree with the values given in \eqref{index-Hr0} provided one takes $\lambda_ap^a=r$, which can be achieved, for example,
by choosing $\lambda_a$ to be the normalized null eigenvector of the quadratic form corresponding to the divisor $\cD_e$
in the notations of Ex. \ref{ex-elliptic}.
\end{itemize}

Let us now combine all the above ingredients into the functions very similar to \eqref{thet-phi-thm}:
\be
\Theta^S_{r,\mu}(\tau,z;\{\Fv_n\})=\sum_{n=1}^r\frac{1}{2^{n-1}}
\sum_{\sum_{i=1}^n \hgam_i=\hgam}
\Fv_n(\{\hgam_i\})
\, \q^{\hf Q_n(\{\hgam_i\})}
\, y^{\sum_{i<j}\gamma_{ij}(c_1(S))}\prod_{i=1}^n H^S_{r_i,\mu_i}(\tau,z),
\label{combTh}
\ee
where the sum goes over all decompositions of the reduced charge $\hgam=(r,\mu- \frac{r}{2}c_1)$,
i.e. with the spectral flow parameter set to zero, the charges
$q_{i}$ are quantized as in \eqref{quant-q} with $r$ replaced by $r_i$, and $\Fv_n$ is a set of functions playing the role 
of kernels of indefinite theta series on $\(\oplus_{i=1}^n\Lambda_{r_i}\)/\Lambda_r$.
In terms of the functions \eqref{combTh}, the generating functions of the refined VW invariants evaluated in the canonical chamber
can be written simply as 
\be
h^S_{r,\mu}=\Theta^S_{r,\mu}\Bigl(\tau,z;\{\Fv_n(c_1(S))\}\Bigr).
\label{whgN}
\ee
The kernels $\Fv_n$ defining these functions depend on a lattice vector 
and are given by
\be
\Fv_n(\{\hgam_i\};v)
=\sum_{\cJ\subseteq \Zv_{n-1}}e_{|\cJ|}\,\delta_\cJ(v)
\prod_{k\in \Zv_{n-1}\setminus \cJ}\Bigl(\sgn(\Gamma_k(v))-\sgn(\Bv_k)\Bigr),
\label{kerg}
\ee
where most of the notations are the same as in \eqref{kern-manyz} except
\be
\begin{split}
	\delta_\cJ(v)=&\,\prod_{k\in \cJ}\delta_{\Gamma_k(v)},
	\qquad
	\Gamma_{k}(v)=\sum_{i=1}^k\sum_{j=k+1}^n \gamma_{ij}(v),
	\\
	\Bv_k=&\, \gamma_{k,k+1}(\nv)+\beta r_k r_{k+1}(r_k+r_{k+1})\, \nv\cdot c_1(S),
\end{split}
\label{defbk}
\ee
and, as usual, $\beta$ encodes the imaginary part of the refinement parameter through $z=\alpha-\tau\beta$.
Note that the kernels $\Fv_n$ have exactly the same structure as the kernels \eqref{kern-manyz}.
It should also be clear that the Jacobi forms $H^S_{r,\mu}$ play the same role as the Jacobi-like forms $\phi^{(\bfr)}_{(\bfhmu,\bfhtmu)}$
in \eqref{thet-phi-thm}. In particular, they ensure the existence of the unrefined limit 
canceling all poles of the indefinite theta series, except the first order pole inherent to the generating functions
of refined invariants. They can be seen as holomorphic modular ambiguities of the solution of the modular anomaly equations,
which are fixed by consistency and matching the known results.

Furthermore, in \cite{Alexandrov:2020dyy} it was shown that the result \eqref{whgN} has a very simple generalization to 
an arbitrary chamber provided it lies in the projection of the K\"ahler cone on the two-dimensional 
plane in the moduli space spanned by the first Chern class and the null vector \eqref{nullv}, 
i.e. $J\in\mbox{\rm Span}(c_1(S),\nv(S))^+$. In this case, it is enough to replace the first Chern class 
in the argument of $\Fv_n$ by polarization $J$. 
Thus, if we denote $h^S_{r,\mu,J}$ to be the generating functions of the refined VW invariants evaluated at $J$, 
one obtains
\be
h^S_{r,\mu,J}=\Theta^S_{r,\mu}\Bigl(\tau,z;\{\Fv_n(J)\}\Bigr).
\label{genJhN}
\ee

Once the generating functions are explicitly known and expressed through indefinite theta series,
it is immediate to find explicit expressions for their modular completions.
Applying the recipe \eqref{replace-alln}, one obtains \cite{Alexandrov:2020dyy}
\be
\whh^{\,S}_{r,\mu,J}(\tau,z)=\Theta^S_{r,\mu}\Bigl(\tau,z;\{\whFv_n(J)\}\Bigr),
\label{complFBJ}
\ee
where the kernels are expressed through the generalized error functions $\Phi_n^E$ \eqref{generrPhiME} as
\be
\whFv_n(\{\hgam_i\};J)=\sum_{\cJ\subseteq \Zv_{n-1}} 
\Phi_{|\cJ|}^E\(\{ \bfv_{\ell}(J)\}_{\ell\in \cJ};\sqrt{2\tau_2}\,(\bfq+\beta\bftet )\)
\prod_{k\in \Zv_{n-1}\setminus \cJ}\Bigl(-\sgn(\Bv_k)\Bigr).
\label{kerhg}
\ee
Here the vectors $\bfq$, $\bftet$ and $\bfv_\ell(J)$ are given by
\be
\bfv_\ell= \sum_{i=1}^\ell\sum_{j=\ell+1}^n\bfv_{ij},
\qquad
\bftet = \sum_{i<j} \bfv_{ij},
\label{def-bfvk}
\ee
Explicit expressions for these vectors can be either read off from \eqref{defbk} 
or found in \cite[\S D]{Alexandrov:2020dyy}.

Note that one of the crucial ingredients of the presented construction is the choice of the null vector $\nv(S)$ \eqref{nullv}.
With mild modifications, a similar construction can be carried out for other choices as well. 
For some of them, the resulting generating functions turn out to be the same as \eqref{genJhN}.
This coincidence has been interpreted in \cite{Alexandrov:2020bwg} as a manifestation of the fiber-base duality 
\cite{Katz:1997eq,Mitev:2014jza}, with the prototypical example given by $\IF_0=\IP^1\times\IP^1$
where the second null vector is $\nv'(\IF_0)=[s]$, which is geometrically indistinguishable from $\nv(\IF_0)=[f]$. 
In more complicated examples involving del Pezzo surfaces, it can be traced back to the Weyl
reflection symmetry of the lattice $\Lambda_{\IB_m}$ \cite{Iqbal:2001ye}.
In all cases, it can be used to generate non-trivial identities between the generalized Appell-Lerch 
functions, which provide an alternative way to express the generating functions.

However, in most cases the other choices of the null vector lead to different functions,
but satisfying the same modular anomaly equations. Of course, the Fourier coefficients of the new functions
are {\it not} VW invariants, but the fact that they possess the same modular properties begs for an explanation.
In particular, one can ask whether they can be interpreted as a new kind of topological invariants of $S$.
One should keep in mind, however, that the coefficients of the generating series are rational numbers,
and whereas in the case of VW invariants they {\it must} produce integer numbers by inverting the formula \eqref{def-bOm},
this is not the case for the new numbers, which might be a serious obstacle in attempts to find their 
mathematical interpretation.

\subsubsection{$\IP^2$} 
\label{subsubsec-P2}

It is natural to ask whether the construction presented above can also be applied to the simplest surface $S=\IP^2$.
In this case $b_2(S)=1$ and therefore we encounter exactly the same problem that we had in \S\ref{subsec-gensol-ac}:
a one-dimensional lattice does not have null vectors.
But we also know a solution to this problem --- lattice extension. 
In the case of VW theory, such a lattice extension can be done in a way that has a geometric interpretation 
as the blow-up of a point into an exceptional divisor. It increases $b_2(S)$ by 1, thereby increasing 
the dimension of the lattice $\Lambda_S$. In particular, the blow-up of $\IP^2$ gives $\IF_1=\IB_1$.
After the blow-up, the lattice has an indefinite signature and has null vectors so that it is amenable
to the above construction.

This procedure has been carried out in \cite{Alexandrov:2020dyy} where it was shown that it leads 
to a version of the blow-up formula \cite{Yoshioka:1996,0961.14022,Li:1998nv}, which relates the generating functions
on two surfaces, $S$ and $\chS$, where the second is the blow-up of the first.
Denoting the exceptional divisor appearing due to the blow-up by $\cD_e$ and 
the obvious lattice embedding by $\iota:\Lambda_S\hookrightarrow \Lambda_{\chS}$,
one finds the following relation
\be
h^S_{r,\mu,J}(\tau,z)=\frac{h^{\chS}_{r,\iota(\mu)+\ell \cD_e,\iota(J)}(\tau,z)}{B_{r,\ell}(\tau,z)}\,,
\label{relhch}
\ee 
where $B_{r,\ell}$ are the blow-up functions \eqref{defBNk}.
An analogous relation can be written for the modular completions as well.
Applying the formula \eqref{relhch} to the case $S=\IP^2$, one obtains 
\be
h^{\IP^2}_{r,\mu}(\tau,z)=
\frac{\Theta^{\IF_1}_{r,\mu \cD_1+\ell \cD_2}\Bigl(\tau,z;\{\Fv_n(\cD_1)\}\Bigr)}{B_{r,\ell}(\tau, z)}\, ,
\label{hP2}
\ee
where on the r.h.s. we used the basis \eqref{baseF1}.

Note that the index $\ell$ on the r.h.s. of the above equations is arbitrary. 
Therefore, the functions $h^{\chS}_{r,\cmu,\iota(J)}$
must satisfy integrability conditions
ensuring the independence of the ratios \eqref{relhch} on this index,
which can be viewed as consistency conditions of the construction.\footnote{In \S\ref{subsec-gensol-ac},
such integrability conditions were avoided by ensuring that the discriminant group of the extended lattice 
$\bbLami{\bfr}$ is equal to the discriminant group of the original lattice $\bfLami{\bfr}$ for any set of charges $\bfr$.
This is why $\bbLami{\bfr}$ had to be much bigger than $\bfLami{\bfr}$.
In our case, this holds only for $r=1$, whereas for generic $r$ the discriminant group after the blow-up
gets an additional factor $\IZ_r$.}
For $S=\IP^2$, they are known to follow at $r=2$ from the periodicity property of the classical Appell function
and at $r=3$ from its generalization proven in \cite{Bringmann:2015}.
For higher ranks they remain still unexplored.
 
In fact, it is possible to rewrite \eqref{relhch} in a ``covariant" form that makes 
obvious the modular properties of the resulting generating functions.
We will write it for their modular completions, but a similar formula holds for the generating functions themselves.
It reads 
\be
\whh^{\,S}_{r,\mu,J}(\tau,z)=  \frac{\eta(\tau)^r}{\prod_{j=1}^r \theta_1(\tau,(2j-1) z)}
	\sum_{\ell=0}^{r-1}\theta_{r,\ell}(\tau,r z)\,\whh^{\,\chS}_{r,\iota(\mu)+\ell \cD_e,\iota(J)}(\tau,z),
\label{compl-blowup}
\ee
where
\be
\theta_{r,\ell}(\tau,z)=
\sum_{k\in r\IZ+\ell+\hf r} \q^{\frac{1}{2r}\, k^2}\, (-y)^k.
\ee
Its specification for $S=\IP^2$ can be obtained as in \eqref{hP2}.

We finish this discussion by noticing that the use of the blow-up relations to obtain the generating functions of VW invariants
is certainly not new. This is precisely how a representation in terms of the generalized Appell-Lerch sums
was derived in \cite{Manschot:2014cca}.
What is new here is that these relations are applied directly to the generating functions.
As was mentioned below \eqref{h10anySref}, usually, one should pass through stack invariants instead.
The main reason for this detour is that the blow-up relations have to be applied on walls of marginal stability where
the VW invariants, in contrast to the stack invariants, are not defined.
For example, in \eqref{hP2}, the polarization $J=\cD_1$ for $\IF_1$ is a wall of marginal stability for
(some of) the VW invariants with $\cmu=\cmu^2 \cD_2$.
A miraculous property of the representations \eqref{genJhN} and \eqref{complFBJ}
is that they continue to be well-defined on the walls!
This happens because the modular completions are actually smooth across the walls and provide
an unambiguous definition of the holomorphic generating functions everywhere in the moduli space, including the walls,
which is realized by a prescription defining the kernels \eqref{kerg} even when some of the Dirac products $\Gamma_k(v)$ vanish.
Of course, this does not mean yet that the VW invariants are defined on the walls.
For example, one can check that the rational invariants extracted from $h^{\IF_1}_{3,0,\cD_1}$
do {\it not} lead to {\it integer} invariants after application of the inverse of the formula \eqref{def-bOm}.
Nevertheless, they correctly reproduce the VW invariants on $\IP^2$ via \eqref{hP2} \cite{Alexandrov:2020dyy}.
Thus, the representation in terms of indefinite theta series obtained by solving the modular anomaly equations 
provides an unexpected new insight into the blow-up relations.

\subsection{Higher supersymmetry}
\label{subsec-SUSY}

Up to this point, our analysis was restricted to $\hf$-BPS states in theories with 8 supercharges like $\cN=2$ supergravity
in four dimensions. However, one may ask whether some of the presented results can be generalized to 
string compactifications with more preserved supercharges.
Remarkably, this is indeed possible and one can write a generalization of the anomaly 
equations considered above that captures the modular behavior of various BPS indices \cite{Alexandrov:2020qpb}.
Of course, in most cases this behavior is already well-known, but it is nice to see that 
there is a single universal framework that describes the modularity of BPS states in all string compactifications.

\subsubsection{Helicity supertraces}

The starting point of the construction is the helicity generating function \cite{Kiritsis:1997gu}
\be
B(\cR,y)=\Tr_{\cR} (-y)^{2J_3}, 
\label{defBRy}
\ee
where $\cR$ is a representation of the supersymmetry algebra and $y=e^{2\pi\I z}$ is a formal expansion parameter
similar to refinement. The coefficients of the Taylor expansion in $z$ of $B(\cR,y)$ are identified 
with {\it helicity supertraces}, which count with sign the short and intermediate multiplets 
in supersymmetric theories \cite{Kiritsis:1997gu}. More precisely, they are given by
\be
B_{2\kk}(\cR)=\(\haf\, y\p_y\)^{2\kk} B(\cR,y)|_{y=1}=\Tr_\cR\Bigl[ (-1)^{2J_3} J_3^{2\kk}\Bigr].
\label{defHS}
\ee
The insertion of each power of $J_3$ in the trace soaks up $2$ fermionic zero modes.
Since each broken supercharge generates a fermionic zero mode and all of them should be soaked up to get a non-vanishing result,
the first helicity supertrace to which a multiplet of $\frac1r$-BPS states in
a 4d theory with $\cN$ extended supersymmetry can contribute non-trivially is $B_{2\kk}$ with
\be
\label{n/N}
\kk=\cN\(1-\frac{1}{r}\).
\ee 
To extract the index $\Omi{\cN|r}(\gamma)$ counting such BPS states of charge $\gamma$ from $B_{2\kk}(\cR)$,
one should substitute $\cR=\cH^{\cN}_{\gamma,j}$, the Hilbert space of states of charge $\gamma$ and spin $j$, 
and factor out the center of mass contribution described by 
the supersymmetry multiplet $\cR_{j,2\kk}$ constructed by acting on a spin $j$ ground state with $2\kk$ oscillators, i.e.
\be
\Omi{\cN|r}(\gamma)=\frac{B_{2\kk}(\cH^{\cN}_{\gamma,j})}{B_{2\kk}(\cR_{j,2\kk})}\, ,
\label{relOm2nHS}
\ee
where $\kk$ is determined by $\cN$ and $r$ through \eqref{n/N}.

A crucial observation is that, on one hand, a similar factorization applied to the full helicity generating function in the $\cN=2$ case
gives rise to the refined BPS indices discussed in \S\ref{subsec-refine}, while on the other hand, the ratio defining them
has a perfect sense for arbitrary $\cN$. In other words, we introduce the refined BPS index in a theory with any number of 
supersymmetries by the ratio
\be
\Omega(\gamma,y)=\frac{B(\cH_{\gamma,j},y)}{B(\cR_{j,2},y)}\, .
\label{relOmrShg}
\ee
The virtue of such an index is that it encodes all BPS indices $\Omi{\cN|r}(\gamma)$. Indeed, taking into account that 
$B(\cR_{j,2\kk},y)=O(z^{2\kk})$, one can show that \cite{Alexandrov:2020qpb}
\be
\Omi{\cN|r}(\gamma)=\frac{(-1)^{\kk-1} }{(2\kk-2)!}\, (y\p_y)^{2\kk-2} \Omega(\gamma,y)|_{y=1}.
\label{relOmN}
\ee

\subsubsection{Conjecture}
\label{subsubsec-conjecture}

Once one has a universal definition of the refined BPS index which works for any number of supersymmetries, 
it is natural to use it to define the refined generating functions as in \eqref{defhDTr}
and expect that these functions have modular properties described by anomaly equations similar to \eqref{exp-whhr}.
However, to make these ideas precise, first, one needs to understand how to incorporate several new features 
absent in the $\cN=2$ case. 

Let us recall that compactifications with $\cN=4$ and $\cN=8$ supersymmetries 
can be obtained by taking $\CY=K3\times T^2$ and $\CY=T^6$, respectively.
An important difference of these manifolds compared to CY threefolds with $SU(3)$ holonomy
is that $b_1(\CY)>0$, which leads to additional scalar and gauge fields in the effective action.
In particular, the electromagnetic charge vector can now be represented as 
$\gamma=(p^0,p^A,q_A,q_0)$ where $A$ runs over $b_2+2b_1$ values.
Hence, the relevant charge lattice is now $(b_2+2b_1)$-dimensional.
But what is the associated quadratic form? It turns out that it can be read off from the prepotential 
governing the couplings of vector multiplets in the effective action at the two-derivative level.
It has a cubic form
\be
\Fcl(X)=-\frac{\kappa_{ABC}X^A X^B X^C}{6X^0}\, ,
\label{Fcl-ext}
\ee
with the tensor $\kappa_{ABC}$ extending the tensor of intersection numbers \cite{Kraus:2006nb}.
Then the natural quadratic form is defined, as usual, as $\kappa_{AB}=\kappa_{ABC}p^C$.

It turns out that in most cases with $\cN>2$ the quadratic form $\kappa_{AB}$ is degenerate
with some number of null eigenvalues. But how to deal with such cases has already been explained in \S\ref{subsec-degen}:
the lattice of electric charges should be restricted to the sublattice orthogonal to the null eigenvectors,
while the weight and index of the refined generating functions change to \eqref{index-Hr0}.

Finally, a genuinely new feature of compactifications with higher supersymmetry is that the quadratic form 
$\kappa_{AB}$ can have signature $(n_+,n_-)$ with both $n_\pm>1$.
This fact may drastically affect the modular anomaly because a naive extension of the existing construction
would lead to divergent theta series. In the following, we simply assume that there is a modification of functions 
$\Er_n$ that takes care about this problem. This will be sufficient for our purposes  
since no explicit expressions in such problematic cases will be required.

Thus, given the refined BPS indices \eqref{relOmrShg}, we define the generating functions $\hr_{p,\mu}(\tau,z)$ \eqref{defhDTr} 
in terms of their rational counterparts \eqref{defbOm}. The difference with the previous definitions is that
now it is the quadratic form $\kappa_{AB}$ that defines the invariant charge $\hq_0=q_0-\hf\, \kappa^{AB}q_{A}q_{B}$.
Then we claim
\begin{conj}\label{conj-superN}
The refined generating functions $\hr_{p,\mu}(\tau,z)$ are higher depth mock Jacobi forms of weight and index \eqref{index-Hr0},
where $\Lambda_p$ is the lattice with the quadratic form $\kappa_{AB}$,
and with modular completions satisfying \eqref{exp-whhr} and the refined version of \eqref{exp-derwh}
where the functions $\rmRirf{\bfr}_{\mu,\bfmu}$ and $\cJr_n$ are constructed using the same lattice and
have a zero of order $n-1$ at $z=0$.
\end{conj}

In fact, we are not really interested in the refined generating functions.
They are just a useful bookkeeping device for the generating functions of the unrefined indices. 
Indeed, using the relation \eqref{relOmN} and the fact that the contribution to the refined index of $\frac1r$-BPS states 
in a theory with $\cN$ extended supersymmetry behaves as $O(z^{2\kk-2})$
where $\kk$ is given by \eqref{n/N}, 
one can obtain a relation between the generating functions of refined and unrefined indices,
generalizing \eqref{unreflim-h} valid in the $(\cN|r)=(2|2)$ case,
\be
\hi{\cN|r}_{p,\mu}(\tau)=\sum_{\hat q_0 \leq \hat q_0^{\rm max}}
\bOmi{\cN|r}_{p,\mu}(\hat q_0)\,\q^{-\hat q_0  }
=\frac{2\I(2\pi)^{3-2\kk} }{(2\kk-2)!}\, \p_z^{2\kk-2}( z \hr_{p,\mu}(\tau,z))|_{z=0}.
\label{relhhr}
\ee
Here we introduced 
\be
\label{defbOmN}
\bOmi{\cN|r}(\gamma) = \sum_{d|\gamma} d^{2\kk-4} \,\Omi{\cN|r}(\gamma/d),
\ee
which is a generalization of \eqref{def-bOm}.
Interestingly, for $\cN>2$, one has $\kk\ge 2$ so that the indices \eqref{defbOmN} are {\it not} rational, 
although still different from $\Omi{\cN|r}(\gamma)$ for non-primitive charges.

The relation \eqref{relhhr} applied to the corresponding completions implies that $\whhi{\cN|r}_{p,\mu}$
are modular forms of weight 
\be 
w(p)=2\cN\(1-\frac{1}{r}\)-\hf\, \rank(\Lambda_p)-3.
\label{weight-hN}
\ee 
Their holomorphic anomaly equation can be derived by applying the differential operator \eqref{relhhr}
to the anomaly equation of $\whhr_{p,\mu}$ and taking the limit $z\to 0$.
Taking again into account that $\Omega(\gamma,y)=O(z^{2\kk-2})$, one finds that the contributions of most BPS states 
to the anomaly simply disappear in the limit and one arrives at the following conclusions:
\begin{itemize}
	\item
	The holomorphic anomaly equation can be non-trivial, and hence the generating functions can be mock modular,
	only for $\frac{1}{\cN}$-BPS states.
	
	\item
	Only $\hf$-BPS states can contribute to the r.h.s. of the holomorphic anomaly equation.
	
	\item
	For $\cN>2$ only the contribution of $\hf$-BPS states with $n=2$ survives the unrefined limit.	
\end{itemize}
If one associates the existence of contributions to the anomaly equation to the existence of bound states, 
these conclusions would translate to the well-known fact that in theories with $\cN>2$ 
the only existing bound states are $\frac{1}{\cN}$-BPS states consisting of two $\hf$-BPS states.

To summarize, the conjecture \ref{conj-superN} that the refined generating functions $\hr_{p,\mu}(\tau,z)$
satisfy the same anomaly equations in theories with any number of extended supersymmetries implies 
that the generating functions $\hi{\cN|r}_{p,\mu}(\tau)$ with $r<\cN$ are vector valued modular forms 
of weight \eqref{weight-hN}, whereas for $r=\cN$ their modular completions satisfy
\be
\p_{\bar\tau}\whhi{\cN|\cN}_{p,\mu}(\tau,\btau)
=\q^{\hf\,\kappa^{AB}q_{A}q_{B}} \sum_{q_0}
\sum_{\gamma_1+\gamma_2=\gamma}\!\!
\cJ_2(\{\hgam_1,\hgam_2\},\tau_2)
\prod_{i=1}^2 \(\bOmi{\cN|2}(\gamma_i)\, \q^{-q_{i,0}}\),
\label{anomeqN}
\ee
where $\gamma=(0,p^A,q_A,q_0)$ and $q_A=\mu_A+\hf\kappa_{AB} p^B$.
Note that we expressed the generating functions on the r.h.s. through the sum over D0-brane charges 
because BPS states differing only by this charge may preserve different number of supersymmetries.

\subsubsection{$\cN=4$}

Let us apply the above results to string compactifications with $\cN=4$ supersymmetry.
They can be realized either as type II string theory on $\CY=K3\times T^2$ 
or as heterotic string theory on $T^6$.

The manifold $\CY$ is characterized by the following data
\be
b_1=2,
\qquad
b_2=23,
\qquad
c_{2,a}p^a=24 p^\flat,
\ee
where the index $\flat$ corresponds to the divisor $\cD_\flat=[K3]$.
Thus, the indices $A,B,\dots$ run over $b_2+2b_1=27$ values $A\in \{\flat,\alpha\}=\{\flat,1,\dots,26\}$,
and the non-vanishing components of the symmetric tensor $\kappa_{ABC}$ are given by
\be
\kappa_{\flat \alpha\beta}=\eta_{\alpha\beta}
=
\(\begin{array}{cc}
	I_{1,1}^{\oplus 5} & 0
	\\
	0 & -C_{16}
\end{array}\),
\qquad
I_{1,1}=\(\begin{array}{cc}
	0 & 1
	\\
	1 & 0
\end{array}\),
\ee
where $C_{16}$ is the Cartan matrix of $E_8\times E_8$.

The symmetries and BPS states are more easily characterized in the heterotic frame.
The full U-duality group is a product of S and T-duality factors, $SL(2,\IZ)\times O(6,22;\IZ)$,
and the electromagnetic charge vector is an $SL(2,\IZ)$ doublet of two vectors under the T-duality group
\be
\gamma=\( \begin{array}{c} Q_I \\ P_I \end{array}\)
=\(\begin{array}{ccc}
	q_0, & -p^\flat, & q_\alpha
	\\
	q_\flat, & p^0, & \eta_{\alpha\beta}p^\beta
\end{array}\),
\label{ch-het}
\ee
where in the second representation we expressed the charge components in terms of the usual type IIA notations.
There are two types of BPS states in this theory.

\paragraph{$\hf$-BPS states}

$\hf$-BPS states are characterized by charges such that $Q\parallel P$ and hence for each of them there is a duality frame where
$P$ can be set to zero. Since BPS indices should be invariant under the action of T-duality on the charges, 
the index $\Omi{4|2}(\gamma)$ depends just on a single quantum number
\be 
n= \hf \, Q^2=\hf\, Q_I \eta^{IJ} Q_J,
\qquad
\eta_{IJ}=\(\begin{array}{cc}
	I_{1,1}^{\oplus 6} & 0
	\\
	0 & -C_{16}
\end{array}\).
\ee 

If we restrict to the D4-D2-D0 $\hf$-BPS states, there are two distinct cases to be considered.
First, if $p^\flat>0$, then all charges in the second line of \eqref{ch-het} must vanish.
Restricting for simplicity to $p^\flat=1$, one obtains the following set of charges and the associated quadratic form
\be
\gamma_1=\(\begin{array}{ccc}
	q_0, & -1, & q_\alpha
	\\
	0, & 0, & 0
\end{array}\),
\qquad 
\kappa_{AB}=\(\begin{array}{cc}
	0 & 0
	\\
	0 & \eta_{\alpha\beta}
\end{array}\).
\label{BPS12-1}
\ee
Taking into account that for this set $\hq_0=-n$, $\hq_0^{\rm max}=1$, while $\rank(\kappa_{AB})=26$
and $\eta_{\alpha\beta}$ is unimodular, one finds that the generating function of $\hf$-BPS indices
corresponding to the magnetic charge $p^A=(1,0,\dots, 0)$ should read
\be
\hi{4|2}_{p}(\tau)=\sum_{n=-1}^\infty \Omi{4|2}(n)\, \q^n
\label{countOm12}
\ee
and be a modular form of weight $-12$ and trivial multiplier system.
This immediately implies that it should be proportional to the inverse discriminant function 
(see Ex. \ref{ex-Delta}),
$\hi{4|2}_{p}(\tau)\sim \Delta^{-1}(\tau)$.
This nicely agrees with the well-known fact that the two functions are actually equal \cite{Dabholkar:1989jt}.

Although the generating function \eqref{countOm12} encodes all $\hf$-BPS indices, it is instructive to see how 
other $\hf$-BPS charges fit our formalism. 
The second possibility to get a D4-D2-D0 $\hf$-BPS state is to take $p^\flat=0$ and other components in the
two lines proportional to each other, i.e.
\be
\gamma_2=\(\begin{array}{ccc}
	\frac{\epsilon}{d_Q}\, q_\flat, & 0, & \frac{\epsilon}{d_Q}\,\eta_{\alpha\beta}p^\beta
	\\
	q_\flat, & 0, & \eta_{\alpha\beta}p^\beta
\end{array}\),
\qquad 
\kappa_{AB}=\(\begin{array}{cc}
	0 & \eta_{\alpha\beta}p^\beta
	\\
	\eta_{\alpha\beta}p^\beta & 0
\end{array}\),
\label{BPS12-2}
\ee
where $\epsilon\in \IZ$ and $d_Q=\gcd(q_\flat, \{p^\alpha\})$.
In this case the charge $q_0$ is not independent being fixed by other charges.
This fact makes the generating series trivial since, instead of a sum over $\hq_0$, there is only one term.
One can show that it has $\hq_0=0$ and hence the corresponding function is a constant. 
Since $\rank(\kappa_{AB})=2$, this agrees with the vanishing of the expected modular weight \eqref{weight-hN}.

\paragraph{$\frac14$-BPS states}

$\frac14$-BPS states are characterized by charges \eqref{ch-het} with $Q$ and $P$ non-parallel,
and their BPS indices depend on three T-duality invariants $(n,m,\ell)=(\hf Q^2,\hf P^2,Q\cdot P)$
and a U-duality invariant, known as torsion \cite{Dabholkar:2007vk}
\be
I(\gamma)=\gcd\{Q_IP_J-Q_J P_I\},
\label{defI}
\ee
so that one can write $\Omi{4|4}(\gamma)=\Omi{4|4}_I(n,m,\ell)$.
The famous result of \cite{Dijkgraaf:1996it} is that the generating function of these indices for $I=1$ 
is a Seigel modular form with respect to $Sp(2,\IZ)$ given by the inverse of the function $\Phi_{10}(\tau,z,\sigma)$
called the Igusa cusp form.
For generic torsion, the indices $\Omi{4|4}_I(n,m,\ell)$ have been found in \cite{Dabholkar:2008zy}
and can be expressed through those with $I=1$:
\be
\Omi{4|4}_I(n,m,\ell)=\sum_{\dd | I} \dd \, \Omi{4|4}_1\(n,\frac{m}{\dd^2},\frac{\ell}{\dd}\).
\label{Om14-I}
\ee

The coefficients $\psi_m(\tau,z)$ of the expansion of $\Phi_{10}^{-1}$ in $\sigma$, 
the variable conjugate to the quantum number $m$,
are Jacobi forms of weight $-10$ and index $m$ with respect to $SL(2,\IZ)$. 
A remarkable fact discovered in \cite{Dabholkar:2012nd} is that they admit a canonical decomposition
\be
\psi_m=\psi^P_m+\psi^F_m,
\label{decomppsi}
\ee
where $\psi^P_m$ contains the ``polar" part of the original function and describes contributions of bound states only, 
whereas all contributions of single centered black holes (immortal dyons) are encoded in $\psi^F_m$, 
which does not have any poles in $z$.
Furthermore, both these functions are mock modular and the modular completion of the generating function of immortal dyons
satisfies
\be
\tau_2^{3/2}\p_{\btau}\whpsi^F_m(\tau,\btau,z)=\frac{\sqrt{m}}{8\pi\I}\, \frac{\Omi{4|2}(m)}{\Delta(\tau)}
\sum_{\ell=0}^{2m-1}\overline{\ths{m}_{\ell}(\tau,0)}\,\ths{m}_{\ell}(\tau,z)\equiv \cA_m(\tau,\btau,z),
\label{holanom}
\ee
where $\ths{m}_{\ell}(\tau,z)$ is the theta series \eqref{deftheta}.

Let us again restrict to D4-D2-D0 states by setting $p^0=0$ in \eqref{ch-het}. 
In addition, for simplicity, we assume $p^\flat=1$ and $p^2>0$. 
Then the T and U-duality invariants are found to be
\be
n=\hf\,q^2-q_0 ,
\qquad
m= \hf\, p^2,
\qquad
\ell=p^\alpha q_\alpha-q_\flat,
\qquad
I(\gamma)=\gcd(\ell,\{p^\alpha\}),
\label{Tinvgen}
\ee
where $q^2=\eta^{\alpha\beta}q_\alpha q_\beta$ and $p^2=\eta_{\alpha\beta}p^\alpha p^\beta$,
while the quadratic form is non-degenerate and is given by
\be
\kappa_{AB}=\(\begin{array}{cc}
	0 & \eta_{\alpha\beta}p^\beta
	\\
	\eta_{\alpha\beta}p^\beta & \eta_{\alpha\beta}
\end{array}\),
\label{qf14}
\ee
with $|\det\kappa_{AB}|=2m$.
Since the indices defining the generating function $\hi{4|4}_{p,\mu}$ are evaluated at the attractor point, 
they count only single-centered black holes\footnote{More precisely, this statement holds for terms with negative $\hq_0$
which count BPS black holes with non-vanishing area. In principle, at the attractor point 
also the scaling solutions \cite{Denef:2007vg,Bena:2012hf} might contribute but, being composed
of at least three constituents, they do not exist in $\cN=4$ theory \cite{Dabholkar:2009dq}.}
and therefore, after multiplication by\footnote{The sign factor is needed to cancel the multiplier system of $\hi{4|4}_{p,\mu}$
and nicely agrees with the presence of the same factor in the definition of the generating function given by the Igusa cusp form, 
which was advocated in \cite{Shih:2005uc}.} 
$(-1)^\ell\ths{m}_{\ell}(\tau,z)$ and summing over $\ell=m-\mu$,  
should coincide with a generalization $\psi_p$ 
of the generating function of immortal dyons $\psi^F_m$ to arbitrary torsion invariant.
According to \eqref{weight-hN}, its weight is expected to be $w(p)+\hf=-10$, while its index is equal to $m$, 
in agreement with the weight and index of $\psi^F_m$.

The most interesting is to compare their anomaly equations.
According to \eqref{anomeqN}, the non-vanishing contributions to the holomorphic anomaly equation 
for $\whpsi_p$ arise only from splits of the charge
$\gamma=\gamma_1+\gamma_2$ where $\gamma_1$ and $\gamma_2$ are both $\hf$-BPS charges.
It is easy to see that this is possible only if one of them belongs to the class \eqref{BPS12-1} and the other to \eqref{BPS12-2}
with $d_Q=I(\gamma)$.
Moreover, all charges of the constituents are fixed in terms of the full charge and a single integer parameter,
so that the lattice one sums over on the r.h.s. of \eqref{anomeqN} is one-dimensional.
Evaluating all the ingredients explicitly and taking into account that $\hi{4|2}_{p}= \Delta^{-1}$, 
one can show \cite{Alexandrov:2020qpb} that the anomaly equation \eqref{anomeqN}
is equivalent to 
\be
\tau_2^{3/2}\p_{\btau}\,\whpsi_p(\tau,\btau,z)
= \sum_{\dd|d_p} \dd \,\cA_{m/\dd^2}(\tau,\btau,\dd z),
\label{mainanomaly}
\ee
where  $d_p\equiv \gcd\{p^\alpha\}$.
If $d_p=1$, which implies the trivial torsion $I(\gamma)=1$, this equation reproduces \eqref{holanom}.
For $d_p>1$, it provides a generalization of the anomaly equation found in \cite{Dabholkar:2012nd} to
the case of a non-trivial torsion. One can also check that it is consistent with the relation \eqref{Om14-I}.

It is worth emphasizing that the function $\psi_p$ for $d_p>1$ comprises contributions of charges with {\it different}
values of the torsion invariant. In fact, it includes states with all $I$ dividing $d_p$.
Therefore, it is different from the generating function of $\frac14$-BPS indices with fixed $I>1$,
which is known to transform properly only under the congruence subgroup $\Gamma^0(I)$ of $SL(2,\IZ)$ \cite{Dabholkar:2008zy}.
Instead, the presented construction automatically produces functions
that transform as (mock) modular or Jacobi forms under the full $SL(2,\IZ)$.
In other words, it tells us how charges with different torsion should be combined together 
in order to form objects with nice modular properties --- if one follows the general prescription,
the result is guaranteed.

\subsubsection{$\cN=8$}

String compactifications with $\cN=8$ supersymmetry are obtained by taking type II string theory on $T^6$.
In this case $b_1=6$ and $b_2=15$, so that the indices $A,B,\dots$ run again over $b_2+2b_1=27$ values.
We will denote the charge components as $p^A=(p^{ij},p_i,\tp_i)$ and $q_A=(q_{ij},q^i,\tq^i)$ where $i=1,\dots,6$
and $q_{ij}$ and $p^{ij}$ are antisymmetric.
The antisymmetric components correspond to the gauge fields coming from the RR sector, while the charges with one index
correspond to the gauge fields arsing from the reduction of the metric and the $B$-field on one-cycles.

The U-duality group is $E_7$ and the charge vector transforms in the irrep $\bf{56}$
of this group. The U-duality orbits are characterized by a single quartic invariant $I_4(\gamma)$,
which implies that the BPS indices depend on a single quantum number. The invariant $I_4(\gamma)$ can be written in terms of 
a cubic invariant of $E_6$ appearing in the reduction $E_7\to E_6\times O(1,1)$ \cite{Pioline:2005vi}
\be 
I_4(\gamma)=4p^0I_3(q)-4 q_0 I_3(p) + 4\p^A I_3(q)\,\p_A I_3(p)-(p^0q_0+p^A q_A)^2,
\ee 
which itself is given by
\be
I_3(p)=\Pf(p^{ij})+p^{ij}p_i \tp_j, 
\qquad 
\Pf(p^{ij})= \frac{1}{48}\, \eps_{ijklmn} p^{ij} p^{kl} p^{mn}.
\ee
In particular, it is the cubic invariant that defines the quadratic form relevant for D4-D2-D0 BPS states,
\be
\kappa_{AB}=\p_A\p_B I_3(p).
\ee
For vanishing NS-charges, it is easy to compute it explicitly 
\be
\kappa_{AB}=\(\begin{array}{ccc}
	\hf\, \eps_{ijklmn}p^{mn} & 0 & 0
	\\
	0 & 0 & p^{rs}
	\\
	0 & p^{rs} & 0
\end{array}\),
\ee 
which gives, in particular, 
\be 
|\det \kappa_{AB}|=2(\Pf(p^{ij}))^9,
\qquad
\hq_0=-\frac{I_4(\gamma)}{4I_3(p)}\, .
\label{hq0-12BPS}
\ee 

The theory has three types of BPS states and there are many charge configurations 
corresponding to D4-D2-D0 bound states realizing them.
In the following we will consider only a few representative examples to demonstrate how the formalism of 
\S\ref{subsubsec-conjecture} reproduces the well-known modular properties of the BPS indices.

\paragraph{$\frac12$-BPS states}

The BPS conditions on charge vectors have been found in \cite{Ferrara:1997ci} and 
can  be written in terms of the quartic invariant and the derivatives $\p_\cI$ with respect to charges
where $\cI$ labels all components of the charge vector $\gamma$.
The $\hf$-BPS condition is the strongest one and reads as
\be  
\p_\cI\p_\cJ I_4(\gamma)|_{{\rm Adj}(E_7)}= 0,
\label{cond12BPS}
\ee 
where ${\rm Adj}(E_7)$ denotes the representation $\bf{133}$ appearing in the decomposition 
$\bf{56}^2=\bf{133}+\bf{1463}$.
Note that it implies the vanishing of $I_4(\gamma)$.
It has been elaborated in full generality in terms of charge components in \cite{Obers:1998fb}.
Instead, let us take the only non-vanishing magnetic charges to be
$p^{12}, p^{34}, p^{56}$. One can show that the condition \eqref{cond12BPS}
implies that there can be at most one non-vanishing magnetic charge, so we take $p^{12}\ne 0$, $p^{34}=p^{56}=0$.
Then the $\hf$-BPS condition leaves only 10 unrestricted electric charges, consistently with the fact that 
$\rank(\kappa_{AB})=10$. Hence, the expected modular weight \eqref{weight-hN} of $\hi{8|2}_{p}$ is equal to 0,
i.e. the generating function of $\hf$-BPS D4-D2-D0 states must be a constant.
And indeed the condition \eqref{cond12BPS} fixes $q_0$ so that $\hq_0=0$.
All this nicely agrees with the fact (see, e.g. \cite{Bianchi:2009mj,Bossard:2016hgy}) 
that there is a single $\hf$-BPS index equal to 1.

\paragraph{$\frac14$-BPS states}

The case of $\frac14$-BPS states is very interesting because it illustrates several non-trivial phenomena.
First, this is the case where BPS indices have a completely different behavior compared to the degeneracies 
counted without the sign insertion.
The latter are counted by
\be
\hi{8|4}_{\rm deg}(\tau)=\frac{1}{16}\prod_{n=1}^\infty\(\frac{1+\q^n}{1-\q^n}\)^8
=\frac{1}{16}\(\frac{\eta(2\tau)}{\eta^2(\tau)}\)^8.
\label{deg14}
\ee
Note that the first term in the Fourier expansion is given by rational number $\frac{1}{16}$ 
because it corresponds to $\hf$-BPS states
and is equal to the ratio of the dimensions of the ultrashort $\hf$-BPS multiplet ($(16)^2$ states)
and the short $\frac14$-BPS multiplet ($(16)^3$ states).
The resulting degeneracies grow exponentially with $n$.
On the other hand, the BPS indices are organized into the generating function \cite[Eq.(2.13)]{Bossard:2016hgy}
\be
\hi{8|4}(\tau)=\frac{E_4(\tau)}{240}+\frac{7}{144}
\label{h84}
\ee
and grow only polynomially.
Let us see how this result can be recovered from the approach used above.

The $\frac14$-BPS condition is given by 
\be 
\p_\cI I_4(\gamma)=0, 
\qquad 
\p_\cI\p_\cJ I_4(\gamma)|_{{\rm Adj}E_7}\ne 0,
\label{cond14BPS}
\ee
which also implies the vanishing of $I_4(\gamma)$.
Let us again restrict to the case where the only non-vanishing magnetic charges are $p^{12}, p^{34}, p^{56}$.
Then the condition \eqref{cond14BPS} requires that at least one of these charges must vanish.
If one considers the most natural possibility $p^{12}, p^{34}\ne 0$, $p^{56}=0$, one finds that
it is similar to the case of $\hf$-BPS states discussed above because the $\frac14$-BPS condition
fixes the D0-brane charge so that $\hq_0=0$ and the generating function reduces to a constant.
This is consistent with the fact that $\rank(\kappa_{AB})=18$, which implies 
the vanishing of the modular weight \eqref{weight-hN}. 

A non-trivial generating function is obtained in a more degenerate case of $p^{12}\ne 0$, $p^{34}=p^{56}=0$
because $q_0$ is then left unrestricted. Since now $\rank(\kappa_{AB})=10$, the modular weight 
of $\hi{8|4}_p$ should be equal to 4.
One can also verify the triviality of the multiplier system and that $\hq_0^{\rm max}=0$, which 
singles out a unique modular form $E_4(\tau)$ satisfying all these requirements, consistently with \eqref{h84}. 
The deviation from the Eisenstein series is due to the fact, already noticed below \eqref{deg14},
that the constant term corresponding to $\hq_0=0$ counts $\hf$-BPS states and not $\frac14$-BPS. 
Importantly, it does {\it not} generate a holomorphic anomaly for the modular completion $\whhi{8|4}_p$ 
because the completing term is holomorphic being just a constant.
Its precise value can be obtained as
\be
\frac{B_{12}(\cR_{0,8})}{B_{12}(\cR_{0,12})}=\frac{1}{12!}(y\p_y)^{12}\Bigl[(1-y)^4 (1-y^{-1})^4\Bigr]_{y=1}
=\frac{19}{360}
\ee
and agrees with \eqref{h84} due to $\frac{1}{240}+\frac{7}{144}=\frac{19}{360}$.

\paragraph{$\frac18$-BPS states}

The last case of $\frac18$-BPS states has been studied in many works 
(see, e.g., \cite{Shih:2005qf,Pioline:2005vi,Sen:2008ta,Sen:2008sp})
and it was found that the $\frac18$-BPS indices are given by 
\be
\Omi{8|8}(\gamma)
= \sum_{s\, :\, \nabla_X F_0\in \IZ} s\, N(s)\, \hc\(\frac{I_3(Q)}{s^2}\, ,\, \frac{J_L}{s}\)
=\sum_{2s|\chi(\gamma)} s\, \hc(I_4(\gamma)/s^2),
\label{Omi88}
\ee
where in the first representation $N(s)$ is the number of common divisors of $X^I$ and $\p_I F_0$ with
\be
F_0(X)=\frac{I_3(X)}{X^0}\, ,
\qquad
X^I=(s, q_A),
\ee
and
\be
Q_A=p^0 q_A+\p_A I_3(p),
\qquad
2J_L=(p^0)^2 q_0+p^0 p^A q_A+2I_3(p).
\ee
The second representation is manifestly U-duality invariant and we refer to \cite{Sen:2008sp}
for the precise definition of the function $\chi(\gamma)$.
The most important ingredient in this formula is provided by $\hc(n,\ell)$, 
the Fourier coefficients of the Jacobi form of weight $-2$ and index 1
\be
\phi(\tau,z)=-\frac{\theta_1(\tau,z)^2}{\eta(\tau)^6}=\sum_{n=0}^\infty \sum_{\ell\in\IZ} \hat c(n,\ell)\, \q^n\, y^\ell.
\label{Jacobi18}
\ee
The coefficients depend actually on a single variable
\be
\hc(n,\ell)=\hc (4n-\ell^2),
\ee
which is used in the second representation in \eqref{Omi88}.
It turns out that in this form they coincide with the Fourier coefficients of the following function
\be
\Phi(\tau)=\frac{\theta_4(2\tau)}{\eta(4\tau)^6}=\sum_{n=-1}^\infty \hc(n)\, \q^n
\, .
\label{fun18tauy}
\ee

However $\Phi(\tau)$ is not the function we are looking for because it is {\it not} modular with respect to  
the full $SL(2,\IZ)$.
The reason is that the Jacobi form \eqref{Jacobi18} has index 1 and hence implies that its coefficients can be
combined into a {\it vector valued} modular form with 2 components by means of the theta expansion 
as in \eqref{thetadecomp}:
\be
\phi(\tau,z)=\sum_{\ell=0,1}(-1)^\ell h_\ell(\tau)\,\ths{1}_{\ell}(\tau,z),
\ee
where $\ths{1}_0(\tau,z)=\theta_3(2\tau,2z)$,  $\ths{1}_1(\tau,z)=\theta_2(2\tau,2z)$
and for convenience we included a sign factor that affects only the multiplier system of $h_\ell(\tau)$.
The two components correspond to odd and even values of 
the quartic invariant $d=4n-\ell^2$ and are given by
\be
h_\ell(\tau)=(-1)^\ell\sum_{d\in 4\IZ -\ell^2} \hc(d)\, \q^{d/4}
=\(\frac{\theta_3(2\tau)}{\eta(\tau)^6}\, ,\,\frac{\theta_2(2\tau)}{\eta(\tau)^6} \),
\label{hell}
\ee
which is most easily obtained by decomposing \eqref{fun18tauy}:
\be
\Phi(\tau/4)=\frac{\theta_4(\tau/2)}{\eta(\tau)^6}=\frac{\theta_3(2\tau)}{\eta(\tau)^6}-\frac{\theta_2(2\tau)}{\eta(\tau)^6}.
\ee
The vector \eqref{hell} does transform as a modular form under the full $SL(2,\IZ)$
as follows from Ex. \ref{ex-theta}.

On the other hand, the general construction of \S\ref{subsubsec-conjecture} implies that
we should consider
\be
h_{p,\mu}(\tau)
=\sum_{I_4\in 4I_3(p) \IZ+I_4(\mu) } \bOmi{8|8}(I_4)\, \q^{I_4/4I_3(p)},
\label{hBPS18}
\ee
where we took into account the relation \eqref{hq0-12BPS} and that the BPS indices depend only on the quartic invariant.
Let us again restrict to the case where all magnetic charges vanish except $p^{12}, p^{34}, p^{56}$
and denote $m=I_3(p)=p^{12}p^{34}p^{56}$. 
Since the $\hf$-BPS condition requires vanishing of at least two of the charges $p^{12}, p^{34}, p^{56}$,
it is impossible to decompose the magnetic charge $p^A$ into two charges giving rise to $\hf$-BPS states.
Therefore, the r.h.s. of the holomorphic anomaly equation \eqref{anomeqN} vanishes 
and the generating function \eqref{hBPS18} must be a vector valued modular form.
Its weight follows from \eqref{weight-hN}. Since the quadratic form is non-degenerate with $\rank(\kappa_{AB})=27$,
it is equal to $-5/2$.
Furthermore, given that $c_{2,a}=0$ and $(p^3)=6I_3(p)=6m$, the most singular term has the power $-\hq_0^{\rm max}=-\frac{m}{4}$,
while due to \eqref{hq0-12BPS}, $\mu_A$ run over $2m^9$ values.
For $m=1$, these properties reproduce those of the 2-dimensional vector \eqref{hell}.
One can also show that the multiplier systems also agree, and since all charges are primitive, 
$\bOmi{8|8}(I_4)=\Omi{8|8}(I_4)$. 
Thus, up to an overall scale, the generating function \eqref{hBPS18} must coincide with $h_\ell(\tau)$.
Of course, for $m>1$, the generating function will be different, but it is constructed 
from the same set of BPS indices $\Omi{8|8}(I_4)$ and hence carries the same information.

\section{Conclusions}
\label{sec-concl}

Mock modularity is a beautiful mathematical structure that represents now a rapidly developing
and expanding subject of mathematical research with numerous and deep relations to theoretical physics.
Its manifestations range from non-compact CFTs \cite{Troost:2010ud}, sigma models \cite{KumarGupta:2018rac} 
and black hole state counting in string compactifications
\cite{Dabholkar:2012nd,Alexandrov:2016tnf,Alexandrov:2018lgp} 
to Vafa-Witten theory \cite{Vafa:1994tf,Dabholkar:2020fde},
Donaldson-Witten theory \cite{Korpas:2017qdo,Korpas:2019cwg}, moonshine phenomenon \cite{Cheng:2011ay},
quantum invariants of three-dimensional manifolds \cite{Cheng:2018vpl,Cheng:2023row} and many other setups.
In this review we concentrated mainly just on one of these manifestations --- mock modularity of 
the generating functions of BPS indices counting states of supersymmetric black holes in Calabi-Yau compactifications
and realized in mathematics as rank 0 generalized DT invariants.
We showed that it governs a universal structure represented by an iterated system of anomaly equations.
It is universal because it turns out to describe many phenomena beyond the original setup.
It remains valid for various degenerations, in the non-compact limit, after inclusion of a refinement,
and even for compactifications with higher supersymmetry where it allows to reproduce most of the known results. 
This universality suggests that the same or a similar structure may govern also the other manifestations
of mock modularity mentioned above, as has been shown, for instance, for Vafa-Witten theory \cite{Alexandrov:2019rth}.

Despite the original argument for modularity of the generating functions of D4-D2-D0 BPS indices
came from the analysis of a CFT living on the brane world-volume \cite{Maldacena:1997de}, 
it is the target space perspective that turned out to be more productive.
In this physical picture, the origin of modularity can be traced back to S-duality of type IIB string theory,
while the mock modularity appears to arise due to wall-crossing, i.e. the existence of bound states whose stability depends on
values of the moduli. On the other hand, a pure mathematical understanding of both these phenomena 
for the generating functions of rank 0 DT invariants for generic CY threefolds
is still absent and only recently first steps have been undertaken in this direction 
in the simplest case of the quintic threefold and unit D4-brane charge \cite{{talkSheshmani}}.

The main application of the mock modular properties of the generating functions expressed by the system of anomaly equations
is the actual computation of these generating series. So far the work in this direction has concentrated on 
non-compact cases (VW theory) and one-parameter CY threefolds. It demonstrated that there is a nice interplay between
mock modularity of rank 0 DT invariants and the holomorphic anomaly of topological strings which compute GV invariants
of the same CY threefold. Computing one set of invariants helps computing the other and vice-versa.
Proceeding in this way, allows us to overcome the limitations of the direct integration approach of solving
topological string theory on compact CYs. 

However, to further pursue this idea  and apply it to higher D4-brane charges and to more general CY threefolds
with more moduli, we need new wall-crossing relations between various topological invariants, 
which would allow us to compute them more efficiently than the currently known relations.
This might be seen as the key open problem of this research program.
The most promising avenue seems to be the study of D6-$\aD$ wall crossing.
However, the existing results in this direction are insufficient for applications.
An intriguing workable prescription was proposed in \cite{VanHerck:2009ww}, but 
it seems to be at odds with the standard mathematical definition of DT invariants and the standard wall-crossing formulas.

Although higher rank DT invariants are not expected to possess modular properties,
one can still ask whether the presented results can help in computing them.
In principle, according to the recent results
\cite{Toda:2011aa,Feyzbakhsh:2020wvm,Feyzbakhsh:2021rcv,Feyzbakhsh:2021nds,Feyzbakhsh:2022ydn},
they all should be expressible through rank 0 invariants.
Unfortunately, these results are not constructive yet, and there are no explicit formulas 
that would allow us to do so.

One could ask whether some of the results can be extended to theories with less supersymmetry 
describing more realistic compactifications. Unfortunately, it is hard to expect such an extension
since not only BPS states do not exist in such theories, but also the fate of S-duality remains unclear \cite{Grimm:2007xm}.
Nevertheless, recently some inflation models have been proposed where modular symmetry plays a crucial role
\cite{Casas:2024jbw,Kallosh:2024ymt}. Mock modularity however did not show up yet in these investigations.

Finally, an almost unexplored subject is the interplay between mock modularity and non-commutativity
suggested by the emergence of a non-commutative star-product structure on the moduli space
(and its twistor space) after the inclusion of a refinement.
What does this non-commutativity mean physically? What does it imply for the low energy effective action?
Why should it be compatible with S-duality? These are just few questions that may be asked about 
this new exciting playground for mock modularity.

\vspace{6pt} 




\acknowledgments{The author is grateful to Sibasish Banerjee, Khalil Bendris, Soheyla Feyzbakhsh, Nava Gaddam, 
	Albrecht Klemm, Pietro Longhi, Jan Manschot, Suresh Nampuri, Boris Pioline and Thorsten Schimannek 
	for many useful discussions and collaboration on the topics presented in this review.}


%

\appendixtitles{yes} 
\appendixstart
\appendix
\numberwithin{equation}{section}

\section{Trees}
\label{ap-trees}

\vspace{-0.4cm}
\begin{figure}[H]
	\isPreprints{\centering}{} 
	\includegraphics[width=16cm]{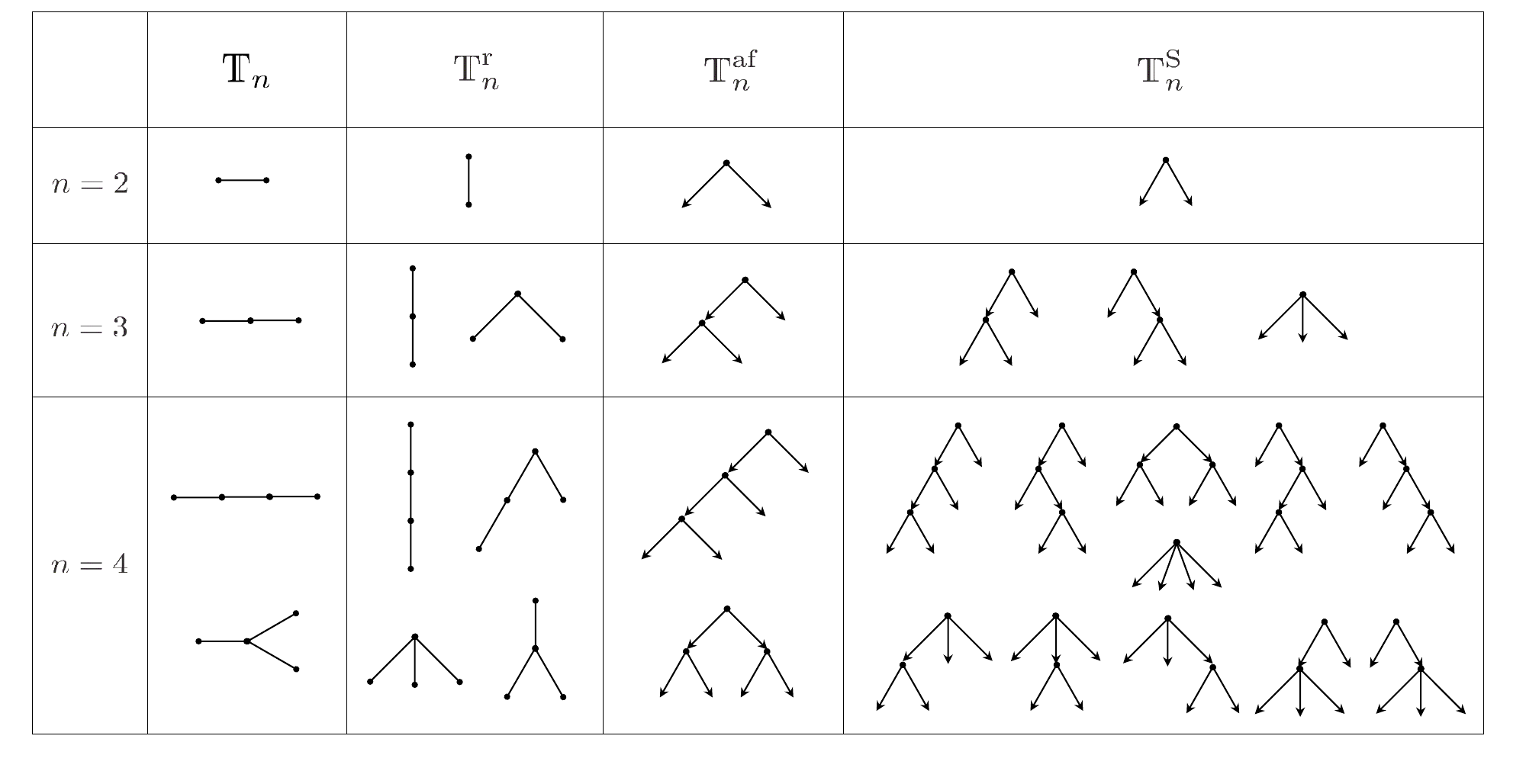}
	\vspace{-0.8cm}
	\caption{Various types of trees mentioned in the review.
		$\IT_n$ is the set of unrooted trees  with $n$ vertices.
		$\IT_n^{\rm r}$ is the set of rooted trees with $n$ vertices.
		$\IT_n^{\rm af}$ is the set of attractor flow trees with $n$ leaves (of which we only draw the different topologies).
		$\IT_n^{\rm S}$ is the set of Schr\"oder trees with $n$ leaves.
		In addition, an important role is played by the set  $\IT_n^\ell$
		of unrooted labeled trees which are obtained from $\IT_{n}$ by assigning different labels to the vertices.\label{fig-table}}
\end{figure}   

\section{Generalized error functions}
\label{ap-generr}

The generalized error functions have been introduced in \cite{Alexandrov:2016enp,Nazaroglu:2016lmr} 
(see also \cite{kudla2016theta,funke2017theta}).
They are defined by
\bea
E_n(\cM;\vu)&=& \int_{\IR^n} \de \vu' \, e^{-\pi\sum_{i=1}^n(u_i-u'_i)^2} \prod_{i=1}^n \sign(\cM^{\rm tr} \vu')_i\, ,
\label{generr-E}
\eea
where $\vu=(u_1,\dots,u_n)$ is $n$-dimensional vector and $\cM$ is $n\times n$ 
matrix of parameters.\footnote{The information carried by $\cM$ is in fact highly redundant. 
	For example, for $n=1$ the dependence on $\cM$ drops out, whereas
	$E_2$ and $E_3$ depend only on 1 and 3 parameters, respectively.}
They generalize the ordinary error function because at $n=1$ one has $E_1(u)=\Erf(\sqrt{\pi}\, u)$.
Their main role is to provide modular completions of theta series with quadratic forms
of indefinite signature.

To get kernels of indefinite theta series, we need however functions depending on a $d$-dimensional vector
rather than an $n$-dimensional one. To define such functions, let $\{\vbbm_i\}$ be a set
of $n$ vectors of dimension $d$, 
and it is assumed that these vectors span a positive definite subspace in $\IR^d$ endowed with 
a bilinear form\footnote{This bilinear form is {\it opposite} to the bilinear form $\star$ used in 
	the beginning of \S\ref{sec-indef} up to Eq. \eqref{gentheta} (see also footnote \ref{foot-sign}). 
	In fact, the sign of the bilinear form affects just the overall sign of $\Phi_n^E$.
	\label{foot-sign-ap}} 
$\ast$, i.e. $\vbbm_i\ast\vbbm_j$ is a positive definite matrix. We also introduce an orthonormal basis 
$\cB=\{\ebbm_i\}$ for this subspace.
Then we set
\be
\Phi_n^E(\{\vbbm_i\};\xbbm)=E_n(\cM;\cB\ast \xbbm),
\qquad
\cM_{ij}=\ebbm_i\ast\vbbm_j.
\label{generrPhiME}
\ee
The detailed properties of these functions can be found in \cite{Nazaroglu:2016lmr}.
The most important of them are the following:
\begin{itemize}
	\item  
	The functions $\Phi_n^E(\{\vbbm_i\};\xbbm)$ do not depend on the choice of the basis $\cB$ and are smooth functions of $\xbbm$.
	\item 
	They solve the Vign\'eras equation \eqref{Vigdif} with $\lambda=0$.
	\item 
	At large $\xbbm$, they reduce to $\prod_{i=1}^n \sgn (\vbbm_i\ast\,\xbbm)$.
    \item 
    Their derivative is given by 
\be 
\p_\xbbm \Phi_n^E(\{\vbbm_i\};\xbbm)=2\sum_{k=1}^n \frac{\vbbm_k}{||\vbbm_k||}\,
e^{-\pi\, \frac{(\vbbm_k\ast\xbbm)^2}{\vbbm_k^2}}\,\Phi_{n-1}^E(\{\vbbm_i\}_{i\ne k};\xbbm).
\ee 
\item 
If one of the vectors is null, it reduces the rank of the generalized error function.
Namely, for $\vbbm_\ell^2=0$, one has
\be
\Phi_n^E(\{\vbbm_i\};\xbbm)=\sgn (\vbbm_\ell\,\ast\xbbm)\,\Phi_{n-1}^E(\{\vbbm_i\}_{i\ne \ell};\xbbm).
\label{Phinull}
\ee
In other words, for such vectors the completion is not required.

\end{itemize}

We also need a second set of functions that can be seen as generalizations of the complementary error function.
They were also defined in \cite{Alexandrov:2016enp,Nazaroglu:2016lmr}, but we will use a slightly generalized version
introduced in \cite{Alexandrov:2019rth} that depends on an additional vector of parameters:
\be 
\hM_n(\cM;\vu,\frb)=\(\frac{\I}{\pi}\)^n |\det\cM|^{-1} \int\limits_{\IR^n-\I (\vu-\frb)}\de\vu'\,
\frac{e^{-\pi \vu'^{\rm tr} \vu' -2\pi \I \vu'^{\rm tr} \vu}}{\prod_{i=1}^n(\cM^{-1}\vu')_i}\, .
\ee
For $n=1$, one recovers the ordinary complementary error function,  $\hM_1(u,0)=-\sgn(u)\Erfc(\sqrt{\pi}|u|)$,
and, more generally, $\hM_n$ are exponentially suppressed for large $\vu$ as 
$\hM_n\sim \frac{(-1)^n}{\pi^n}\, |\det\cM|^{-1}\frac{e^{-\pi\vu^{\rm tr}\vu}}{\prod_{i=1}^n(\cM^{-1}\vu)_i}$.
In turn, these functions are used to define 
\be
\hPhi_n^M(\{\vbbm_i\};\xbbm,\bbbm)=\hM_n(\cM;\cB\cdot \xbbm,\cB\ast \bbbm),
\label{def-hPhiM}
\ee
where the matrices $\cM$ and $\cB$ are the same as in \eqref{generrPhiME}, 
which can be used as kernels of indefinite theta series.
More precisely, $\hPhi_n^M$ can be seen as iterated Eichler integrals providing modular completions of such theta series.
Note that, in contrast to $\Phi_n^E$, they are smooth only away from real codimension-1 loci in $\IR^d$.

An important fact is that the functions $\Phi_n^E$ and $\hPhi_n^M$ can be expressed through each other and sign functions.
More precisely, they satisfy the following identity \cite[Prop. 1]{Alexandrov:2019rth}:
\be
\Phi_n^E(\{\vbbm_i\};\xbbm)=\sum_{\cI\subseteq \Zv_n}\hPhi_{|\cI|}^M(\{\vbbm_i\}_{i\in\cI};\xbbm,\bbbm)
\prod_{j\in \Zv_n\setminus \cI}\sign(\vbbm_{j\perp \cI}\ast(\xbbm-\bbbm)),
\label{expPhiE-mod}
\ee
where the sum goes over all possible subsets (including the empty set) of the set $\Zv_{n}=\{1,\dots,n\}$, 
$|\cI|$ is the cardinality of $\cI$,
and $\vbbm_{j\perp \cI}$ denotes the projection of $\vbbm_j$ orthogonal to the subspace spanned by $\{\vbbm_i\}_{i\in\cI}$.
This identity should also make clear the role of the vector $\bbbm$: if $\xbbm=\sqrt{2\tau_2}(\kbbm+\bbbeta)$ as
in \eqref{gentheta}, 
choosing $\bbbm=\sqrt{2\tau_2} \bbbeta$, one obtains that the sign functions in \eqref{expPhiE-mod} are independent of $\bbbeta$.
This is important for the refined construction in \S\ref{subsec-refine} where the constant contributions $\Efrf_n$ \eqref{Efref}
are also $\beta$-independent.

\section{Lattice factorization and theta series}
\label{ap-fact}

In this appendix we recall some basic facts about the decomposition of a lattice into sublattices
and the corresponding factorization of theta series (see \cite{CSbook} for more details).

Let $\bbLambda$ be a lattice of dimension $d$ and $\bbLambda^{(a)}$, $a=1,\dots,n$, are orthogonal sublattices of dimensions $d_a$
such that $\sum_{a=1}^n d_a=d$. The problem that we want to address here is that in general 
$\oplus_{a=1}^n\bbLambda^{(a)}$ is only a sublattice of $\bbLambda$. While all elements of $\bbLambda$ can be decomposed 
as linear combinations of elements of $\bbLambda^{(a)}$, the decompositions may involve rational coefficients.
In such situation, to get the full lattice from the sublattices, one has to introduce the so called {\it glue vectors}.
They are given by the sum of representatives of the discriminant groups $\ID^{(a)}=(\bbLambda^{(a)})^*/\bbLambda^{(a)}$
which at the same time belongs to the original lattice, i.e. 
$\glueg_\Asf=\oplus_{a=1}^n\gluegi{a}_\Asf \in\bbLambda$ where $\gluegi{a}_\Asf\in \ID^{(a)}$.
The number of the glue vectors is equal to 
\be
n_g=\left|\frac{\prod_{a=1}^n\det\bbLambda^{(a)}}{\det\bbLambda}\right|^{1/2}\, ,
\label{Ng}
\ee 
where $\det\bbLambda=|\bbLambda^*/\bbLambda|$ is the order of the discriminant group 
and is equal to the determinant of the matrix of scalar products of the basis elements.
The decomposition formula of the lattice $\bbLambda$ then reads
\be 
\bbLambda=\bigcup\limits_{\Asf=0}^{n_g-1} \[\mathop{\oplus}\limits_{a=1}^n \(\bbLambda^{(a)}+\gluegi{a}_\Asf\)\].
\label{lat-glue}
\ee 

The main application of the lattice decomposition \eqref{lat-glue} is a factorization of theta series.
Let us consider a general indefinite theta series as in \eqref{gentheta} 
with a kernel having a factorized form
$\Phi(\xbbm)=\prod_{a=1}^n\Phi_a(\xbbm^{(a)})$
where the upper index $^{(a)}$ on a vector denotes its projection to $\bbLami{a}$.
Then the lattice factorization formula \eqref{lat-glue} implies that
one can split the sum in the definition of the theta series into 
$n$ sums coupled by the additional sum over the glue vectors so that one arrives 
at the following identity for theta series
\be
\vth_{\bbmu}(\tau, \zbbm;\bbLambda, \Phi, \pbbm)=\sum_{\Asf=0}^{n_g-1}
\prod_{a=1}^n \vth_{\bbmu^{(a)}+\gluegi{a}_\Asf}(\tau, \zbbm^{(a)};\bbLami{a}, \Phi_a, \pbbm^{(a)}).
\label{factortheta-gen}
\ee

\section{Degenerate case with non-degenerate quadratic form}
\label{ap-null}

In this appendix we compute the holomorphic anomaly of the modular completion \eqref{exp-derwh-lim} 
in the degenerate case where $(p_0^3)=0$.

Before we specialize \eqref{exp-derwh-lim} to the degenerate case, let us rewrite it in a more explicit form by solving
the constraint on the D2-brane charges 
\be 
q_{1,a}+q_{2,a}=\mu_a+\hf\, r^2\kappa_{0,ab}p_0^b
\label{constr-2ch}
\ee 
where $\kappa_{0,ab}=\kappa_{abc}p_0^c$ and $q_{i,a}$ are decomposed as in \eqref{decomp-spfl} with $p_i^a=r_ip_0^a$.
In terms of the spectral flow parameters $\eps_i^a$, the constraint \eqref{constr-2ch} takes the form
\be 
\kappa_{0,ab}\(r_1\eps_1^b+r_2\eps_2^b\)=\Delta\mu_a+r_1r_2\kappa_{0,ab}p_0^b,
\label{constr-eps}
\ee 
where
\be
\Delta\mu_a=\mu_a-\mu_{1,a}-\mu_{2,a}.
\label{Deltamu}
\ee
An immediate consequence of this relation is that $\Delta\mu\in r_0\Lambda_0$ where $r_0=\gcd(r_1,r_2)$
and the lattice $\Lambda_0$ is endowed with the quadratic form $\kappa_{0,ab}$.
Furthermore, let $\rho_1^a$ and $\rho_2^a$ be any two integer valued vectors satisfying 
\be 
\kappa_{0,ab}\(r_1\rho_1^b+r_2\rho_2^b\)=\Delta\mu_a+r_1r_2\kappa_{0,ab}p_0^b.
\ee 
Then the general solution to \eqref{constr-eps} is given by
\be
\eps_1^a=\rho_1^a+ \frac{r_2}{r_0}\,\eps^a,
\qquad
\eps_2^a=\rho_2^a-\frac{r_1}{r_0}\,\eps^a,
\qquad 
\eps^a\in \IZ.
\ee 
Substituting this result into the spectral flow decomposition of $q_{i,a}$, 
it is easy to compute  
\be
\gamma_{12}=r_0 p_0^a k_a,
\qquad
Q_2(\hgam_i,\hgam_2)= 
-\frac{k^2}{\rr_{12}}\, ,
\ee
where we denoted $\rr_{12}=rr_1r_2/r_0^2$ and
\be
\begin{split} 
k_a(\eps)=&\, \rr_{12}\kappa_{0,ab}\eps^b+\mu_{12,a},
\qquad
k^2=\kappa_0^{ab}k_a k_b,
\\
\mu_{12,a}=&\, \frac{1}{r_0}\(r_2\mu_{1,a}-r_1\mu_{2,a}
+r_1r_2\kappa_{0,ab}\(\rho_1^b-\rho_2^b +\hf\, (r_1-r_2)p_0^b\)\).
\end{split}
\label{def-param12}
\ee
As a result, the holomorphic anomaly takes the form
\be
\p_{\bar\tau}\whh_{rp_0,\mu}=\frac{1}{8\pi\I(2\tau_2)^{3/2}}
\sum_{r_1+r_2=r}r_0\sqrt{\rr_{12}}\sum_{\mu_1,\mu_2}\delta_{\Delta\mu\in r_0\Lambda_0}\, 
\vth^{(r_1,r_2)}_{p_0,\mu_{12}}\,
\whh_{r_1p_0,\mu_1}\whh_{r_2p_0,\mu_2},
\label{anom-collin}
\ee
where $\vth^{(r_1,r_2)}_{v,\mu}$ is a theta series similar to 
(the complex conjugate of) the Siegel theta series \eqref{def-vthSp}: 
\be 
\vth^{(r_1,r_2)}_{v,\mu}(\tau,\btau)=(-1)^{r_0p_0^a\mu_a}\sqrt{v^2}\sum_{k\in \rr_{12}\Lambda_0+\mu} 
e^{-2\pi\tau_2 \,\frac{k_v^2}{ \rr_{12}} -\pi\I\tau\, \frac{k^2}{\rr_{12}}}
\label{def-vth12}
\ee 
and we also used the notation $k_v=v^a k_a/\sqrt{v^2}$.

Now we are ready to study what happens if $p_0^2=(p_0^3)=0$.
It is clear that it is sufficient to analyze only the theta series \eqref{def-vth12}.
Since at $v=p_0$ it is ambiguous, we define it at this point through the following limit
\be 
\vth^{(r_1,r_2)}_{p_0,\mu}:=\lim_{\veps\to 0}\vth^{(r_1,r_2)}_{p_\veps,\mu}
\ee 
where $p_\veps=p_0+\veps v_1$ and $v_1$ is any vector with a positive scalar product $p_0\cdot v_1\equiv \sp >0$.
Our goal will be to evaluate this limit explicitly.

To this end, it is convenient first to factorize the theta series \eqref{def-vth12} into a product of two
theta series: one defined by the two-dimensional lattice $\Latc_0=\Span\{p_0,v_1\}$
and the second defined by its orthogonal complement in $\Lambda_0$ which we denote $\Latp_0$.
Since $\Lambda_0$ has signature $(1,b_2-1)$ and $p_0$ is a null vector, it is clear that $\Latc_0$ has signature $(1,1)$,
while $\Latp_0$ has signature $(0,b_2-2)$.
The idea behind this factorization is that only the first theta series will carry a non-holomorphic dependence
so that we reduce our problem to a two-dimensional one. 
(Of course, if $\Lambda_0$ is two-dimensional, $\Latp_0$ does not arise.)
 
The problem however is that generically $\Latc_0\oplus\Latp_0$ does {\it not} coincide with $\Lambda_0$:
not all elements of $\Lambda_0$ can be obtained by linear combinations with integer coefficients of
the elements of $\Latc_0$ and $\Latp_0$. 
A way to deal with this problem is explained in appendix \ref{ap-fact}:
one should introduce the so called glue vectors $\glueg_\Asf$, $\Asf=0,\dots,n_g-1$, where
the number of the glue  vectors is determined by the cardinalities of the discriminant groups \eqref{Ng}.
Due to this, since $\det\Latc_0=\sp^2$, from practical reasons 
it is convenient to choose $v_1$ that minimizes the scalar product $\sp$.

Applying the factorization formula \eqref{factortheta-gen} to our case, one finds
\be 
\vth^{(r_1,r_2)}_{v,\mu}(\tau,\btau)=
\sum_{\Asf=0}^{n_g-1}\vth^{\parallel}_{v,\mupr+\rr_{12}\gluegl_\Asf}(\tau,\btau)\,
\vth^{\perp}_{\mu^\perp+\rr_{12}\gluegp_\Asf}(\tau),
\label{def-vth12-fact}
\ee 	
where the superscripts $^\parallel$ and $^\perp$ denote the projections on $\Latc$ and $\Latp$, respectively, and 
\bea 
\vth^{\parallel}_{v,\mu}(\tau,\btau)&=&(-1)^{r_0p_0^a\mu_{a}}\sqrt{v^2}\sum_{k\in \rr_{12}\Latc_0+\mu} 
e^{-2\pi\tau_2 \,\frac{k_v^2}{ \rr_{12}} -\pi\I\tau\, \frac{k^2}{\rr_{12}}},
\label{vth12paral}
\\
\vth^{\perp}_{\mu}(\tau) &=&\sum_{k\in \rr_{12}\Latp_0+\mu} e^{-\pi\I\tau\, \frac{k^2}{\rr_{12}}}.
\label{vth12perp}
\eea	
Representing $\Latc_0=\{\ell_0 p_0+\ell_1 v_1,\ \ell_0,\ell_1\in\IZ\}$ and expanding $\mu=\mu_0p_0+\mu_1v_1$, 
one obtains 
\be  
k_{p_\veps}^2=\frac{((\rr_{12}\ell_1+\mu_1)(\sp+\veps v_1^2)+\veps \sp(\rr_{12}\ell_0+\mu_0))^2}{2\veps \sp+\veps^2 v_1^2}\, .
\ee 
Therefore, for $v=p_\veps$, the summand in \eqref{vth12paral} is exponentially vanishing in the small $\veps$ limit unless 
$\ell_1=-\mu_1/\rr_{12}$.
In particular, this implies that $\mu_1$ must belong to $\rr_{12}\IZ$.
Taking into account that under this restriction on $\ell_1$ one has $k^2=0$, 
in the leading order we remain with
\be
\vth^{\parallel}_{p_\veps,\mu}\approx
\delta^{(\xi\rr_{12})}_{p_0\cdot \mu}\,
\sqrt{2\xi\veps}\sum_{\ell_0\in\IZ}
e^{-\frac{\pi \tau_2 \xi\veps}{\rr_{12}}\(\rr_{12} \ell_0+\mu_0\)^2} ,
\label{thetaanom-eps}
\ee
where $\delta^{(n)}_x$ is defined in \eqref{defdelta}.
Now one can replace the integer variable $\ell_0$ by $\sqrt{\veps}\ell_0$ so that in the limit $\veps\to 0$,
the summation becomes equivalent to an integral and we arrive at the following result
\be
\vth^{\parallel}_{p_0,\mu}=
\delta^{(\xi\rr_{12})}_{p_0\cdot \mu}\,
\sqrt{2\xi}\int_{\IR} \de x\,
e^{-\pi\tau_2 \xi\rr_{12} x^2}
=\delta^{(\xi\rr_{12})}_{p_0\cdot \mu}\,\sqrt{\frac{2}{\rr_{12}\tau_2}}\, .
\label{thetaanom-int}
\ee
Combining it with \eqref{anom-collin} and \eqref{def-vth12-fact}, one reproduces the formula \eqref{anom-null} 
given in the main text.

\section{Hirzebruch and del Pezzo surfaces}
\label{ap-surface}

The del Pezzo surface $\IB_m$ is the blow-up of $\IP^2$ over $m$ generic points.
It has $b_2(\IB_m)=m+1$ and a basis of $H_2(\IB_m,\IZ)$ is given by the hyperplane class of $\IP^2$ and
the exceptional divisors of the blow-up denoted, respectively, by $\cD_1$ and $\cD_2,\dots,\cD_{m+1}$.
In this basis the intersection matrix and the first Chern class are given by
\be
C_{\alpha\beta}= \diag (1,-1,\dots,-1),
\qquad
c_1(\IB_m) = 3\cD_1 -\sum_{\alpha=2}^{m+1}\cD_\alpha.
\label{dataBk}
\ee

The Hirzebruch surface $\IF_m$ is a projectivization of the $\cO(m)\oplus \cO(0)$ bundle over $\IP^1$.
It has $b_2(\IF_m)=2$ and in the basis given by the curves corresponding to the fiber $[f]$ and the section of the bundle $[s]$,
the intersection matrix and the first Chern class are the following
\be
C_{\alpha\beta} = \begin{pmatrix} 0 & 1 \\ 1 & -m \end{pmatrix},
\qquad
c_1(\IF_m) = (m+2)[f]+2[s].
\label{dataFk}
\ee
In fact, $\IF_1=\IB_1$, i.e. it is the blow-up of $\IP^2$ at one point,
and by changing the basis to
\be
\cD_1=[f]+[s],
\qquad
\cD_2=[s],
\label{baseF1}
\ee
one brings $C_{\alpha\beta}$ and $c_1(\IF_1)$ to the form \eqref{dataBk} with $m=1$.

Finally, note that, for all above surfaces as well as for $\IP^2$, the signature of the intersection matrix is $(1,b_2(S)-1)$ and
\be
c_1^2(S)=10-b_2(S).
\label{c1b2}
\ee

\isPreprints{}{
\begin{adjustwidth}{-\extralength}{0cm}
} 

\reftitle{References}




\isPreprints{}{
\end{adjustwidth}
} 
\end{document}